\documentclass[12pt,a4paper]{article}
\usepackage{jheppub}
\usepackage{multirow}
\usepackage{chemarrow}
\usepackage{amsmath,amssymb,amsfonts,mathtools}
\usepackage{float}
\usepackage{subfig}
\usepackage{scalefnt}
\usepackage{wrapfig}
\usepackage{fancybox}
\usepackage{ifsym}
\newcommand{\bea}{\begin{eqnarray}\displaystyle}
\newcommand{\eea}{\end{eqnarray}}

\newlength{\arrow}
{\setlength{\fboxsep}{15pt}
\setlength{\mylength}{\linewidth}%
\addtolength{\mylength}{-2\fboxsep}%
\addtolength{\mylength}{-2\fboxrule}%
\Sbox
\minipage{\mylength}%
\setlength{\abovedisplayskip}{0pt}%
\setlength{\belowdisplayskip}{0pt}%
\equation}%
{\endequation\endminipage\endSbox
\[\fbox{\TheSbox}\]}

\usepackage{epsfig}
\usepackage{tikz}
\usetikzlibrary{decorations.pathmorphing}

\usepackage{float}

\def\2{{1\over2}}
\let\<=\langle \let\>=\rangle
\def\new#1\endnew{{\bf #1}}
\def\ifundefined#1{\expandafter\ifx\csname#1\endcsname\relax}
\ifundefined{draftmode}\else    \input draftmode        \fi
\let\Msize=\footnotesize             
\def\BM{\Msize\begin{matrix}}           \def\EM{\end{matrix}}
\def\MN M:#1 #2 N:#3 #4 {{(#1_{#2},#3_{#4})}}
\def\MNH M:#1 #2 N:#3 #4 H:#5,#6 [#7]{{(#1_{#2},#3_{#4})^{#5,#6}_{#7}}}

\def\dd{\mathrm{d}}

\DeclareMathOperator{\ch}{c}

\newcommand{\pmi}{\phantom{-}}

\newcommand{\be}{\begin{equation}}
\newcommand{\ee}{\end{equation}}

\title{Strings of Minimal 6d SCFTs}
\author[\ast]{Babak Haghighat,}
\author[\dag]{Albrecht Klemm,}
\author[\ast]{Guglielmo Lockhart,}
\author[\ast]{Cumrun Vafa}
\affiliation[\ast]{Jefferson Physical Laboratory, Harvard University, Cambridge, MA 02138, USA}
\affiliation[\dag]{Bethe Center for Theoretical Physics, Universit\"at Bonn, Nussallee 12, D-53115 Bonn}

\abstract{
We study strings associated with minimal 6d SCFTs, which by definition have only one 
string charge and no Higgs branch. These theories are labelled by a number $n$ 
with $1\leq n\leq 8$ or $n=12$. Quiver theories have previously been proposed which 
describe strings of SCFTs for $n=1,2$. For $n>2$ the strings interact with the bulk 
gauge symmetry. In this paper we find a quiver description for the $n=4$ string using  
Sen's limit of F-theory and calculate its elliptic genus with localization techniques. 
This result is checked using the duality of F-theory with M-theory and topological string 
theory whose refined BPS partition function captures the elliptic genus of the SCFT strings. 
We use the topological string theory to gain insight into the elliptic genus for other 
values of $n$.}

\begin{document}
\maketitle


\section{Introduction}
Six-dimensional superconformal theories have light strings as their basic building blocks.  One approach
to a better understanding of these theories involves unlocking the mysteries associated with these strings.  In particular
one would like to describe the free propagation of such strings and the degrees of freedom on their worldsheet.
Recently, many advances have been made in our understanding of 6d SCFTs \cite{Heckman:2013pva,Gaiotto:2014lca,DelZotto:2014hpa} including in many cases an effective
description of their strings' worldsheet QFT \cite{Haghighat:2013gba,Haghighat:2013tka,Kim:2014dza}.  The goal of this paper is to study the strings associated with minimal 6d SCFTs.  
These are $(1,0)$ SCFTs in six dimensions which have only one string charge (i.e. a one dimensional tensor branch), and are non-Higgsable.  They are labelled by an integer $2\leq n \leq 12$
excluding $n=9,10,11$, and are realized within F-theory as elliptic fibrations over a base  ${\cal O}(-n) \rightarrow \mathbb{P}^1$ \cite{Morrison:2012np}. It is also natural to include the $n=1$ case here although strictly speaking it is not part of the non-Higgsable family.
The cases $n=2,3,4,6,8,12$ also arise in the F-theory context as simple orbifolds of $ \mathbb{C}^2\times T^2 $ \cite{Witten:1996qb} where we 
rotate each plane of $\mathbb{C}^2$ by an $n$-th root of unity $\omega$ and compensate by rotating the elliptic fiber by $\omega^{-2}$.

 For the $ n = 1 $ and $ n = 2 $ cases,  quivers have been found which describe the worldsheet dynamics of the corresponding strings \cite{Haghighat:2013gba,Haghighat:2013tka,Kim:2014dza}. 
 The $n=1$ case corresponds to the exceptional CFT with $E_8$ global symmetry describing an M5 brane near the M9  boundary wall \cite{Witten:1995gx,Ganor:1996mu,Klemm:1996hh,Seiberg:1996vs}; 
 the $n=2$ case, on the other hand, corresponds to the $(2,0)$ SCFT of type $A_1$.  In this paper we extend this list by finding the quiver for the $n=4$ case.  
 This is one of the orbifold cases for which the elliptic fiber can have arbitrary complex modulus $\tau$, as the only symmetry required in the fiber is $\mathbb{Z}_2$, 
 which does not fix the modulus of the torus. To find the quiver describing the strings of this theory, we use Sen's limit of F-theory \cite{Sen:1997gv}, 
 which corresponds to taking the modulus of the torus $\tau_2 \gg 1$. Following this approach we are able in particular to compute the elliptic genus of these strings, 
 which we do explicitly for the first few string numbers. 
 
If we compactify the theory on a circle, the elliptic genus computes the BPS degeneracies of the wrapped strings. 
Following the duality between F-theory and M-theory and the relation between M-theory, BPS counting in five dimensions, and topological strings, we
find that the elliptic genera are encoded in the topological string partition function defined on the corresponding elliptically-fibered Calabi-Yau, 
similar to the observation for the $n=1$ case in~\cite{Klemm:1996hh}. Within topological string theory the genus zero BPS invariants 
can be easily calculated using mirror symmetry even for high degree $k$ in the base, which corresponds to $k$ times wrapped strings. 
However, since the boundary conditions are only known to some extent~\cite{Huang:2013yta}, the higher genus theory cannot be completely solved with the 
generalized holomorphic anomaly equations; on the other hand, the elliptic genus computation provides the all genus answer. 
In particular we can use this relation to successfully check our answer for the elliptic genus of the $n=4$ strings.

For the other values of $n$, no worldsheet description of the associated strings is known.  For these cases we employ topological string 
techniques to obtain BPS invariants of the corresponding geometry, which can be related to an expansion of the elliptic genus of small 
numbers of strings for specific values of fugacities.

The organization of this paper is as follows: In Section \ref{sec:minimalSCFTs} we review the classification of minimal 6d SCFTs 
and their F-theory realization. We also review the quivers describing the worldsheet dynamics of the strings of the $n=1$ and $n=2$ models. 
In Section \ref{sec:D4Quiver} we derive the quiver for the strings of the $ n = 4 $ theory by exploiting its orbifold realization. Furthermore, using 
the quiver we obtain an integral expression for the elliptic genus of $k$ strings which we evaluate explicitly for the cases $k=1$ and $k=2$. We 
then discuss how one can extract from this the BPS degeneracies associated to the strings.
In Section \ref{sec:CYgeometry} we construct explicitly the elliptic Calabi-Yau manifolds corresponding to $n=1,\dots,12$ as hypersurfaces in 
toric ambient spaces, solve the topological string theory and calculate the genus zero BPS invariants associated 
to these Calabi-Yau manifolds, from which one can obtain BPS degeneracies associated to the strings. In Appendix \ref{LG}  we give a description of the local mirror 
geometry for some of these elliptic Calabi-Yau threefolds in terms 
of non-compact Landau-Ginzburg models.

\section{Minimal SCFTs in six-dimensions}
\label{sec:minimalSCFTs}

Six-dimensional SCFTs can be classified in the context of F-theory by considering compactifications on an elliptically fibered Calabi-Yau threefold $X$ with non-compact base $B$. In the case where all fiber components are blown down the fibration $\pi : X \rightarrow B$ can be described in terms of the Weierstrass form
\begin{equation}
	y^2 = x^3 + fx + g,
\end{equation}
where $f$ and $g$ are sections of the line bundles $\mathcal{O}(-4 K_B)$ and $\mathcal{O}(-6 K_B)$. The discriminant locus, along which the elliptic fibers are singular, is a section of $\mathcal{O}(-12 K_B)$ and has the following form:
\begin{equation}
	\Delta = 4 f^3 + 27 g^2.
\end{equation}
The discriminant locus corresponds to the location of seven-branes in the system. More precisely, each component of the discriminant locus is identified with a seven-brane wrapping a divisor $\Sigma \subset B$. Each seven-brane supports a gauge algebra $g_{\Sigma}$ which is determined by the singularity type of the elliptic fiber along $\Sigma$ \cite{Morrison:1996na,Morrison:1996pp}.

In the maximally Higgsed phase (that is, when all hypermultiplet vevs that can be set to non-zero value are turned on) one can classify the resulting models in terms of the base geometry $B$ only \cite{Morrison:2012np}. Non-Higgsability requires that the divisor $\Sigma \subset B$ be rigid. This implies that $ \Sigma $ must be a $\mathbb{P}^1$ curve with self-intersection $-n < 0$ for a positive number $n$ (in the following we will refer to this as a $ (- n) $ curve), and the local geometry is the bundle $\mathcal{O}(-n) \rightarrow \mathbb{P}^1$. Furthermore, it can be shown that $n$ is only allowed to take the values $1 \leq n \leq 8$ or  $n = 12$ \cite{  Morrison:1996na,Morrison:1996pp,Morrison:2012np}. In the $n=1$ case, corresponding to the E-string $(1,0)$ SCFT \cite{Witten:1995gx,Ganor:1996mu,Seiberg:1996vs}, the discriminant vanishes along the non-compact fiber over isolated points on the $\mathbb{P}^1$. In this case instead of a gauge symmetry one finds an $E_8$ global symmetry.
In the $n=2$ case the fiber is everywhere non-singular, and one finds the $ A_1 $ $(2,0)$ SCFT which corresponds to the world-volume theory of M5 branes in flat space. For $n>2$, the seven-branes wrap the compact $ \mathbb{P}^1 $, and therefore the 6d SCFT has non-trivial gauge symmetry. In the non-Higgsable case this gauge symmetry is completely determined by the integer $n$. We summarize the list of possibilities in the following table:

\begin{center}
	\begin{tabular}{|c|c|c|c|c|c|c|c|}
		\hline
		\textrm{7-brane} & 3 & 4 & 5 & 6 & 7 & 8 & 12 \\
		\hline
		$g_{\Sigma}$ & $SU(3)$ & $SO(8)$ & $F_4$ & $E_6$ & $E_7$ & $E_7$ & $E_8$ \\
		\hline
		Hyper & -- & -- & -- & -- & $\frac{1}{2}\mathbf{56}$ & -- & -- \\
		\hline
	 \end{tabular}
\end{center}

In the $n=7$ case, one finds that in addition to $E_7$ gauge symmetry the 6d theory also contains a half-hypermultiplet. The cases $n=9$, $n=10$ and $n=11$ lead to $E_8$ gauge symmetry but additionally contain ``small instantons"; these cases can be reduced to chains of the more fundamental geometries summarized in the table, as discussed in \cite{Heckman:2013pva}.

These geometries (excluding the cases $n=1,5,7$) can equivalently be realized as orbifolds of the form $(T^2\times \mathbb{C}^2)/\mathbb{Z}_n$, $n=2,3,4,6,8,12$. Here, $\mathbb{Z}_n$ acts on the $z_i$, $i=1,2,3,$  coordinates of $T^2$ and 
$\mathbb{C}^2$ as 
\begin{equation}
\left( \begin{array}{c}
z_1\\
z_2\\
z_3 \end{array}\right)\mapsto
\left( \begin{array}{ccc}
\omega^{-2}&&\\
& \omega & \\
&& \quad\omega  \end{array}\right)
\left( \begin{array}{c}
z_1\\
z_2\\
z_3 \end{array}\right)\ ,
\label{Z3}
\end{equation}
 with $\omega^n=1$. This construction will be in particular useful when we study the $ n = 4 $ SCFT, as it will enable us to find a weak coupling description for the corresponding model.
 
\subsection{Strings of the $\mathcal{O}(-1)\rightarrow \mathbb{P}^1$ and $\mathcal{O}(-2)\rightarrow \mathbb{P}^1$ models}
 
\begin{figure}[h!]
  \centering
	\includegraphics[width=0.8\textwidth]{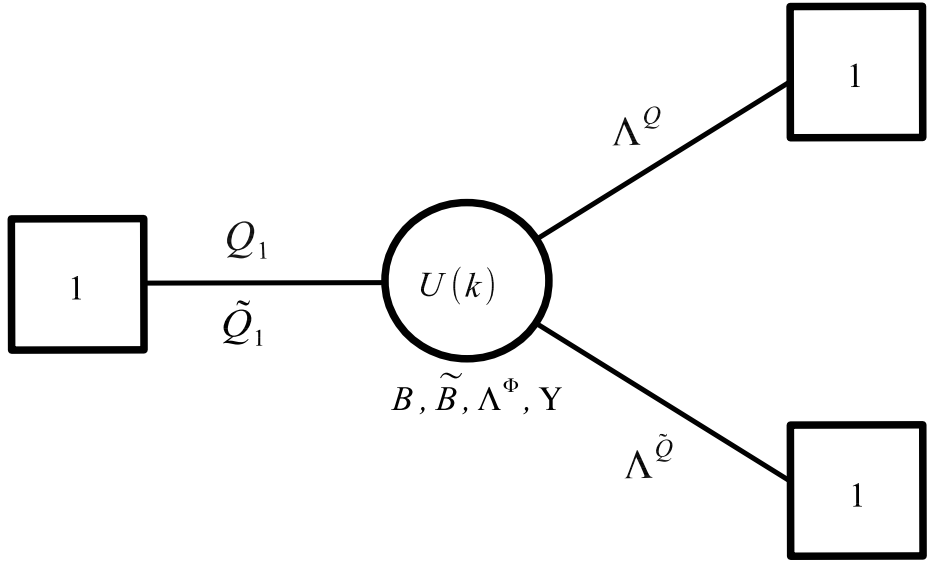}
  \caption{The quiver describing the $\mathcal{O}(-2) \rightarrow \mathbb{P}^1$ strings.}
  \label{fig:O(-2)quiver}
\end{figure}

\begin{figure}[ht]
  \centering
	\includegraphics[width=0.6\textwidth]{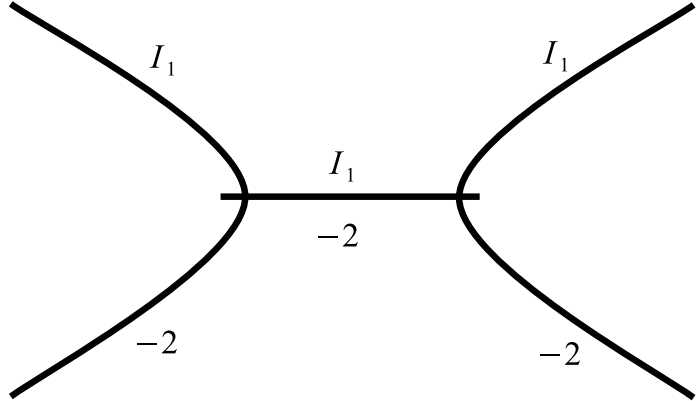}
  \caption{Configuration of $ (-2) $ curves that gives rise to the $ O(-2)\to \mathbb{P}^1 $ local geometry. We have also indicated the degeneration of the elliptic fiber over each curve that gives rise to the M-string geometry.}
    \label{fig:O(-2)branes}
\end{figure}

Let us next discuss the strings that appear on the tensor branch of  6d SCFTs. From the point of view of F-theory these strings arise from D3 branes which wrap the $\mathbb{P}^1$ curve in the base $B$ in the limit of small $ \mathbb{P}^1 $ size. Let us first review the `M-strings' that arise in the $ n = 2 $ case. Since in this case the orbifold acts trivially on the torus, its modulus $ \tau $ can be taken to be arbitrary, and in particular one can take the weak coupling limit $ \tau \to i\infty $ and study this system from the point of view of Type IIB string theory compactified on $ B $. It turns out \cite{Haghighat:2013gba, Haghighat:2013tka} that the dynamics of M-strings are captured  by the two-dimensional quiver gauge theory depicted in Figure \ref{fig:O(-2)quiver}. For $ k $ strings, this quiver describes a two-dimensional  $ \mathcal{N} = (4,0)$ theory with gauge group $ U(k) $ and the following $ (2,0) $ field content:  $Q$ and $\widetilde Q$ are chiral multiplets in the fundamental representation of $U(k)$, while $\Lambda^Q$ and $\Lambda^{\widetilde Q}$ are fundamental Fermi multiplets. Furthermore, the Fermi multiplet $\Lambda^{\phi}$ and vector multiplet $\Upsilon$ combine into a  $(4, 0)$ vector multiplet, and the adjoint chiral multiplets $ B,\widetilde B $ combine into a $ (4, 0) $ hypermultiplet. One intuitive way to see how this comes about is to look at the  configuration of $(-2)$ curves which captures the local geometry $ \mathcal{O}(-2)\to\mathbb{P}^1 $ \cite{DelZotto:2014hpa} and is pictured in Figure \ref{fig:O(-2)branes}.

The left and right $(-2)$ curves are non-compact, whereas the curve in the middle is a compact $\mathbb{P}^1$. Choosing the elliptic fiber to be trivial would lead upon circle compactification to $ U(2) $ $ \mathcal{N} = 4 $ gauge theory; it is in fact possible to deform this theory to $ \mathcal{N}=2^* $ by letting the elliptic fiber degenerate over each curve to an $I_1$ singularity (that is, by wrapping a D7 brane over each curve). D3 branes wrapping the compact $ (-2) $ curve give rise to the strings of the resulting 6d SCFT, and upon circle compactification their BPS degeneracies then capture the BPS particle content of the $ U(2) $ $ \mathcal{N}=2^* $ theory. It is easy to understand how the field content of the quiver in Figure \ref{fig:O(-2)quiver} arises from strings that end on the D3 branes: D3-D3 strings give rise to a $ (4,4) $ vector multiplet in the adjoint of $ U(k) $ consisting of the (2, 0) multiplets $ \Upsilon, \Lambda^\Phi, B, \tilde B $; strings stretching from the D3 branes to the D7 brane wrapping the same compact $\mathbb{P}^1$ give rise to the chiral multiplets $Q$ and $\tilde Q$; finally, strings stretching between the D3 branes and the D7 branes that wrap the non-compact $(-2)$ curves give rise to the Fermi multiplets $\Lambda^Q$ and $\Lambda^{\widetilde Q}$. Whether D3-D7 strings give rise to chiral or Fermi multiplets is determined by the number of dimensions that are not shared by the D3 and D7 branes (four for the D3-D7 strings leading to $ Q,\widetilde Q $, eight for the ones leading to $ \Lambda^Q,\Lambda^{\widetilde Q} $).

\begin{figure}[t]
  \centering
	\includegraphics[width=0.8\textwidth]{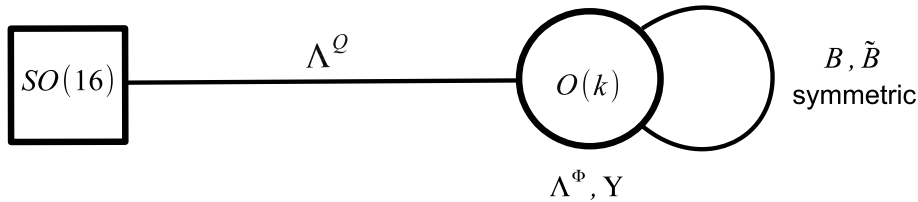}
  \caption{The quiver for $\mathcal{O}(-1) \rightarrow \mathbb{P}^1$ strings.}
  \label{fig:O(-1)quiver}
\end{figure}

Recently, a quiver gauge theory was also found that describes the dynamics of E-strings, corresponding to the $\mathcal{O}(-1) \rightarrow \mathbb{P}^1$ case \cite{Kim:2014dza}. In terms of $ (2,0) $ multiplets, the theory of $ k $ E-strings was found to have the following field content: a vector multiplet $ \Upsilon $ and a Fermi multiplet $ \Lambda^\phi $ in the adjoint representation of $ O(k) $, two chiral multiplets $ B,\tilde B $ in the symmetric representation of $ O(k) $, and a Fermi multiplet $ \Lambda^Q $ in the bifundamental representation of $ O(k) $ and of a $ SO(16) $ flavor group, which enhances to $ E_8 $ at the superconformal point. The relevant quiver is shown in Figure \ref{fig:O(-1)quiver}.

\subsection{From strings of 6d SCFTs to topological strings}

In cases where a quiver gauge theory description is available for the strings of minimal six-dimensional SCFTs, one can use the methods of \cite{Gadde:2013dda, Benini:2013nda, Benini:2013xpa} to compute the elliptic genus for an arbitrary number of strings. The elliptic genus will depend on the complex structure $ \tau $ of the torus as well as a number of fugacities corresponding to various $ U(1) $ symmetries enjoyed by the two-dimensional quiver theory. In particular, it will always depend on two parameters $ \epsilon_1, \epsilon_2 $ that correspond to rotating the $ \mathbb{C}^2 $ transverse to the strings' worldsheet in the six-dimensional worldvolume of the SCFT. In addition to this, the elliptic genus will depend on a number of fugacities $m_i$ parametrizing the Cartan of the flavor symmetry group of the worldsheet theory. In the F-theory picture these fugacities correspond to K\"ahler parameters of the resolved elliptic fiber of the Calabi-Yau.

The elliptic genus encodes detailed information about the spectrum of the strings. Being able to reproduce this information with an alternative method is therefore an important check of the validity of the quiver theory. This can be achieved by exploiting duality between F-theory and M-theory \cite{Vafa:1996xn,Morrison:1996na}, and in particular the relation between D3 branes on one side and M2 branes on the other \cite{Klemm:1996hh}. This duality relates F-theory on $ X\times T^2\times \mathbb{R}^4 $ (where $ X $ is an elliptically fibered Calabi-Yau threefold) to M-theory on $ X \times S^1\times\mathbb{R}^4 $; under this duality the complex structure $ \tau $ of the $ T^2 $ gets mapped to the K\"ahler parameter of the elliptic fiber of $ X $ on the M-theory side.  D3 branes wrapping the base $ \mathbb{P}^1 $ as well as $ T^2 $ correspond to strings wrapped on the torus. It turns out \cite{Klemm:1996hh} that the BPS states of a configuration of $ k $ strings with $ m $ units of momentum along a circle get mapped to BPS M2 branes wrapping the base $ \mathbb{P}^1 $ $ k $ times and the elliptic fiber $ m $ times; furthermore, if a string BPS state has nonzero flavor symmetry charges, the corresponding BPS M2 brane will also wrap additional curves in $ X $.

The precise relation between the counting of BPS states on the two sides turns out to be \cite{Haghighat:2013gba}:
\begin{equation}
	Z^{\textrm{top}}(X,\epsilon_1,\epsilon_2, \tau, t_b, m_i) = Z_0(\tau,\epsilon_1,\epsilon_2,m_i) \left(1+ \sum_{k=1}^{\infty} Q^k Z_k(\epsilon_1,\epsilon_2,\tau,m_i)\right),
\end{equation}
where $ Z^{\textrm{top}}(\epsilon_1,\epsilon_2, \tau, t_b, m_i) $ is the topological string partition function that counts BPS configurations of M2 branes on the M-theory side (or, equivalently, 5d BPS states of the theory arising from M-theory compactification on $ X $), and $ Z_k $ is the elliptic genus of $ k $ strings of the six-dimensional SCFT. Furthermore $t_b$ is the K\"ahler class of the base $\mathbb{P}^1$ and $Q$ is proportional to $e^{-t_b}=Q_b$\footnote{The proportionality factor will be a combination of K\"ahler classes in the resolution of the elliptic fiber and its exact form can be determined by requiring $Q$ to be invariant under the monodromy associated to $SL(2,\mathbb{Z})$, as in \cite{Candelas:1994hw}.}. In other words, the topological string partition function is given by a sum over elliptic genera of the six-dimensional strings, except for a simple piece $ Z_0 $ which captures contributions coming from vector multiplets and can be obtained straightforwardly.

In the next section we will discuss the case of the $ \mathcal{O}(-4) $ theory and determine the quiver describing its strings. Furthermore, we will find an integral expression for the elliptic genera of these strings; we will evaluate these integrals explicitly for one and two strings and present an answer in a form from which BPS degeneracies may be readily extracted. In Section \ref{sec:O(-4)} we will compute the topological string partition function of the corresponding Calabi-Yau geometry and extract BPS invariants which can be shown to agree with the elliptic genus computations.

\section{Quiver for the $\mathcal{O}(-4) \rightarrow \mathbb{P}^1$ model}
\label{sec:D4Quiver}

We now turn to the strings of the $ n = 4 $ (1,0) SCFT in 6d and construct a quiver theory that describes their dynamics. Recall that the six-dimensional theory is obtained by compactifying F-theory on the following orbifold geometry:
\begin{equation}
	CY_3 = \left(T^2 \times \mathbb{C} \times \mathbb{C}\right) / \mathbb{Z}_4,
\end{equation}
where the orbifold action $\mathbb{Z}_4$ on the complex coordinates $ (z_1,z_2,z_3) $ of $ T^2\times \mathbb{C}\times \mathbb{C} $ is given by:
\begin{equation}
	\left(\begin{array}{c}z_1\\z_1\\z_3\end{array}\right) \mapsto \left( \begin{array}{ccc}
\omega^{-2}&&\\
& \omega & \\
&& \quad\omega  \end{array}\right)
 \left(\begin{array}{c}z_1\\z_2\\z_3\end{array}\right),
\end{equation}
and $\omega = i$. To obtain a F-theory construction in terms of a non-compact elliptic Calabi-Yau one has to first blow up the singularity at $\mathbb{C}^2/\mathbb{Z}_4$. The resulting space is described by the bundle
\begin{equation}
	\mathcal{O}(-4) \longrightarrow \mathbb{P}^1,
\end{equation}
with the singular elliptic fiber $T^2/\mathbb{Z}_2$ over the $\mathbb{P}^1$ base. The resolution of this fiber leads to the $I_0^*$ fiber in the Kodaira classification of elliptic fibrations. In fact, one can obtain an infinite family of six-dimensional theories by taking the singular fiber to be of type $ I^*_{p} $, with $ p \geq 0 $. Lowering $ p $ corresponds in physical terms to Higgsing. This geometry can be equivalently viewed in the weak coupling limit as a type IIB orientifold of the $\mathbb{C}^2/\mathbb{Z}_2$ singularity \cite{Sen:1997gv}. In this limit the singular elliptic fiber over $\mathbb{P}^1$ can be interpreted as the presence of $4+p$ D7-branes wrapping the $\mathbb{P}^1$ together with an orientifold 7-plane. This gives rise to a $\mathcal{N}= (1,0)$ $SO(8+2p)$ gauge theory in the six non-compact directions parallel to the branes. Furthermore, D3-branes wrapping the $\mathbb{P}^1$ give rise to strings in the six-dimensional theory. 

In the following we study the worldsheet theory of these strings and obtain a quiver gauge theory description for it. The particular orientifold we are interested in has been studied in some detail in \cite{Uranga:1999mb} and we shall describe it here briefly. The theory we want to study is type IIB theory on $\mathbb{C}^2/\mathbb{Z}_2$, modded out by $\Omega \Pi$, where $\Omega$ is world-sheet parity and $\Pi$ acts as
\begin{equation}
	\Pi : z_1 \rightarrow z_2, \quad z_2 \rightarrow - z_1,
\end{equation}
with $z_1$, $z_2$ parametrizing the two complex planes in $\mathbb{C}^2/\mathbb{Z}_2$. The D7-branes wrapping the $\mathbb{P}^1$ can, in the singular limit,be thought of as D5-branes probing $\mathbb{C}^2/\mathbb{Z}_2$ together with an orientifold 5-plane at $z_1=z_2=0$. Similarly, D3-branes become D1-branes whose worldvolume theory we wish to determine. We start by describing the brane system probing $\mathbb{C}^2$ and successively add the $\mathbb{Z}_2$ orbifold and  $\mathbb{Z}_2$ orientifold actions. Before the orbifolding, the theory living on the D1-branes is a $\mathcal{N} = (4,4)$ $U(k)$ gauge theory with one adjoint and $N$ fundamental hypermultiplets, where $N$ denotes the number of D5-branes \cite{Haghighat:2013tka}. To summarize, we have the following massless field content on the worldvolume:
\begin{equation}
	\begin{array}{ccc}
		\textrm{multiplet} & \textrm{bosons} & \textrm{fermions} \\
		\hline
		 \textrm{vector} & b^{AY}, A_{\pm \pm} & \psi_-^{A' Y}, \psi_+^{A A'} \\
		 \textrm{adjoint hyper} & b^{A' \tilde A'} & \psi_-^{A \tilde A'}, \psi_+^{\tilde{A}' Y}\\
		 \textrm{fundamental hyper} & H^{A'} & \chi^A_-, \chi^Y_+
	\end{array}
\end{equation}
where the indices $(A' \tilde{A}')$ represent the fundamental representations of the two $SU(2)$ groups rotating the directions $X^2,X^3,X^4,X^5$ (the directions orthogonal to D1 but parallel to D5) while $(A,Y)$ are indices for the $SU(2)$'s rotating $X^6,X^7,X^8,X^9$ (the directions orthogonal both to D1 and D5). Next, we embed the $\mathbb{Z}_2$ orbifold action generated by
\begin{equation}
	\left(\begin{array}{cc}\omega & ~ \\ ~ & \omega^{-1}\end{array}\right), \quad \omega \in \mathbb{Z}_2
\end{equation}
into $SU(2)_Y$ and hence obtain the following action on fields with $Y$-index
\begin{equation} \label{eq:orbifold}
	(b^{AY}, \psi_+^{\tilde{A}' Y}, \psi_+^{\tilde{A}' Y}, \chi_+^Y) \mapsto (\omega b^{AY}, \omega\psi_-^{A' Y}, \omega \psi_+^{\tilde A' Y}, \omega \chi^Y_+).
\end{equation}
The resulting theory has $\mathcal{N} = (4,0)$ supersymmetry and its field content can equally well be described in terms of $\mathcal{N} = (2,0)$ chiral superfields $\Sigma$, $\Phi$, $B$, $\tilde B$, $Q$, $\tilde Q$, a $(2,0)$ gauge superfield $\Upsilon$, and $ (2, 0) $ Fermi superfields $\Lambda^\Phi,\Lambda^B,\Lambda^{\widetilde B},\Lambda^{Q},\Lambda^{\widetilde Q}$.
The decomposition of $ \mathcal{N} = $ (4,0) fields in terms of $ (2,0)  $ components is as follows:
\begin{equation}
	b^{A Y} \leftrightarrow (\Sigma, \Phi), \quad b^{A' \tilde{A}'} \leftrightarrow (B,\tilde B), \quad H^{A'} \leftrightarrow (Q,\tilde Q),
\end{equation}
\begin{equation}
	\psi_+^{AA'} \leftrightarrow (\Lambda^{\Phi},\Upsilon),\quad \psi_+^{\tilde{A}' Y} \leftrightarrow (\Lambda^B,\Lambda^{\tilde{B}}),\quad \chi_+^Y \leftrightarrow (\Lambda^Q, \Lambda^{\tilde{Q}}).
\end{equation}
Following \cite{Douglas:1996sw}, the theory one obtains after the orbifold (\ref{eq:orbifold}) is a quiver gauge theory whose gauge nodes correspond to the nodes of the affine $A_1$ quiver (for more general $\mathbb{Z}_{N}$ orbifolds one would obtain the affine $A_{N-1}$ quiver). The fields that do not carry a $Y$  index are localized at each node, while those with a $Y$ index connect adjacent nodes \cite{Haghighat:2013tka}.  In order to turn this D1 quiver into a D3 quiver one needs furthermore to turn off D1 brane charge and instead introduce D3 branes wrapped around blow-up cycles of the resolved $A_1$ singularity. This transformation corresponds to removing the last node of the inner quiver as well as all links ending on it. Correspondingly, in the case of the $A_1$ singularity which is of interest here, the single remaining $U(k)$ gauge node contains an $\mathcal{N}=(4,0)$ vector multiplet $(\Upsilon,\Lambda^{\phi}$) and an adjoint hypermultiplet $(B,\tilde B)$.

Next, we come to the orientifolding. Orientifolds of orbifolds were discussed in \cite{Uranga:1999mb}. There it was found that for $\mathbb{C}^2/\mathbb{Z}_n$ orbifolds the gauge group in the $\omega^l$-twisted ($l=0, \cdots, n-1$) D-brane sector is of $Sp$-type if $l$ is even and of $SO$-type if $l$ is odd. This implies for our case that we have an orthogonal gauge group on the D5-branes in the untwisted sector and a symplectic one on the D5-branes of the $\omega$-twisted sector. Furthermore, anomaly cancellation in six dimensions fixes the ranks of the gauge groups such that the allowed configurations are $SO(8+2p)\times Sp(p)$ \cite{Uranga:1999mb}. This corresponds to having $4+p$ D5-branes together with an O5-plane at the Orbifold singularity. Uplifting this to F-theory one finds that the $p=0$ case is obtained from the $I_0^*$ fiber while the $p>0$ cases come from $I_p^*$ fiber types. 

\begin{figure}[here!]
  \centering
	\includegraphics[width=0.6\textwidth]{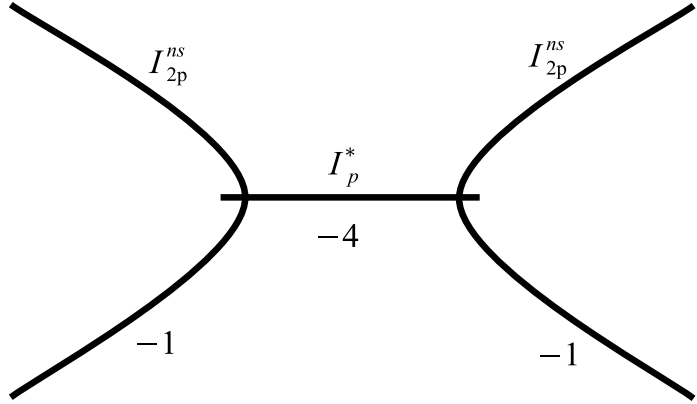}
  \caption{The local geometry that gives rise to the $SO(8+2p) $ 6d SCFT.}
  \label{fig:O(-4)branes}
\end{figure}

In fact, in the F-theory setup the six-dimensional theory has $SO(8+2p)$ gauge group and \textit{two} $Sp(p)$ \textit{flavor} nodes. The situation here is analogous to the $\mathcal{O}(-2) \rightarrow \mathbb{P}^1$ case: the two flavor nodes correspond to non-compact D7 branes intersecting the compact curve as shown in Figure \ref{fig:O(-4)branes} (see for example \cite{DelZotto:2014hpa} for more details about the geometry).

\begin{figure}[here!]
  \centering
	\includegraphics[width=0.8\textwidth]{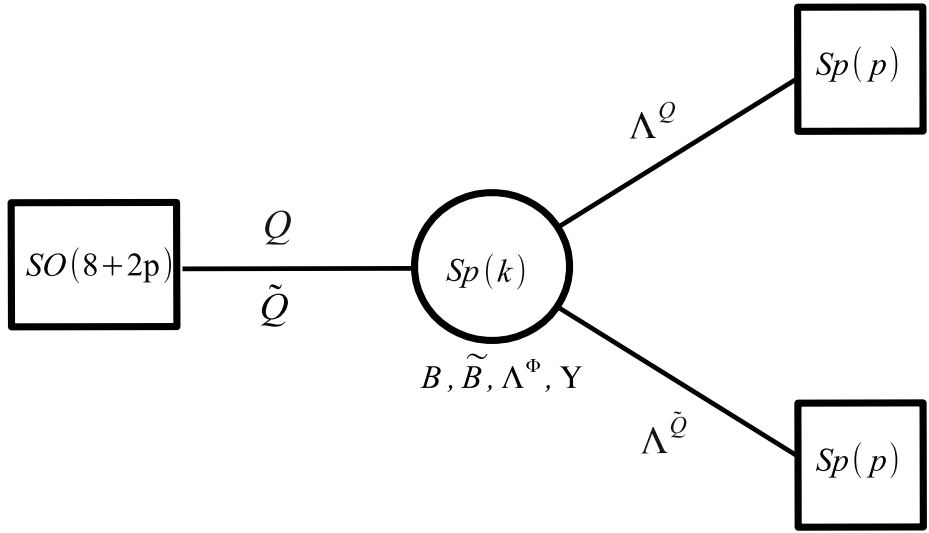}
  \caption{The $\mathbb{C}^2/\mathbb{Z}_4$ quiver.}
  \label{fig:C2modZ4}
\end{figure}

From the point of view of the two-dimensional theory living on the strings the $SO(8+2p)$ gauge node and the $Sp(p)$ flavor nodes descend to flavor nodes. Furthermore, orientifolding implies that $ (\Upsilon, \Lambda^\phi) $ transform in the symmetric (that is, adjoint)  representation of $Sp(k)$\footnote{Orientifolding amounts to projecting the gauge group from $U(2k)$ to $Sp(k)$.}, while $ (B,\widetilde B) $ transform in the antisymmetric representation \cite{Gimon:1996rq}. It is interesting to note that the introduction of two $Sp(p)$ nodes is also necessary from gauge anomaly cancellation in two dimensions which will be reviewed later.  The resulting two-dimensional quiver is the one depicted in Figure \ref{fig:C2modZ4}. 

The various fields in the quiver have different charges with respect to the two $U(1)_{\epsilon_1}\times U(1)_{\epsilon_2} \subset SO(4)$  that rotates the $X^2, X^3, X^4, X^5$ directions. We denote the fugacities by $ \epsilon_1,\epsilon_2 $, as they are the same parameters that appear in the Nekrasov partition function.
For completeness we also present the charges of the fields of the quiver under the different $U(1)$'s and gauge groups; these charges are obtained directly by the orbifolding construction, as in \cite{Haghighat:2013tka}.
\begin{center}
	\begin{tabular}{|c|c|c|c|c|c|c|c|}
	\hline
		~ & $\Lambda^{\Phi}$ & $B$ & $\tilde{B}$ & $Q$ & $\tilde{Q}$ & $\Lambda^Q$ & $\Lambda^{\tilde Q}$\\
		\hline
		$Sp(k)$ & symmetric & anti-symmetric & anti-symmetric & $\square$ & $\overline{\square}$ & $\square$ & $\overline{\square}$\\
		\hline
		$U(1)_{\epsilon_1}$ & $-1$ & $1$ & $0$ & $\frac{1}{2}$ & $\frac{1}{2}$ & $0$ & $0$\\
		\hline
		$U(1)_{\epsilon_2}$ & $-1$ & $0$ & $1$ & $\frac{1}{2}$ & $\frac{1}{2}$ & $0$ & $0$ \\
		\hline
	\end{tabular} 
\end{center} 
We have arrived at the conclusion that the theory for $ k $ strings is an $ Sp(k) $ gauge theory with a (2,0) vector multiplet $ \Upsilon $ and a Fermi multiplet $ \Lambda^\Phi $ in the adjoint (i.e. symmetric) representation, two chiral multiplets $ B,\widetilde B $ in the antisymmetric representation, and two chiral multiplets $Q, \widetilde Q $, each in the bifundamental representation of $ Sp(k) \times SO(8+2p) $. If $p > 0$ one also has Fermi multiplets $ \Lambda^Q,\Lambda^{\widetilde Q} $ in the bifundamental of $Sp(k) \times Sp(p)$. One can pick a basis $ \{e_i\}_{i=1}^k $of the weight lattice of $ Sp(k) $ in which the fundamental representation has weights $ \pm e_i \,(i=1,\dots,k) $. In this basis, the symmetric representation has weights $ e_i\pm e_j \, (\forall i, j) $, while the antisymmetric representation has weights $ e_i\pm e_j \, (i \neq j) $. We also pick $\, \pm m_i $ to be the Cartan parameters dual to the weights of the fundamental representation of $SO(8+2p)$, and $ \pm \mu_i $ and $ \pm \tilde\mu_i $ to be the Cartan parameters for the two $ Sp(p) $ flavor groups.

Let us next comment on gauge anomaly cancellation in two dimensions. The contribution of chiral fermions running in the loop to the anomaly is proportional to the index of their representation $T(R)$ defined as:
\begin{equation}
	\textrm{Tr}(T^a_R T^b_R) = T(R) \delta^{ab}.
\end{equation} 
Furthermore, left-moving fermions contribute with a positive sign to the anomaly while right-moving ones contribute with a negative sign. Thus, for our particular quiver we obtain the following result:
\begin{eqnarray}
	a(L) - a(R) & = & T_B(\textrm{anti-sym}) + T_{\tilde B}(\textrm{anti-sym}) + (n + 4) (T_{Q}(\square) + T_{\tilde Q}(\square) \nonumber \\
	~              & ~ & - T_{\Upsilon}(\textrm{sym}) - T_{\Lambda^{\phi}}(\textrm{sym}) - n (T_{\Lambda^Q}(\square) + T_{\Lambda^{\tilde Q}}(\square) \nonumber \\
	~              & = & 2k - 2 + 2k - 2 + (n + 4) (1 + 1) - (2k+2) - (2k+2) - n (1 + 1) \nonumber \\
	~              & = & 0, \nonumber \\
\end{eqnarray}
where use has been made of the identities
\begin{equation}
	T(\textrm{anti-sym}) = 2k - 2, \quad T(\textrm{sym}) = 2k + 2, \quad T(\square) = 1,
\end{equation}
and the fact that the fundamental fields transform in real representations and therefore only have half the number of degrees of freedom.

\subsection{Localization computation}
Having written down the field content of the two-dimensional theory of $ k $ strings, it is straightforward to compute its elliptic genus, following the localization computation of \cite{Benini:2013xpa,Benini:2013nda}. The elliptic genus is given by a contour integral of a one-loop determinant:
\begin{equation} \label{eq:Zkint}
	Z_{k\text{ strings}} = \int Z_{k\text{ strings}}^{1-loop}(z_i,m_j,\mu_i,\tilde \mu_i, \epsilon_1,\epsilon_2),
\end{equation}
where $ Z_{k\text{ strings}}^{1-loop} $ is a $ k $-form on the $ k $ complex-dimensional space of flat $ Sp(k) $ connections on $ T^2 $, which is a complex torus parametrized by variables $ \zeta_i = \frac{1}{2\pi i}\log z_i  $, and the contour of integration is determined by the Jeffrey-Kirwan prescription \cite{JeffreyKirwan}. The one-loop determinant is obtained by multiplying together the contributions of all multiplets and takes the following form:
\[ Z_{k\text{ strings}}^{1-loop} = Z_{\Upsilon}\,Z_{\Lambda^\Phi}\,Z_{B}Z_{\tilde B}\, Z_{Q}Z_{\tilde Q}\,\,Z_{\Lambda^Q}\,Z_{\Lambda^{\tilde Q}},\]
where\footnote{We use the following definitions for the Dedekind eta function $ \eta(\tau) $ and Jacobi theta function $ \theta_1(z, \tau) $:
\begin{equation}
\eta(\tau) = q^{1/24}\prod_{j=1}^\infty (1-q^j);\qquad \theta_1(z,\tau) = -i q^{1/8} \sqrt{z}\prod_{j=1}^\infty (1-q^j)(1-z^{-1} q^j)(1-z^{-1}q^j),
\end{equation}
where $q=e^{2\pi i \tau}$.}
\begin{align} Z_{\Upsilon} &= \left(\prod_{i=1}^k\frac{d \zeta_i\, \theta_1'(0)\theta_1(z_i^2)\theta_1(z_i^{-2})}{\eta^3}\right)\left(\prod_{i<j}^k\prod_{s_1=\pm 1,\,s_2=\pm 1}\frac{\theta_1(z_i^{s_1}z_j^{s_2})}{\eta}\right)\\
Z_{\Lambda^\Phi} &= \left(\prod_{i=1}^k\frac{\theta_1(d t)\theta_1(dt z_i^2)\theta_1(dt z_i^{-2})}{\eta^3}\right)\left(\prod_{i<j}^k\prod_{s_1=\pm 1,\,s_2=\pm 1}\frac{\theta_1(dt z_i^{s_1}z_j^{s_2})}{\eta}\right)\\
Z_{B}Z_{\widetilde B} &= \left(\frac{\eta^2}{\theta_1(d)\theta_1(t)}\right)^k\left(\prod_{i<j}^k\prod_{s_1=\pm 1,\,s_2=\pm 1}\frac{\eta^2}{\theta_1(d z_i^{s_1}z_j^{s_2})\theta_1(t z_i^{s_1}z_j^{s_2})}\right)\\
Z_{Q} Z_{\tilde Q} &= \prod_{i=1}^k\prod_{j=1}^{p+4} \frac{\eta^4}{\theta_1(\sqrt{dt}z_i Q_{m_j})\theta_1(\sqrt{dt}z_i^{-1} Q_{m_j})\theta_1(\sqrt{dt}z_i Q_{m_j}^{-1})\theta_1(\sqrt{dt}z_i^{-1} Q_{m_j}^{-1})}\\
Z_{\Lambda^Q} Z_{\Lambda^{\tilde Q}} &= \prod_{i=1}^k\prod_{j=1}^{p} \frac{\theta_1(z_i Q_{\mu_j})\theta_1(z_i^{-1} Q_{\mu_j})\theta_1(z_i Q_{\tilde \mu_j})\theta_1(z_i^{-1} Q_{\tilde \mu_j})}{\eta^4},
\end{align}
and $ Q_{m_i} = e^{2\pi  i m_i} $, $ Q_{\mu_i} = e^{2\pi i \mu_i} $, $ Q_{\tilde \mu_i} = e^{2\pi i \tilde\mu_i}$, $ d = e^{2\pi i \epsilon_1} $, $ t = e^{2\pi i \epsilon_2} $. The integral itself is then obtained by computing a sum over Jeffrey-Kirwan residues of the one-loop determinant:
\begin{equation}
Z_{k \text{ strings}} = \frac{1}{\vert\text{Weyl}[Sp(k)]\vert}\sum_{\alpha} \text{JK-res}(\alpha, \mathfrak{q}) Z_{k\text{ strings}}^{1-loop},\label{eq:resSum}\end{equation}
where $ \alpha $ labels poles of $ Z_{k\text{ strings}}^{1-loop} $ and the role of $ \mathfrak{q} $ will be clarified shortly.
In the following sections we will compute the residue sum for one and two strings, in which case the evaluation of Jeffrey-Kirwan residues turns out to be straightforward and we do not need to resort to the full-fledged formalism.

\subsubsection*{One string}
For a single string, the one-loop determinant is given by a one-form:

\begin{align}
Z_{1\text{ string}}^{1-loop} &=d\zeta \frac{2\pi i \eta^2 \theta(z^2)\theta(z^{-2})\theta(d t)\theta(d t z^2)\theta(d t z^{-2})}{\theta(d) \theta(t)\eta^3}\nonumber\\
&\times  \prod_{i=1}^{p}\frac{\theta(Q_{\mu_i} z)\theta(Q_{\mu_i}^{-1} z)\theta(Q_{\tilde\mu_i} z)\theta(Q_{\tilde\mu_i}^{-1} z)}{\eta^4}\nonumber\\
&\times \prod_{i=1}^{4+p} \frac{\eta^4}{\theta(\sqrt{dt}Q_{m_i} z)\theta(\sqrt{dt}Q_{m_i}^{-1} z)\theta(\sqrt{dt}Q_{m_i} z^{-1})\theta(\sqrt{dt}Q_{m_i}^{-1} z^{-1})}.\label{eq:1string1loop}\end{align}

One first needs to identify the singular loci of the integrand. Each of the theta functions in the second line of \eqref{eq:1string1loop} determines a (0-dimensional) singular hyperplane within the one complex dimensional space $ \mathfrak{M}_{1\text{ string}} $ spanned by $ \zeta = \log{z} $, for a total of $ 4\cdot(4+p) $ distinct singular points at
\begin{equation} \pm\zeta + \frac{\epsilon_1+\epsilon_2}{2} \pm m_i = 0, \,\, i=1,\dots 4+p.\footnote{Since $ \zeta\sim\zeta+1\sim\zeta + \tau $, each theta function leads to a single pole.}\label{eq:1stringhyperplanes}\end{equation}
To determine which poles contribute to the residue sum, one needs to consider the normal vectors to the singular hyperplanes. In this case, the normal vector is simply $ \pm\partial_\zeta $, where the sign is the one multiplying $ \zeta $ in \eqref{eq:1stringhyperplanes}. The data that enters the Jeffrey-Kirwan residue computation corresponds of two quantities: the position of the pole in the $ \zeta $ plane and a choice of a vector $ \mathfrak{q} \in \text{T}\mathfrak{M}_{1\text{ string}} $. In this case, we can choose either $ \mathfrak{q} = \pm\partial\zeta $; let us pick $ \mathfrak{q} = -\partial_\zeta $. For two-dimensional theories, it can be argued that once the sum over residues is performed the answer is independent of the choice of $ \mathfrak{q} $. Next, one picks the poles satisfying the property that $ \mathfrak{q} $ lies within the one-dimensional cone spanned by the vector normal to the corresponding hyperplane. In this trivial example one finds that only the following poles contribute to the integral:
\begin{equation} -\zeta+\frac{\epsilon_1+\epsilon_2}{2}\pm m_i = 0.\end{equation}
Evaluating the Jeffrey-Kirwan residues in this situation corresponds to summing over the ordinary residues at these poles. Summing over the eight residues and dividing by $ \text{Weyl}[Sp(1)]=\mathbb{Z}_2 $ leads to the following answer:
\begin{eqnarray}
	Z_{\text{1 string}} & =& \frac{1}{2} \frac{\eta^2}{\theta(d)\theta(t)}\times\sum_{i=1}^{4+p}\left[\frac{\theta(d t Q_{m_i}^2)\theta(d^2 t^2 Q_{m_i}^{2})}{\eta^2}\prod_{j\neq i}\prod_{s=\pm 1}\frac{\eta^2}{\theta(Q_{m_i}Q_{m_j}^s)\theta(dt Q_{m_i}Q_{m_j}^s)}\right. \nonumber \\
	~                           & ~ &\hspace{1.1in} \times\prod_{j=1}^{p}\prod_{s=\pm1}\frac{\theta(\sqrt{d t}  Q_{m_i} Q_{\mu_j}^s)\theta(\sqrt{d t}  Q_{m_i}Q_{\tilde\mu_j}^s)}{\eta^2} +(Q_{m_i}\to Q_{m_i}^{-1})\Bigg]. \nonumber \\ \label{eq:Z1}
\end{eqnarray}
Note some features of this expression:  The existence of theta functions in the denominator which depend on $SO(8+2p)$ fugacities suggests that the $SO(8+2p)$ continues to be carried by some bosonic degrees of freedom in the IR.  Also, the fact that
the expressions include a mixture of $\epsilon_i$ (captured by $t,d$) and $m_i$ suggests a non-trivial structure
for the theory which makes it unlikely to correspond to a free theory in the IR.  It would be interesting to identify the non-trivial $(4,0)$ CFT whose elliptic genus is given by the above expression.  Perhaps ideas similar to the ones employed in \cite{Gadde:2014ppa} can be used to do this.

In Section \ref{sec:BPSnumbers} we will explain how to extract from this expression the BPS degeneracies corresponding to a single string; for the $ p = 0 $ case, one finds a precise match with the BPS invariants of the geometry that engineers the $ \mathcal{O}(-4)\to \mathbb{P}^1$  SCFT, to be discussed in Section \ref{sec:CYgeometry}.

\subsubsection*{Two strings}

\begin{figure}[h!]
\begin{center}
\includegraphics[width=3.2in]{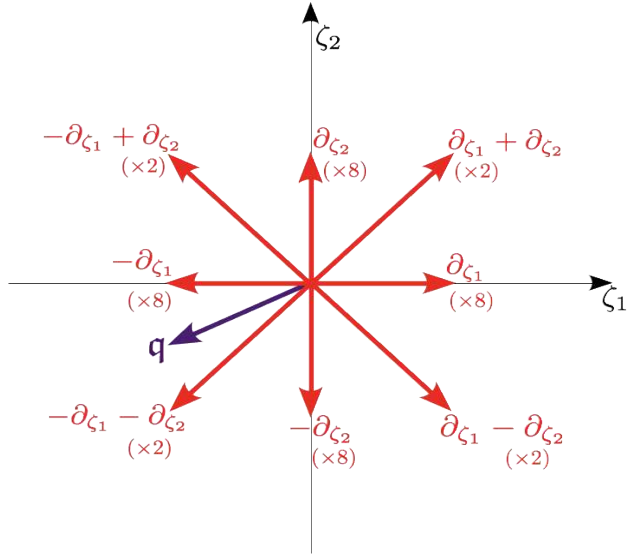}
\end{center}
\caption{Singular hyperplane configuration for the two-string elliptic genus. The vectors normal to the singular hyperplanes are displayed, along with the multiplicity with which they occur. Our choice of $ \mathfrak{q} $ is also displayed here.}
\label{fig:2stringfigure}
\end{figure}

The computation for two strings proceeds analogously; first, one should identify the hyperplanes in the two-dimensional space $ \mathfrak{M}_{2\text{ strings}} $ along which the denominator of $ Z_{2\text{ strings}}^{1-loop} $ vanishes. There are $ 8(p+5) $ such hyperplanes:
\begin{eqnarray} \pm\zeta_j + \frac{\epsilon_1+\epsilon_2}{2} \pm m_i &= 0, \,\, i=1,\dots 4+p,\,\, j=1,2;\\
\pm\zeta_1 \pm\zeta_2+\epsilon_1&=0;\\
\pm\zeta_1 \pm\zeta_2+\epsilon_2&=0,\end{eqnarray}
where $ \zeta_i = \log(z_i) $. For concreteness, let us focus from now on to the case where $ p = 0 $, keeping in mind that the computation for arbitrary $ p $ proceeds analogously. We display the vectors normal to the hyperplanes, as well as our choice of $ \mathfrak{q} \in \text{T}\mathfrak{M}_{2\text{ strings}} $, in Figure \ref{fig:2stringfigure}.
The next step is to identify the points at which hyperplanes intersect. The computation of Jeffrey-Kirwan residues is simplified by the fact that for generic values of $ m,\epsilon_1,\epsilon_2 $ at most two hyperplanes intersect at the same time. The poles whose residues contribute to the elliptic genus are those for which $ \mathfrak{q} $ lies within the cone spanned by the vectors normal to the corresponding hyperplanes. For example, since $ \mathfrak{q} $ lies in the cone spanned by $ -\partial_{\zeta_1} $ and $ -\partial_{\zeta_1}-\partial_{\zeta_2} $, but not in the one spanned by $ -\partial_{\zeta_1} $ and $ -\partial_{\zeta_1}+\partial_{\zeta_2} $, the residue evaluated at
\begin{equation} -\zeta_1 +\frac{\epsilon_1+\epsilon_2}{2}+m_1 = 0; \qquad -\zeta_1-\zeta_2+\epsilon_1 = 0,\end{equation}
will contribute, while the one at 
\begin{equation} -\zeta_1 +\frac{\epsilon_1+\epsilon_2}{2}+m_1 = 0; \qquad -\zeta_1+\zeta_2+\epsilon_1 = 0\end{equation}
will not. Following this prescription, one arrives at the following list of poles whose residues contribute to the computation:

\begin{alignat}{10}
\alpha_{i,j,s}:&{}&\quad\zeta_1 &= \frac{\epsilon_1+\epsilon_2}{2}+ s m_i, \quad \zeta_2\, &={}&\,\zeta_1 +\epsilon_j;&& \nonumber\\&{}&&&&&\hspace{-1in}(i = 1,\dots, 4,\, j = 1,2,\, s = \pm1)\label{eq:list1}\\
\alpha'_{i,j,s}:&{}&\quad\zeta_2 &= \frac{\epsilon_1+\epsilon_2}{2}+ s m_i, \quad \zeta_1\, &={}&\,\zeta_2 +\epsilon_j;&& \nonumber\\&{}&&&&&\hspace{-3in}(i = 1,\dots, 4,\, j = 1,2,\, s = \pm1)\label{eq:list2}\\
\alpha''_{i,j,s}:&{}&\quad-\zeta_2 &= \frac{\epsilon_1+\epsilon_2}{2}+ s m_i, \quad \zeta_1\, &={}&\,-\zeta_2 +\epsilon_j;&& \nonumber\\&{}&&&&&\hspace{-3in}(i = 1,\dots, 4,\, j = 1,2,\, s = \pm1)\label{eq:list3}\\
\beta_{i,j,s}:&{}&\quad\zeta_1 &= \frac{\epsilon_1+\epsilon_2}{2}+s m_i, \quad \zeta_2\, &={}&\,-\zeta_1 +\epsilon_j;&& \nonumber\\&{}&&&&&\hspace{-3in}(i = 1,\dots, 4,\, j = 1,2,\, s = \pm1)\label{eq:list4}\\
\gamma_{i,j,s_1,s_2}:&{}&\zeta_1 &=\frac{\epsilon_1+\epsilon_2}{2}+ s_1 m_i, \quad \zeta_2\, &={}&\,\frac{\epsilon_1+\epsilon_2}{2}+ s_2 m_j.&& \nonumber\\&{}&&&&&\hspace{-3in}(i=1,\dots,4,\, j\neq i,s_1=\pm1,\, s_2=\pm1)\label{eq:list5}\end{alignat}
The prescription outlined above also picks up some additional poles, but they do not contribute to the elliptic genus since the numerator of $ Z_{2\text{ strings}}^{1-loops} $ turns out to vanish for them. Therefore, the elliptic genus of two strings is obtained by summing over the residues that correspond the 112 poles listed in Equations (\ref{eq:list1})--(\ref{eq:list5}). In practice, one can exploit $ Sp(2) $ Weyl symmetry to show that the residues of poles $ \eqref{eq:list2} $ and $ \eqref{eq:list3} $ are identical to the ones of $ \eqref{eq:list1} $. For the same reason, one can set $ j < i $ in \eqref{eq:list5} and multiply the corresponding 24 residues by a factor of $ 2 $.

After these considerations, we are ready to write down the elliptic genus of two strings:
\begin{equation} Z_{2\text{ strings}} = \frac{1}{8}\left[3\sum_{\alpha_{i,j,s}} \text{Res}_{\alpha_{i,j,s}}Z_{2\text{ str.}}^{1-l.}+\sum_{\beta_{i,j,s}} \text{Res}_{\beta_{i,j,s}}Z_{2\text{ str.}}^{1-l.}+2\sum_{\substack{\gamma_{i,j,s_1,s_2}\\j<i}} \text{Res}_{\gamma_{i,j,s_1,s_2}}Z_{2\text{ str.}}^{1-l.} \right],\label{eq:Z2string}\end{equation}
where  we have divided by an overall factor of 8 $ = \vert\text{Weyl}[Sp(2)]\vert $, and the residues have the following explicit form:
\begin{align} \text{Res}_{\alpha_{i,1,s}}Z_{2\text{ strings}}^{1-loop} &= \frac{\theta_1(d^2 t\, Q_{m_i}^{2s})\theta_1(d^3 t\, Q_{m_i}^{2s})\theta_1(d^3 t^2\, Q_{m_i}^{2s})\theta_1(d^2 t^2\, Q_{m_i}^{2s})}{\theta_1(d)\theta_1(t)\theta_1(d^2)\theta_1(t/d)}\nonumber\\
&\times \prod_{j\neq i}\prod_{r=\pm 1}\frac{\eta^4}{\theta_1(Q_{m_i}^s Q_{m_j}^r)\theta_1(d\,Q_{m_i}^s Q_{m_j}^r)\theta_1(dt\, Q_{m_i}^s Q_{m_j}^r)\theta_1(d^2t\,Q_{m_i}^s Q_{m_j}^r)};
\end{align}
\begin{align} \text{Res}_{\alpha_{i,2,s}}Z_{2\text{ strings}}^{1-loop} &= \frac{\theta_1(d t^2\, Q_{m_i}^{2s})\theta_1(d t^3\, Q_{m_i}^{2s})\theta_1(d^2 t^3\, Q_{m_i}^{2s})\theta_1(d^2 t^2\, Q_{m_i}^{2s})}{\theta_1(d)\theta_1(t)\theta_1(t^2)\theta_1(d/t)}\nonumber\\
&\times \prod_{j\neq i}\prod_{r=\pm 1}\frac{\eta^4}{\theta_1(Q_{m_i}^s Q_{m_j}^r)\theta_1(t\, Q_{m_i}^s Q_{m_j}^r)\theta_1(dt\, Q_{m_i}^s Q_{m_j}^r)\theta_1(dt^2\, Q_{m_i}^s Q_{m_j}^r)};
\end{align}
\begin{align} \text{Res}_{\beta_{i,1,s}}Z_{2\text{ strings}}^{1-loop} &= \frac{\theta_1(d^2 t)\theta_1(t\, Q_{m_i}^{2s})\theta_1(t/d\, Q_{m_i}^{2s})\theta_1(d^2\, Q_{m_i}^{-2s})\theta_1(dt^2\, Q_{m_i}^{2s})\theta_1(d^2 t^2\, Q_{m_i}^{2s})}{\theta_1(d)\theta_1(t)^2\theta_1(d^2)\theta_1(t/d)\theta_1(Q_{m_i}^{2s})}\nonumber\\
&\times \prod_{j\neq i}\prod_{r=\pm 1}\frac{\eta^4}{\theta_1(Q_{m_i}^s Q_{m_j}^r)\theta_1(d\,Q_{m_i}^s Q_{m_j}^r)\theta_1(t\, Q_{m_i}^s Q_{m_j}^r)\theta_1(dt\,Q_{m_i}^s Q_{m_j}^r)};
\end{align}
\begin{align} \text{Res}_{\beta_{i,2,s}}Z_{2\text{ strings}}^{1-loop} &= \frac{\theta_1(d t^2)\theta_1(d\, Q_{m_i}^{2s})\theta_1(d/t\, Q_{m_i}^{2s})\theta_1(t^2\, Q_{m_i}^{-2s})\theta_1(d^2t\, Q_{m_i}^{2s})\theta_1(d^2 t^2\, Q_{m_i}^{2s})}{\theta_1(d)^2\theta_1(t)\theta_1(t^2)\theta_1(d/t)\theta_1(Q_{m_i}^{2s})}\nonumber\\
&\times \prod_{j\neq i}\prod_{r=\pm 1}\frac{\eta^4}{\theta_1(Q_{m_i}^s Q_{m_j}^r)\theta_1(d\,Q_{m_i}^s Q_{m_j}^r)\theta_1(t\, Q_{m_i}^s Q_{m_j}^r)\theta_1(dt\,Q_{m_i}^s Q_{m_j}^r)};
\end{align}
\begin{align} \text{Res}_{\gamma_{i,j,s_1,s_2}}Z_{2\text{ strings}}^{1-loop} &= \frac{\theta_1(d t\, Q_{m_i}^{2s_1})\theta_1(d t\, Q_{m_j}^{2s_2})\theta_1(d^2t^2 Q_{m_i}^{2s_1})\theta_1(d^2t^2 Q_{m_j }^{2s_2})}{\theta_1(d)^2\theta_1(t)^2}\frac{\theta_1(d^2 t^2 Q_{m_i}^{s_1}Q_{m_j}^{s_2})}{\theta_1(Q_{m_i}^{s_1}Q_{m_j}^{s_2})}\nonumber\\
&\times \eta^8\Bigg[\theta_1(d\,Q_{m_i}^{s_1}Q_{m_j}^{s_2})\theta_1(t\,Q_{m_i}^{s_1}Q_{m_j}^{s_2})\theta_1(d\,Q_{m_i}^{s_1}Q_{m_j}^{-s_2})\theta_1(t\,Q_{m_i}^{s_1}Q_{m_j}^{-s_2})\nonumber\\
&\times\theta_1(d\,Q_{m_i}^{-s_1}Q_{m_j}^{s_2})\theta_1(t\,Q_{m_i}^{-s_1}Q_{m_j}^{s_2})\theta_1(d^2 t\,Q_{m_i}^{s_1}Q_{m_j}^{s_2})\theta_1(d t^2\,Q_{m_i}^{s_1}Q_{m_j}^{s_2})\Bigg]^{-1}\nonumber\\
&\times \prod_{k\neq i,j}\prod_{r=\pm 1}\frac{\eta^4}{\theta_1(Q_{m_i}^{s_1} Q_{m_k}^r)\theta_1(Q_{m_j}^{s_2} Q_{m_k}^r)\theta_1(d t\, Q_{m_i}^{s_1} Q_{m_j}^r)\theta_1(dt\, Q_{m_j}^{s_2} Q_{m_j}^r)}.
\end{align}
After summing over the 56 residues, one is left with a weight zero meromorphic elliptic function with modular parameter $ \tau $ and six elliptic parameters ($ \epsilon_{1},\epsilon_{2}  $ and the SO(8) fugacities $ (m_1,\dots,m_4) $). In Section \ref{sec:O(-4)} we will check the validity of our answer in the unrefined limit $ \epsilon_2 = -\epsilon_1 $ by verifying that it exactly reproduces the genus 0 BPS invariants of the Calabi-Yau threefold that engineers the six-dimensional theory under consideration. If $ \epsilon_1,\epsilon_2 $ are left arbitrary, Equation \eqref{eq:Z2string} can be used to compute arbitrary genus refined BPS invariants of this geometry.

For higher numbers of strings the computation of the elliptic genus from Equation (\ref{eq:resSum}) proceeds analogously, but for simplicity and clarity  of exposition we limit our discussion to the cases of one and two strings.

\subsection{Modular anomaly}
In this section we wish to study the behavior of the elliptic genus under $ SL(2,\mathbb{Z}) $ transformations 
\begin{equation} \gamma: (t_b, \tau,m_i,\mu_i,\epsilon_i) \to \left(t_b,\,\frac{a \tau+b}{c\tau + d}, \, \frac{m_i}{c\tau+d}, \, \frac{\mu_i}{c\tau+d},\, \frac{\epsilon_i}{c\tau+d}\right),\quad  \begin{pmatrix}a & b\\ c & d\end{pmatrix}\in SL(2,\mathbb{Z}).\end{equation}
The modular properties of the elliptic genus can be best understood starting from the integral expression (\ref{eq:Zkint}), where the integration variables $ \zeta_i $ also transform as elliptic parameters: $\zeta_i \to \zeta_i/(c\tau+d)$. Each $ (2,0) $ multiplet contributes to the integrand a factor of the form $\left(\frac{\theta_1(y,\tau)}{\eta(\tau)}\right)^{\pm 1}$.
Recall that under $ S $ and $ T $ transformations
\begin{equation}\frac{\theta_1(e^{2\pi i\zeta},\tau+1)}{\eta(\tau+1)} = \frac{e^{\pi i/4}\theta_1(e^{2\pi i\zeta},\tau)}{e^{\pi i/12}\eta(\tau)} = e^{\pi i/6} \frac{\theta_1(e^{2\pi i\zeta},\tau)}{\eta(\tau)},\end{equation}
\begin{equation}\frac{\theta(e^{2\pi i\zeta/\tau},-1/\tau)}{\eta(-1/\tau)} = \frac{e^{-3\pi i/4}e^{\frac{\pi i}{\tau}\zeta^2} \theta_1(e^{2\pi i\zeta},\tau)}{e^{-\pi i/4} \eta(\tau)}=e^{-\pi i/2}e^{\frac{\pi i}{\tau}\zeta^2}\frac{\theta_1(e^{2\pi i\zeta},\tau)}{\eta(\tau)}.\end{equation}
Using this, one can easily check that the elliptic genus of $ k $ strings is invariant under $ \tau \to \tau + 1 $:
\begin{equation} Z_{k\text{ strings}}(\tau+1) = Z_{k\text{ strings}}(\tau);\end{equation}
on the other hand, under  $ \tau \to -1/\tau $ one can show that the integrand picks up a $ z_i $-independent phase:
\begin{align} 
\frac{Z_{k\text{ strings}}(-1/\tau)}{Z_{k\text{ strings}}(\tau)} = \exp\bigg[-\frac{\pi i}{\tau}&\bigg(\epsilon_1\epsilon_2\left(4k^2-2 k\right) -\left(\epsilon_1+\epsilon_2\right)^2k(2+p)\nonumber\\
& \hspace{.5in}- 4k \sum_{j=1}^{4+p} m_j^2+2k\sum_{j=1}^p (\mu_j^2+\widetilde\mu_j^2)\bigg)\bigg] .\label{eq:zmodular}
\end{align}
In other words, $ Z_{k\text{ strings}} $ transforms as a modular function up to an anomalous phase factor. The origin of this factor can be easily understood by considering the following representation of the theta function:
\begin{equation} \theta(z,\tau) = \eta(\tau)^3 (2\pi \zeta) \exp\left(\sum_{k\geq 1}\frac{B_{2k}}{(2k)(2k)!}E_{2k}(\tau)(2\pi i \zeta)^{2k}\right),\end{equation}
where the dependence on the modular parameter $ \tau $ is expressed in terms of the Eisenstein series
\begin{equation} E_{2k}(\tau) =\frac{1}{2 \zeta(2k)}\sum_{(m,n)\in \mathbb{Z}^2\backslash (0,0)}\frac{1}{m\tau + n},\qquad k \geq 1
\end{equation}
and $ \zeta(z) $ is the Riemann zeta function. For any $ k \geq 2 $, $ E_{2k}(\tau) $ is a modular form of weight $ 2k $. On the other hand, $E_2(\tau)$ transforms anomalously:
\begin{equation} E_2\left(\frac{a\tau+b}{c\tau+d}\right) = (c\tau+d)^2 E_2(\tau)-\frac{6 i c}{\pi}(c\tau + d).\end{equation}
In other words, the phase factors appearing in Equation \eqref{eq:zmodular} are completely determined by the $ E_2(\tau) $-dependence of the integrand, and in lieu of \eqref{eq:zmodular} we might as well have written:
\begin{align} \partial_{E_2}Z_{k\text{ strings}} = -\frac{1}{24}(2\pi)^2\bigg( & \epsilon_1\epsilon_2\left(4k^2-2 k\right) -\left(\epsilon_1+\epsilon_2\right)^2k(2+p) \nonumber\\
& \hspace{.4in} - 4k \sum_{j=1}^{4+p} m_j^2+2k\sum_{j=1}^p (\mu_j^2+\widetilde\mu_j^2)\bigg)Z_{k \text{ strings}}.\label{eq:modularanomaly}\end{align}
This expression is very similar to the E-string and M-string modular anomaly equations found in \cite{Hosono:1999qc,Haghighat:2013gba,Huang:2013yta}: in the E-string ($ \mathcal{O}(-1)\to \mathbb{P}^1 $) case, one has:
\begin{align*} \frac{1}{Z_E^{d}}\partial_{E_2}Z_E^d &=-\frac{1}{24}(2\pi)^2(\epsilon_1\epsilon_2(k^2+k) -k(\epsilon_1+\epsilon_2)^2 +k(\sum_i m_i^2)),\end{align*}
while in the M-string ($\mathcal{O}(-2)\to \mathbb{P}^1$) case, one finds:
\begin{align*}\frac{1}{Z_M^{k}}\partial_{E_2}Z_M^d &= -\frac{1}{12}(2\pi)^2(\epsilon_1\epsilon_2k^2 -\frac{k}{4}(\epsilon_1+\epsilon_2)^2 +k \,m^2).
\end{align*}
As discussed in Section \ref{sec:minimalSCFTs}, in all these cases the elliptic genera of the strings capture part of the topological string partition function of the corresponding Calabi-Yau $ X $:
\begin{equation} Z^{\textrm{top}}(X) = Z_{0}(X)\cdot \left(1 + \sum_{k=1}^\infty Q^k Z_{k\text{ strings}}(X) \right).\end{equation}
In all cases, $ X $ is elliptically fibered, and the topological string partition function is expected to be invariant under modular transformations (\ref{eq:zmodular}). However, this is in contradiction with the fact that $ Z_{k\text{ strings}} $ is only invariant up to a phase. The resolution to this apparent contradiction is well known: in the topological string expression the second Eisenstein series $ E_2(\tau) $ should be replaced by its modular completion
\begin{equation}\widehat{E}_2(\tau,\overline\tau) = E_2(\tau) -\frac{6i}{\pi(\tau-\overline\tau)},\end{equation}
which under the $ SL(2,\mathbb{Z}) $ action transforms as follows:
\begin{equation} \widehat{E}_{2}\left(\frac{a\tau+b}{c\tau+d},\frac{a\overline\tau+b}{c\overline\tau+d}\right) = (c\tau+d)^2\widehat{E}_2(\tau,\overline\tau).\end{equation}
This implies that the topological string partition function is a well defined modular function of $ \tau $, but no longer depends holomorphically on it:
\begin{equation} \partial_{\overline\tau} Z_{top}(X) = \frac{6i}{\pi(\tau-\overline\tau)^2}\partial_{\widehat{E}_2} Z_{top}(X) \neq 0.\end{equation}

\subsection{Refined BPS invariants}
\label{sec:BPSnumbers}

\begin{figure}[here!]
  \centering
	\includegraphics[width=0.4\textwidth]{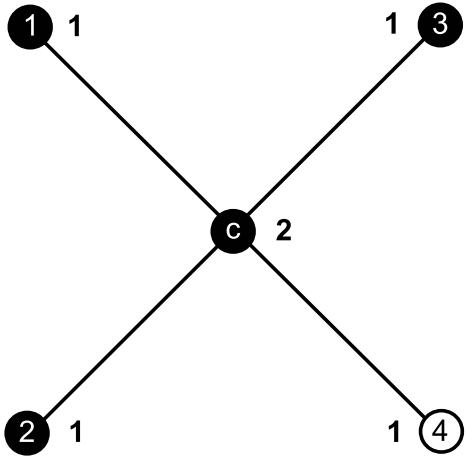}
  \caption{Extended Dynkin diagram for $D_4$.}
  \label{fig:affD4}
\end{figure}

Let us now explain how to extract refined BPS invariants from the elliptic genera of $k$ strings $Z_k$, again specializing to the case $ p = 0 $ where the global symmetry on the worldsheet is just $ SO(8) $. In order to proceed note that the full partition function of the topological string is given by
\begin{equation} \label{eq:topstringexp}
	Z^{\textrm{top}} = e^F = Z_0\left(1+\sum_{k=1}^{\infty} Q^k Z_k \right),
\end{equation}
where $Z_k$ is the elliptic genus of $k$ strings and $Q$ is a combination of exponentiated K\"ahler moduli of the elliptic Calabi-Yau geometry to be determined later.  Furthermore,  we perform the following change of basis which replaces the mass parameters $Q_{m_i}$ by the parameters $Q_i$ corresponding to a choice of simple roots of $SO(8)$:
\begin{equation}
	Q_{m_1} = Q_1 Q_c \sqrt{Q_2} \sqrt{Q_3}, \quad Q_{m_2} = Q_c \sqrt{Q_2} \sqrt{Q_3}, \quad Q_{m_3} = \sqrt{Q_2} \sqrt{Q_3}, \quad Q_{m_4} = \frac{\sqrt{Q_3}}{\sqrt{Q_2}}. \color{white}\sum_{{j=1}}\color{black}
\end{equation}
In addition to these parameters, let us also define the parameter $Q_4$ corresponding to the affine node of the extended Dynkin diagram of type $D_4$ as shown in Figure \ref{fig:affD4}.
The elliptic genera $Z_k$ can then be expanded in positive powers of $Q_1,Q_2,Q_3,Q_4$ and $Q_c$ upon replacing $Q_{\tau} = e^{2\pi i \tau}$ by the following combination:
\begin{equation}
	Q_{\tau} = Q_1 Q_2 Q_3 Q_4 Q_c^2,
\end{equation}
where the powers are determined by the Coxeter labels of the nodes in Figure \ref{fig:affD4}.

Taking the logarithm of Equation (\ref{eq:topstringexp}), the free energy $F$ can be expanded as:
\begin{equation}
	F = \log Z = \log(Z_0) + Z_1 Q + \left(-\frac{1}{2} Z_1^2 + Z_2\right)Q^2 + \left(\frac{Z_1^3}{3} - Z_1 Z_2 + Z_3\right) Q^3 + \mathcal{O}(Q^4).
\end{equation}
In order to make contact with the computation of the Calabi-Yau BPS invariants from Section \ref{sec:CYgeometry} we identify $Q$ with the combination
\begin{equation} \label{redefqb}
	Q = Q_b \frac{Q_4}{Q_1 Q_2 Q_3 Q_c^2},
\end{equation}
where $Q_b = e^{-t_b}$ and $t_b$ is the K\"ahler class of the base of the elliptic fibration.
The refined BPS invariants are encoded in the free energy $F$ as follows \cite{Gopakumar:1998jq,Hollowood:2003cv}
\begin{equation} \label{eq:nBPS}
	\small{F = \sum_{\stackrel{j_l,j_R=0}{m=1}}^{\infty} \sum_{\beta \in H_2(M,\mathbb{Z})} \frac{n_{\beta}^{j_L,j_R}}{m} \frac{(-1)^{2(j_L + j_R)} \sqrt{d^m t^m} \left(\sum_{n=-j_L}^{j_L} (d/t)^{m n} \right)\left(\sum_{n=-j_R}^{j_R} (d t)^{m n} \right) e^{m (\beta,\underline{t})}}{(1-d^m)(1-t^m)}},
\end{equation}
where $\underline{t}$ denote the K\"ahler moduli of the Calabi-Yau. The above free energy encodes BPS degeneracies $n^{j_L,j_R}_{\beta}$ of short multiplets of the five-dimensional quantum field theory arising from circle compactification of the six-dimensional SCFT. In this context the labels $j_L$ and $j_R$ refer to the spins of the two $SU(2)$ subgroups of the little group $SO(4)$ in the decomposition $SO(4) = SU(2)_L \times SU(2)_R$ and $\beta$ labels the string charge as well as the various flavor charges. Denoting by $\underline{t_f}$ the collection of the K\"ahler moduli of the resolved elliptic fiber and making use of the expansion
\begin{equation}
	F(\epsilon_1,\epsilon_2,\underline{t}) = F_0(\epsilon_1,\epsilon_2,\underline{t_f}) + \sum_{i=1}^{\infty} F_i(\epsilon_1,\epsilon_2,\underline{t_f}) Q^i,
\end{equation}
we find
\begin{eqnarray}
	F_1(\epsilon_i, m_i, \tau) & = & Z_1(\epsilon_i, m_i, \tau) \nonumber \\
	F_2(\epsilon_i,m_i,\tau)  &  = & Z_2(\epsilon_i, m_i,\tau) - \frac{1}{2} Z_1(\epsilon_i, m_i, \tau)^2 \nonumber \\
	\vdots & ~ & ~ \nonumber \\
\end{eqnarray}
Since for $F_1$ there is no multi-wrapping, we can set $m=1$ in Equation (\ref{eq:nBPS}) and extract the invariants $n_{\beta}^{j_L,j_R}$ immediately from the expression (\ref{eq:Z1}) for the elliptic genus of one string. Let us specify $\beta$ in terms of the following basis of $H^2(X,\mathbb{Z})$: $J_b, J_1, J_2, J_3, J_4, J_c$; that is, we write:
\begin{equation}
	n^{j_L,j_R}_\beta = n^{j_L,j_R}_{n_b,n_1,n_2,n_3,n_4,n_c}.
\end{equation}
In the following tables we present a sample of invariants for some specific choices of low degree curves.
\begin{table}[H]
\centering
\footnotesize{\begin{tabular} {|c|c|c|c|c|c|c|c|c|} 
\hline 
$2j_L \backslash 2j_R$ & 0 & 1 & 2 & 3 & 4 & 5 & 6 & 7 \\ \hline
 0                & 0 & 33 & 0 & 28 & 0 & 9 & 0 & 1 \\ \hline
1                 & 2 & 0 & 3 & 0 & 1 & 0 & 0 & 0 \\ \hline
\end{tabular}} 
\vskip 3pt   $n^{j_L,j_R}_{1,2,2,1,1,3}$ 
\end{table}
\begin{table}[H]
\centering
\footnotesize{\begin{tabular} {|c|c|c|c|c|c|c|c|c|} 
\hline 
$2j_L \backslash 2j_R$ & 0 & 1 & 2 & 3 & 4 & 5 & 6 & 7 \\ \hline
0                                  & 0 & 28 & 0 & 42 & 0 & 29 & 0 & 9 \\ \hline
 1                                 & 1 & 0 & 3 & 0 & 3 & 0 & 1 & 0 \\ \hline
\end{tabular}} 
\vskip 3pt   $n^{j_L,j_R}_{1,2,2,1,1,4}$ 
\end{table}
\begin{table}[H]
\centering
\footnotesize{\begin{tabular} {|c|c|c|c|c|c|c|c|c|} 
\hline 
$2j_L \backslash 2j_R$ & 0 & 1 & 2 & 3 & 4 & 5 & 6 & 7 \\ \hline
0                                   & 0 & 10 & 0 & 11 & 0 & 6 & 0 & 1 \\ \hline
1                                   & 0 & 0 & 0 & 0 & 0 & 0 & 0 & 0 \\ \hline
\end{tabular}} 
\vskip 3pt   $n^{j_L,j_R}_{1,3,3,1,1,3}$ 
\end{table}
\begin{table}[H]
\centering
\footnotesize{\begin{tabular} {|c|c|c|c|c|c|c|c|c|} 
\hline 
$2j_L \backslash 2j_R$ & 0 & 1 & 2 & 3 & 4 & 5 & 6 & 7 \\ \hline
0                                  &  0 & 41 & 0 & 47 & 0 & 28 & 0 & 9 \\ \hline
1                                  & 2 & 0 & 4 & 0 & 3 & 0 & 1 & 0 \\ \hline
\end{tabular}} 
\vskip 3pt   $n^{j_L,j_R}_{1,3,3,1,1,4}$ 
\end{table}
Analogously, we can extract all refined invariants for two strings, that is for base wrapping number $n_b = 2$. For example:
\begin{table}[H]
\centering
\footnotesize{\begin{tabular} {|c|c|c|c|c|c|c|c|c|} 
\hline 
$2j_L \backslash 2j_R$ & 0 & 1 & 2 & 3 & 4 & 5 & 6 & 7 \\ \hline
0                                   & 0 & 1 & 0 & 2 & 0 & 1 & 0 & 0 \\ \hline
1                                   & 0 & 0 & 0 & 0 & 0 & 0 & 0 & 0 \\ \hline
\end{tabular}} 
\vskip 3pt   $n^{j_L,j_R}_{2,1,1,0,0,1}$ 
\end{table}
\begin{table}[H]
\centering
\footnotesize{\begin{tabular} {|c|c|c|c|c|c|c|c|c|} 
\hline 
$2j_L \backslash 2j_R$ & 0 & 1 & 2 & 3 & 4 & 5 & 6 & 7 \\ \hline
 0                                  & 0 & 2 & 0 & 2 & 0 & 1 & 0 & 0 \\ \hline 
1                                   &  0 & 0 & 0 & 0 & 0 & 0 & 0 & 0 \\ \hline 
\end{tabular}} 
\vskip 3pt   $n^{j_L,j_R}_{2,2,2,0,0,1}$ 
\end{table}
In order to extract unrefined invariants from these one has to sum over the right-moving spin of the multiplets as follows:
\begin{equation}
	n^{j_L}_{\beta} = \sum_{j_R} (-1)^{2j_R} (2 j_R + 1) n^{j_L,j_R}_{\beta}.
\end{equation}
Furthermore, in order compare with the genus expansion of the topological string, the $SU(2)_L$ representations have to be organized into
\begin{equation}
	I^n_L = \left[(\frac{1}{2}) + 2(0)\right]^{\otimes n},
\end{equation}
and the BPS invariants $ n^g_{\beta} $ can be obtained by comparing the two sides of the identity
\begin{equation}
	\sum n^{j_L,j_R}_{\beta} (-1)^{2j_R} (2j_R + 1)[j_L] = \sum_g n^{g}_{\beta} I^g_L.
\end{equation}
The expansion coefficients in $I_L^n = \sum_j c_j^{2n} [j/2]$ can be found for example in \cite{Huang:2010kf}. Using these results we can compute unrefined invariants. For $n_b = 1$, for example, for the curves considered above one has:
\begin{eqnarray}
	n^0_{1,2,2,1,1,3} = -272 & \quad & n^1_{1,2,2,1,1,3} = 16, \nonumber \\
	n^0_{1,2,2,1,1,4} = -534 & \quad & n^1_{1,2,2,1,1,4} = 32, \nonumber \\
	n^0_{1,3,3,1,1,3} = -108 & \quad & n^1_{1,3,3,1,1,3} = 0, \nonumber \\
	n^0_{1,3,3,1,1,4} = -582 & \quad & n^1_{1,3,3,1,1,4} = 36, \nonumber \\
\end{eqnarray}
and for the $ n_b = 2 $ curves considered above we obtain
\begin{eqnarray}
	n^0_{2,1,1,0,0,1} = -16 & \quad & n^1_{2,1,1,0,0,1} = 0, \nonumber \\
	n^0_{2,2,2,0,0,1} = -18 & \quad & n^1_{2,2,2,0,0,1} = 0. \nonumber \\
\end{eqnarray}
For these classes all invariants with $g \geq 2$ vanish.

Following the procedure outlined above we have extracted an extensive list of BPS invariants corresponding to one and two strings. The genus zero invariants can be computed independently by employing the mirror symmetry and topological string techniques presented in the next section. When comparing the elliptic genus results to the topological string computation presented in Section \ref{sec:O(-4)}  we find a perfect agreement.

\section{The Calabi-Yau geometries with elliptic singularities}
\label{sec:CYgeometry}

In this section we construct the local elliptic Calabi-Yau geometries 
corresponding to the different minimal 6d SCFTs. Our strategy will be 
to first find a minimal compact elliptic Calabi-Yau 3-fold  with the 
right type of elliptic fiber degeneration over the rigid divisor $\Sigma$ in the base $B$  
and subsequently take the local limit by decompactifying the normal direction to 
$\Sigma$ in $B$. The resulting space will be the non-compact Calabi-Yau 3-fold.

Non-compact Calabi-Yau  geometries played an important 
role in the development of topological string theory, which can 
frequently be completely solved on these geometries, by well 
understood relations to matrix models, integrable models and gauge 
theories. One wide class of examples are the non-compact toric 
Calabi-Yau spaces; another one with some overlap to the first consists of the local (almost) Fano varieties $\mathcal{O}(-K_S)\rightarrow S$, where $S$ is an 
(almost) Fano variety. That is, one considers a rigid divisor $S$ in 
the Calabi-Yau 3-fold and decompactifies the normal direction to $S$. 
In these cases local mirror symmetry leads to a mirror curve. 
In the present case, instead, we decompactify the normal direction to a 
rigid divisor $\Sigma$ in the base $B$ of an elliptic 
Calabi-Yau 3-fold. Then, the mirror geometry does not 
reduce to a curve. In cases where there is an orbifold description one can describe the local mirror geometry as a non-compact Landau-Ginzburg model 
as we exemplify for the $\mathbb{Z}_3$ orbifold in Appendix~\ref{LG}.

\subsection{The local geometries}

The new local geometries we consider arise in Calabi-Yau threefolds, 
where we zoom close to the elliptic singularity of an elliptic fibration 
over a divisor $\Sigma$ in a two-dimensional base $B$. 
The divisor $\Sigma$ corresponds to the 7-brane locus in F-theory with 
gauge symmetry $g_{\Sigma}$  and the exceptional divisors  that resolve the 
elliptic singularity intersect with the negative Cartan matrix $C_{\hat g_{\Sigma}}$ 
of the affine Lie algebra ${\hat g_{\Sigma}}$ associated to $g_{\Sigma}$.    

We consider non-Higgsable singularities; in other words, the divisor $\Sigma$ 
has to be rigid and the Calabi-Yau space has no complex structure 
deformations which could resolve the singularity.  The simplest 
example for a non compact threefold of this type are elliptic fibrations 
over $B=({\cal O}(-n)\rightarrow \mathbb{P}^1)$.
We will start with a compact threefold $M_3$ constructed as 
elliptic fibration over a Hirzebruch surface $B = \mathbb{F}_n$~\cite{Morrison:1996pp}.
The  $(-n)$ section of the rational fibration of $\mathbb{F}_n$ 
is then the rigid gauge symmetry divisor $\Sigma$, 
with ${\cal O}(-n)$ as its normal bundle. 

This setup allows us to solve the topological string using mirror symmetry 
with normalizable intersections and instanton actions. By decompactifying 
the normal direction we can easily  decouple six-dimensional gravity.   

\subsubsection*{Hirzebruch surfaces as base}
\label{sec:Hirzebruch}

Let us recall that the Hirzebruch surfaces $\mathbb{F}_n$ are rational $\mathbb{P}^1$ 
fibrations over $\mathbb{P}^1$, where $n$ parametrizes the twisting of the 
fiber. They can be constructed torically or as gauged linear sigma models 
with four chiral fields $\Phi_i$, $i=1,\ldots, 4$ and two $U(1)$'s under which the fields have charges 
$l^{(1)}_i$ and $l^{(2)}_i$. We also use the description in terms of a toric fan 
in which each field $\Phi_i$  corresponds to a primitive vector $\nu_i$ spanning the fan 
in the integer lattice $\mathbb{Z}^2$  and summarize the base data as follows:
\begin{equation} 
 \label{datafn} 
 \begin{array}{cc|rrr|rrl|} 
    \multicolumn{2}{c}{Div } & \multicolumn{3}{c}{\nu^*_i}    &l^{(1)}& l^{(2)}\\
    D_0=K    &&  1&      0&   0&         -2 & n-2&     \\  
    D_1=S    &&  1&      0&   1&         1&   0&        \\ 
    D_2=F    &&  1&      1&   0&         0&   1&          \\ 
    D_3=S'   &&  1&      0&   -1&         1&  -n&         \\ 
    D_4=F    &&  1&     -1&   -n&        0&   1&         \\  
  \end{array} \ . 
\end{equation} 
Here we added the inner point  $\nu_0=(0,0)$ and promoted the 
points $\nu_i$ to $\nu_i^*=(1,\nu_i)\in \mathbb{Z}^3$. This is useful for 
describing the non-compact Calabi-Yau as the anti-canonical bundle over 
$\mathbb{F}_n$ ($\Phi_0$ is the noncompact direction), but it could be 
omitted for the discussion of the compact Hirzebruch surface. 
Each point $\nu_i$ correponds to a toric divisor $D_i=\{\Phi_i=0\}$ and within 
the surface the homological relations between these divisors are 
$S=S'+n F$. The nonvanishing intersections are 
$S^2=n$, $F S=1$ and $(S')^2=-n$; therefore, $S'$ 
becomes the gauge theory divisor.   

Geometrically the  $l^{(k)}$, $k=1,2$ represent curve classes $[C_k]$ 
and the intersection with the toric divisors $D_i$ is given by  
\be
[C_k]\cdot [D_i]=l^{(k)}_i \ .  
\label{intersectionscurves}
\ee
The $l^{(k)}$ are also called Mori vectors; in the present example, $k=1$ represents   
the base $\mathbb{P}^1$ while $k=2$ represents the fiber  $\mathbb{P}^1$ of 
$\mathbb{F}_n$.

\subsubsection*{The elliptic fiber types}
\label{sec:elliptic fibers}

Next, we want to construct the relevant Calabi-Yau spaces as elliptic fibrations over the Hirzebruch surfaces $\mathbb{F}_n$. From these we will finally obtain the local $\mathcal{O}(-n)\rightarrow \mathbb{P}^1$ models by taking the size of the $\mathbb{P}^1$ fiber specified by the class $F$ of the Hirzebruch surface $\mathbb{F}_n$ to infinity. However, it turns out that there are multiple ways to realize the elliptic fiber singularity of the appropriate type leading to different Mordell-Weyl groups. In this paper we will be interested in a rank one Mordell-Weyl group, that is an elliptic fibration with a single section. In the following we describe how to achieve this desired fibration structure.

Generically, for $n > 2$, the situation is such that the discriminant vanishes on the base $\mathbb{P}^1$ called $S'$ (we therefore have $\Sigma = S'$) and on isolated points of the fiber $\mathbb{P}^1$ denoted by $F$. Let us describe the non-compact geometry which arises when we take the size of $F$ to be infinite. We will denote curves on which the discriminant vanishes point-wise by $(-1)$-curves borrowing the terminology from E-strings. In fact, the analogy goes even further in that a subset of the $E_8$ Weyl symmetry of E-strings can act on sections of this elliptic fibration. The exact subset is determined by the elliptic fiber singularity type at the intersection points of the non-compact limit of $F$ with $S'$. We will denote the corresponding Kodaira group by $g_E$ which should not be confused with $g_{\Sigma}$ discussed in Section \ref{sec:minimalSCFTs} which labels the fiber degeneration on $S'$. A consistency condition is that $g_{\Sigma}$ should be a subgroup of $g_E$, that is $g_{\Sigma} \subset g_E$.  In order to restrict to one section for each model we only consider the fiber type  $g_{E}=E_8$ in this paper which leads to the following schematic picture of curve configurations:
\begin{figure}[here!]
  \centering
	\includegraphics[width=0.6\textwidth]{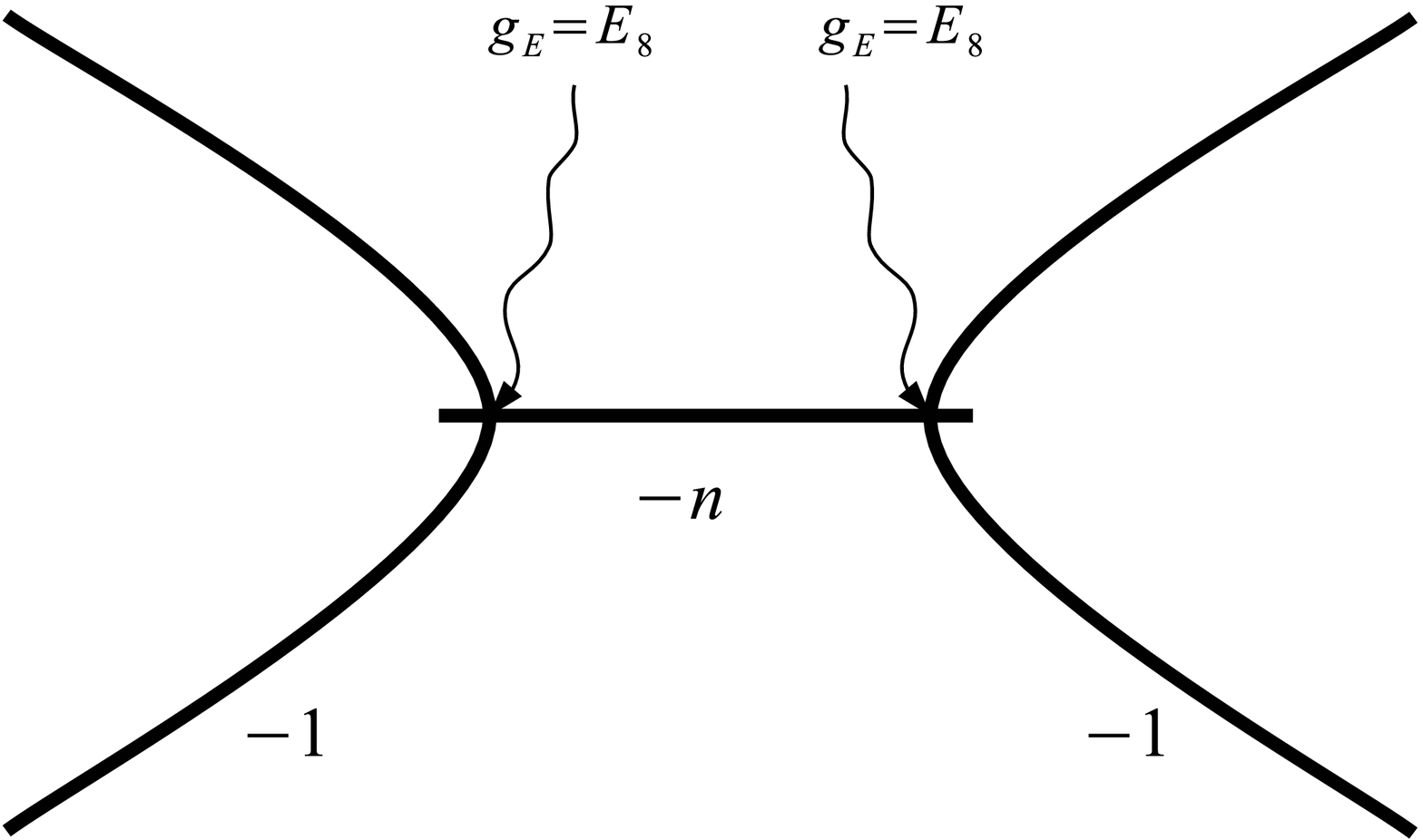}
\end{figure} 

For other choices of $g_E$ (simplest cases are $g_{E}=E_n$, $n=3,\ldots,8$, where 
$\{E_n\}_{n=3}^8=A_1\times A_2,A_4,D_4,E_6,E_7,E_8$) the subset of $E_8$ which becomes 
the Mordell-Weyl group of the elliptic Calabi-Yau is determined by the commutant of the Weyl group $g_E$ in the Weyl group of $E_8$.  Note that gluing together the two non-compact $(-1)$-curves gives back the compact Calabi-Yau $M_3$ with base $B=\mathbb{F}_n$. $M_3$ can equivalently be viewed as a K3 fibration over the $(-n)$-curve as the elliptic fibration over $F$ has second Chern class $24$ due to the two $E_8$-type degenerations of the elliptic fiber shown in the above figure. For pure 6d gauge theories the Euler number of $M_3$  is given purely in terms of group theory 
data as~\cite{Klemm:1996ts} 
\be 
\chi_{pG}(M_3)=-2 C(g_{E}) \int_{B} c_1^2(B)- {\rm rank}(g_{\Sigma}) C(g_{\Sigma}) \int_{\Sigma} c_1(\Sigma)\ .
\label{Eulernumber}  
\ee
Since we are interested in having a single section, we take $g_E = E_8$.  In this case the generic elliptic fiber can be given as a degree $6$ hypersurface in the weighted projective space $\mathbb{P}^2(1,2,3)$. For the different models labeled 
by $n$ the corresponding dual Coxeter number $C(g_{\Sigma})$ and the Euler numbers of the minimal compact Calabi-Yau manifolds 
are given in Table \ref{tab:euler}. 

\begin{table}[h!]
\begin{center}
	\begin{tabular}{|c|c|c|c|c|c|c|c|c|c|c|c|}
		\hline
	         \textrm{7-brane} 
                &n=1,2& 3 & 4 & 5 & 6 & 7& 8 & 9& 10& 11& 12 \\ \hline
		$g_{\Sigma}$ & - & $A_2$ & $D_4$ & $F_4$ & $E_6$ &$E_7^{(\frac{1}{2}HM)}$&  $E_7$ & $E^{(3)}_8$ & $E^{(2)}_8$ & $E^{(1)}_8$& $E_8$ \\
		$C_{g}$ & - & $3$ & $8$ & $12$ & $12$ & $18$ & $18$ & $30$ & $30$ & $30$ &  $30$  \\ 
		$-\chi(M_3)$& 480 & 492 & 528 & 576 & 624 & 676& 732 &780&840 &900 &960\\
                $h_{11}(M_3) - 1 $& 2 & 4 & 6 & 6 & 8 & 9 & 9 &13 &12 &11 &10\\
		\hline
	 \end{tabular}
\end{center}
\caption{Table of Coxeter numbers $C_{g}$ and Euler numbers $\chi(M_3)$ for the different minimal SCFT Calabi-Yau threefolds.}
\label{tab:euler}
\end{table}

Note that, compared to the local geometry associated to the 6d SCFT, the compact geometry 
leads to a larger number of hypermultiplets and  one additional vector multiplet. 

In the table, we also list the $E_8$ cases with a non-zero number $n_I=12-n$ of small instantons as $E_8^{(n_I)}$. 
Each instanton corresponds to an additional tensor multiplet in the 6d theory. Each of the latter contains one 
additional modulus, so $h_{11}$ increases  by $n_I$. The 6d anomaly cancellation condition~\cite{Green:1984bx,Erler:1993zy} 
 moreover enforces the relation $\#\text{HM} -\#\text{VM}=273- 29 n_I$, which implies $\chi(M_3)=\chi_{pG}(M_3) - 2 C_{{E_8}_E} n_I$. 
This yields the Hodge  numbers  $E_8^{(1,2,3)}$ in Table \ref{tab:euler}. The corresponding toric hypersurfaces are specified in Section~\ref{E8}.

\subsubsection*{Tate's algorithm for elliptic fiber singularities and toric constructions}

If the  generic fiber is the elliptic  curve $X_6(3,2,1)$, the elliptically fibered Calabi-Yau threefold 
over a base $B$ for this fiber type takes the Tate form   
\be
 y^2 + x^3 + a_6(\underline {u}) z^6 + a_4(\underline {u})  x z^4 + a_3(\underline {u})  y z^3 + a_2 
 (\underline {u}) z^2 x^2 + a_1(\underline {u})  z x y= 0 \ ,
 \label{tateform}
\ee
where coordinates on the base  $B$ are denoted generically by ${\underline{u}}$.

The construction of gauge singularities inside an elliptically fibered Calabi-Yau $n$-fold  
with the necessary toric data to solve the topological string proceeds as 
follows~\cite{Kachru:1997bz, Bizet:2014uua}. One constructs reflexive 
polyhedra such that  the Calabi-Yau is given by the anti-canonical hypersurface $W_{\Delta}(Y)=0$ in the corresponding toric variety, with $W_{\Delta}(Y)$ being in a generic Tate form (\ref{tateform}).
Then one chooses the divisor $\Sigma$ in $B$, restricts the coefficients of $W_{\Delta}(Y)$ so that at $\Sigma$ one has the suitable Tate singularity~\cite{Tate}, constructs the Newton polytope $\Delta_r$ to the restricted polyhedron and its dual $\Delta^*_r$, and finally resolves all non-toric divisors by modifying $\Delta_r$, without changing the 
singularity at $\Sigma$.

For instance, if $B$ is a $\mathbb{P}^1$ fibration over $\Sigma$, one splits the coordinates of the Tate form into $\{Y_k\}=\{z,x,y,u_1,u_2,w,v\}$, where 
$u_i$ are coordinates of $\Sigma$ and $w,v$ are homogeneous 
coordinates of the $\mathbb{P}^1$ fiber. For example, for $w=0$ the whole $\Sigma$ becomes a gauge divisor and by setting the  coefficients of the monomials in 
$a_i({\underline {u}},v,w,{\underline z})$ to zero (that is, choosing a specialization of the complex structure moduli), 
one can put the  $a_i({\underline {u}},v,w,{\underline z})$  in the following form:
\begin{equation}
a_1 = \alpha_1 w^{[a_0]} \, , \quad 
a_2 = \alpha_2 w^{[a_2]} \, , \quad 
a_3 = \alpha_3 w^{[a_3]} \, , \quad 
a_4 = \alpha_4 w^{[a_4]} \, , \quad 
a_6 = \alpha_6 w^{[a_6]} \, ,
\label{ai}  
\end{equation}
where $\alpha_i({\underline {u}},v,w,{\underline z})$ are of order 
zero in $w$. Choosing this leading behavior at $w=0$ leads by Tate's 
algorithm to singular fibers and hence results in a gauge group along $\Sigma$. The association 
of the leading powers of $[a_i]$ with the singularity is given by Tate's 
algorithm~\cite{Tate}. The discussion applies to the Hirzebruch surfaces 
as bases $B$ which can be viewed as a $\mathbb{P}^1$ fibration over $\Sigma=\mathbb{P}^1$. As already mentioned the divisor $\Sigma$ becomes $S'$ in this case.  

In the following sections we construct the minimal compact Calabi-Yau threefolds 
with the prescribed local geometries as hypersurfaces in a toric 
ambient space. Minimal means that they have just  one additional modulus, 
whose decompactification leads to the local geometry. The cases of 
main interest have the Euler number (\ref{Eulernumber}) as indicated in Table \ref{tab:euler}.       

\subsection{Solution of the topological string on the toric hypersurface Calabi-Yau spaces}

Generically the Calabi-Yau geometries under consideration can be described as anti-canonical hypersurfaces $H$ 
given by 
\be 
W_{\Delta}(Y)=\sum_{\nu_i\in\Delta} a_i \sum_{\nu^*_k\in \Delta^*} Y_k^{\langle \nu_i, \nu_k^*\rangle +1}=0\
\label{eq:cyconstraint} 
\ee
in $\mathbb{P}_{\Delta^*}$, where $(\Delta,\Delta^*)$ are  reflexive polyhedra. 

We denote by $\nu^*_i\in\mathbb{Z}^4 $ the relevant points of $\Delta^*$ whose complex 
hull in ${\mathbb{R}}^4$ is $\Delta^*$ and by $l^{(k)}$ the charges or Mori vectors, which fulfill    
\begin{equation} 
\sum_{i} l^{(k)}_i {\bar \nu}^*_i=0 \ , 
\end{equation}
where ${\bar \nu}^*_i=(1,\nu_i^*)$. The Mori vectors span the Mori cone, which is dual to the 
K\"ahler cone. The possible choices of Mori cones constitute the secondary 
fan whose data are encoded in the possible star triangulations of $\Delta^*$. 
Some of them are redundant, because the Calabi-Yau manifold still has the 
same Mori cones. Others correspond truly to different topological phases 
of the gauged linear sigma model. According to the theorem of C.T.C. Wall~\cite{CTCWall} 
the topological type of the Calabi-Yau threefold $M_3$ is fixed by the independent 
Hodge numbers, which for an $SU(3)$ holonomy manifold are $h_{11}$ and $h_{21}$, 
the triple intersection numbers $J_i \cdot J_j\cdot J_k$ and evaluation of the second Chern 
class on the basis $J_i$ of divisors dual to the basis of the K\"ahler cone. Only 
the $J_i \cdot J_j\cdot J_k$ change in a non-trivial way in the transitions.
Given the $l^{(k)}$ and the C.T.C. Wall topological data one can use 
toric mirror symmetry~\cite{Batyrev:1994hm} to predict the genus 
zero BPS numbers for all toric hypersurfaces following~\cite{Hosono:1993qy}.
We review the formalism that leads to the genus zero BPS numbers in 
Appendix \ref{LG}, see (\ref{omega0})-(\ref{periodsatinfinty}).

\subsubsection{ ${\cal O}(-n)\rightarrow \mathbb{P}^1$ geometries with $n=1,2$ }
\label{f1e8} 
The cases $n=1,2$ have only Kodaira fibers of type $I_1$ over codimension 
one in the base and hence no gauge theory divisor. 

To fix the notation used in the following sections we review the $n=1$ case, 
which is of particular interest  as the local geometry is the $\frac{1}{2}$K3 on which F-theory compactification 
yields the E-string theory. The refined BPS spectrum of the $E$-string has an  
interpretation as the refined stable pair invariants on the local 
geometry. The data associated to this geometry is summarized by the following table:
    
\begin{equation}  
 \begin{array}{ccrrrrr|rrr|rrr|} 
    \multicolumn{2}{c}{Div.} &\multicolumn{5}{c}{{\bar \nu}^*_i}     &l_I^{(e)}& l_I^{(f)}& l_I^{(b)}& l_{II}^{(e')}& l_{II}^{(h)}& l_{II}^{(-b)} \\ 
    D_0    &&     1&     0&  0&   0&   0&         -6&   0& 0  &   -6&   0& 0    \\ 
    D_1    &&     1&     -1&  0&   0&   0&         2&   0& 0    &  2&   0& 0     \\ 
    D_2    &&     1&     0&   -1&  0&   0&        3&   0& 0     &  3&   0& 0      \\  
    S'    &&     1&     2&   3&   0&   -1&       0&   1& -1    &  -1&   0& 1     \\ 
    K    &&     1&     2&   3&   0&    0&       1&   -2&-1    &  0&   -3& 1      \\ 
    F    &&     1&     2&   3&   -1&    -1&      0&   0& 1   &  1&   1& -1    \\
    S    &&     1&     2&   3&   0&    1&       0&   1& 0    &  0&   1& 0  \\ 
    F    &&     1&     2&   3&   1&    0&       0&   0& 1      &  1&  1& -1   \\ 
  \end{array} \ . 
  \label{F1case} 
\end{equation} 
The polyhedron $\Delta^*$ has two star triangulations, denoted by subscripts $I$ and $II$, which lead to different 
Mori cones in the secondary fan. Such  different Mori or K\"ahler cones can be understood as 
different geometrical phases of the 2d sigma model, which can have non-geometrical phases as well~\cite{Witten:1993yc}.

We give the C.T.C. Wall data for the first phase, namely phase $I$, which corresponds to the E-string geometry. 
Both phases have $h_{21}=243$ and $h_{11}=3$, and hence Euler number $\chi=-480$. The topological data in the 
phase marked with $I$ in (\ref{F1case}) are encoded  in            
\be
{\cal R}_I=8 J_e^3 + 3 J_e^2 J_f + J_e J_f^2 + 2 J_e^2 J_b + J_e J_f J_b\ ,
\ee
whose coefficients are the classical triple intersection numbers $c_{ijk}=\int J_iJ_jJ_k$. 
The evaluation of the second Chern class is $\int c_2 J_e=92, \int c_2 J_f=36,$  and $\int c_2 J_b=24$. We can see from 
(\ref{intersectionscurves}) that $l_I^{(b)}$ corresponds 
to the section $[C_b]=[S']$ of the base in $\mathbb{F}_1$, the $(-1)$ curve, while $l_I^{(f)}$ 
corresponds to the fiber $[C_f]=[F]$ in $\mathbb{F}_1$, a $(0)$ curve. Over the $(-1)$ curve one 
has a $\frac{1}{2}$-K3, which is the divisor $J_f$  dual to $[C_f]$ in $M_3$,  while over $[C_f]$ one has an elliptically 
fibered K3, which is the divisor $J_b$ dual to $[C_b]$ in $M$.  According to Oguiso's criterion~\cite{Oguiso} we 
see that the latter is a fibration of the geometry $M$ as the K3 does not intersect $J_b^2=0$ and 
$\int c_2J_b=24$. The class $[C_e]$ represents the elliptic fiber. 

The $E$-string partition function 
has the structure $Z=\exp(\lambda^{2g-2} F^{(g)}(Q_\tau,Q_b))$ where the free energies have the form   
$F^{(g)}=\sum_{n=0}^\infty \tilde F_n^{(g)}(Q_\tau) Q_b^n$.
Here $Q_\tau=\exp(2 \pi i \tau)$ with $\tau$  the modular parameter and  $\tilde F_n^{(g)}(Q_\tau)=
\frac{Q_\tau^{\frac{n}{2}}}{\eta(Q_\tau)^{12n }} P_n^{(g)}$ with $P_n^{(g)}(\hat E_2,E_4,E_6)$ an 
almost holomorphic modular form of weight $2g-6n-2$, e.g. $P_1^{(0)}=E_4$ etc.  
One has hence to redefine the K\"ahler parameters so that  $Q=Q_b Q_{\tau}^{\frac{1}{2}}$ 
and 
\be 
F^{(g)}=\sum_{n=0}^\infty F_n^{(g)}(Q_\tau) Q^n.\ 
\ee
In the above formula $F_n^{(g)}(Q_\tau)$ are truly $SL(2,\mathbb{Z})$ invariant coefficients. The 
analogous  redefinition has been made in (\ref{redefqb}) for the $D_4$ string. The  analysis 
of the monodromies of  $M_3$ that yield an $SL(2,\mathbb{Z})$ action on $\tau$ and a 
non-trivial decoupling limit fix the combination $Q$. This was discussed in detail 
in~\cite{Candelas:1994hw} in a similar context and applies to geometries discussed 
below.

The second phase is obtained by flopping the base $[C_b]$ out of the half K3, which becomes thereby 
an elliptic pencil, the del Pezzo surface with degree one. The latter can be obtained by eight 
blow ups of $\mathbb{P}^2$ and is called therefore  $d_8\mathbb{P}^2$. Note that the intersections 
are 
$$
{\cal R}_{II}=8 J_{e'}^3 + 3 J_{e'}^2 J_h + J_{e'} J_h^2 + 9 J_{e'}^2 J_{-b} + 3 J_{e'} J_{h} 
J_{-b} + J_{h}^2 J_{-b} + 9 J_{e'} J_{-b}^2 + 3 J_{h} J_{-b}^2 + 9 J_{-b}^3,
$$
while the evaluation of the second chern class is given by $\int c_2 J_i=\{92,36,102\}$. 
The transformation of the basis $l_{II}^{(e')}=l_{I}^{(e)}+ l_I^{(b)}$, 
$l_{II}^{(h)}=l_I^{(f)}+ l_I^{(b)}$ and  $l_I^{(-b)}=- l_I^{(b)}$ gives already almost the  
intersection ring ${\cal R}_{II}$ except that one gets  $8 J_{-b}^3$ instead of 
$9 J_{-b}^3$, i.e. in a coordinate independent formulation one observes that the 
triple intersection of the divisors dual to the rational curve that gets flopped 
increases by $+1$. This can be argued in general in various ways, see e.g.~\cite{Witten:1996qb}.         

The case $n=2$ has only one phase:
\begin{equation}  
 \begin{array}{ccrrrrr|rrr|} 
    \multicolumn{2}{c}{Div.} &\multicolumn{5}{c}{{\bar \nu}^*_i}     &l_I^{(e)}& l_I^{(f)}& l_I^{(b)} \\ 
    D_0    &&     1&     0&  0&   0&   0&         -6&   0& 0      \\ 
    D_1    &&     1&     -1&  0&   0&   0&         2&   0& 0    \\
    D_2    &&     1&     0&   -1&  0&   0&        3&   0& 0     \\
    S'    &&     1&     2&   3&   0&   -1&       0&   1& -2    \\
    K    &&     1&     2&   3&   0&    0&       1&   -2& 0    \\
    F    &&     1&     2&   3&   -1&    -1&      0&   0& 1   \\
    S    &&     1&     2&   3&   0&    1&       0&   1& 0    \\
    F    &&     1&     2&   3&   1&    0&       0&   0& 1      \\
  \end{array} \ . 
  \label{F1caseb} 
\end{equation} 
In this phase one has a $K3$  and an elliptic fibration 
and the intersection ring is in general 
\be 
{\cal R} = 8 J_e^3 + 4 J_e^2 J_f  + 2 J_e J_f^2 +  2 J_e^2 J_b + J_e J_f J_b, 
\ee 
with 
\be 
\int c_2 J_e=92, \qquad   \int c_2 J_f=48,\qquad   \int c_2 J_b=24 \ .
\ee
The $n=2$ geometry corresponds to the $ A_1 $  $\mathcal{N} = (2,0) $ SCFT; by making the elliptic fiber singular over the $(-2)$ curve, one obtains the M-string geometry which was studied in detail in \cite{Haghighat:2013gba,Haghighat:2013tka}.

\subsubsection{ ${\cal O}(-3)\rightarrow \mathbb{P}^1$ geometry with $\hat A_2$ resolution}
\label{f3geom}

The easiest example with a non-Higgsable gauge symmetry is the $A_2$ case, which has the 
following polyhedron $\Delta^*$:
\begin{equation} 
 \label{tab:dataf3} 
 \begin{array}{ccrrrr|rrrrrl|rrrrr|} 
    \multicolumn{2}{c}{Div.} &\multicolumn{4}{c}{\nu_i^*}     &l^{(1)}& l^{(2)}& l^{(3)} & l^{(4)} & l^{(5)} & &  l_{T^2}&  l_{\mathbb{P}^2}^{(1)}& l_{\mathbb{P}^2}^{(2)}&l_{\mathbb{P}^2}^{(3)}&  l_{de} \\ 
    D_0    &&         0&  0&   0&   0&         -1&   0& 0&  0& 0& & -6 &   -3  &0 & 0  & 0      \\ 
    D_1    &&         -1&  0&   0&   0&         0&   1& 0&  -1& 0&  &2 &   1 & 1 &  0 &  0  \\ 
    D_2    &&         0&   -1&  0&   0&        1&   0& 0&   0&-1&   & 3&    3&  0&   0  &0     \\ 
    D_3    &&        1&   1&   0&  -1&        -1&  0& 0&   0&  2&  & 0&   -3&  0&   0 & 0   \\ 
    D_4    &&         1&   2&   0&  -1&        1&  -3& 0&  3&  0&   & 0&   0 &  -3&  0&  0    \\ 
    S'    &&      2&   3&   0&   -1&       0&   0& 1& -3&  1&       & 0&  0 &  0&  -3  & 0   \\ 
    K    &&       2&   3&   0&    0&       0&   0&-2&  1&  0&        & 1&  0 &  0&  1 & -\frac{5}{3}    \\ 
    F    &&        2&   3&   -1&    -3&       0&   1& 0&  0& -1&     & 0&  1 &  1&  1 & \frac{1}{3}  \\
    S    &&        2&   3&   0&    1&       0&   0& 1&  0&   0&       & 0&   0&  0&  0&   1  \\ 
    F    &&         2&   3&   1&    0&       0&   1& 0&  0&  -1&       & 0&   1&  1&  1&  \frac{1}{3}  \\ 
  \end{array} \ . 
\end{equation}

We study the basis\footnote{The intersection ring and an alternate basis appropriate to the Landau-Ginzburg description is given in Appendix \ref{LG}.} 
which is appropriate to exhibit the curve classes that exhibit the affine $\hat A_2$ singularity over the divisor $S'$, which are depicted in the 
figure below: 
\begin{center}
\resizebox{3cm}{!}{\begin{tikzpicture}{xscale=1,yscale=1}  
\draw[thick] (xyz polar cs: angle=31.5,radius=2) --(xyz polar cs: angle=147.5,radius=2);
\node at  (xyz polar cs: angle=33,radius=2.3)   {$\bf 1$\ $1$} ;
\draw [fill,thick]  (xyz polar cs: angle=30,radius=2)  circle [radius=.1]; 
\draw[thick] (xyz polar cs: angle=152.5,radius=2) --(xyz polar cs: angle=267.5,radius=2); 
\node at  (xyz polar cs: angle=153,radius=2.4)   {$\bf 1$\ $2$} ;
\draw [fill,thick]  (xyz polar cs: angle=150,radius=2)   circle [radius=.1]; 
\draw[thick] (xyz polar cs: angle=272.5,radius=2) --(xyz polar cs: angle=27.5,radius=2); 
\node at  (xyz polar cs: angle=270,radius=2.3)   {$\bf 1$\ $3$} ;
\draw [fill,thick]  (xyz polar cs: angle=270,radius=2)   circle [radius=.1]; 
\end{tikzpicture}} 
\end{center}
This basis corresponds to the following choice of vectors:
\be
\begin{array}{rl} 
l_{\hat A_2}^{1}=4 l^{(1)}+l^{(2)}+l^{(5)}=&(-4, 1, 3, -2, \pmi 1, \pmi 1, 0, 0, 0, 0),\\ 
l_{\hat A_2}^{2}=l^{(1)}+l^{(2)}+l^{(5)}=&(-1, 1, 0, \pmi 1, -2, \pmi 1, 0, 0, 0, 0),\\
l_{\hat A_2}^{3}=l^{(1)}+l^{(2)}+l^{(4)}+l^{(5)}=&(-1,0, 0, \pmi 1, \pmi 1, -2, 1, 0, 0, 0),\\ 
l_b=-(l^{(1)}+l^{(5)})=&(\pmi  1, 0, 0, -1, -1, -1, 0, 1, 0, 1 ), \\
l_{de}=5 l^{(1)}+\frac{8}{3}l^{(2)}+l^{(3)}+ \frac{10}{9} l^{(4)}+\frac{8}{3}l^{(5)}=
&(-5, \frac{14}{9},\frac{7}{3}, \   \frac{1}{3}, \   \frac{1}{3}, \   \frac{1}{3}, -\frac{8}{9}, 0, 1, 0))\ . 
\end{array}
\ee
Note that we have flopped the $\mathbb{P}^1$ represented by the vector $l^{(1)}+l^{(5)}$ 
in the Mori cone  in order to arrive at the appropriate $\mathbb{P}^1$ base for the affine  
$\hat A_2$ singularity. The $\mathbb{P}^1$ base is represented by the Mori vector 
$l_b$, which intersects the three rational components of the degenerate elliptic 
curve with $(-1)$. The decompactification direction can be specified as a rational 
element in $H_2(M_3)$ so that the intersection form of the compact two-dimensional part 
becomes 
\be 
\frac{1}{3^3} J_b \sum_{i,j=1}^3 C_{ij} J_{\hat A_2}^{(i)} J_{\hat A_2}^{(j)}\ .
\ee

We note that the Coxeter labels $a^i$ have the property that for $C_{ij}$ the affine Cartan 
matrix  
\be
\sum_{j=0}^r a^i C_{ij}=0\ . 
\ee   
A first check on our identification is therefore that the $\mathbb{P}^1$ curve classes 
called $l^{(i)}_{\hat D_4}$  add up to the class of the elliptic fiber with the Coxeter lables 
indicated at the affine Dynkin diagram, i.e.     
$$
l_{T^2}= {\bf 1}  l^{(1)}_{\hat A_2}+ {\bf 1} l^{(2)}_{\hat A_2}+ {\bf 1} l^{(3)}_{\hat A_2} \ .
$$
This  is geometrically required, because the curve class of the elliptic fiber has self intersection $0$. 
We list in the following some of the BPS invariants $n^{(0)}_{d_b,d^{1}_{\hat A_2},d^{2}_{\hat A_2},d^{3}_{\hat A_2}}$, 
where the degree in the decompactified direction is zero. The number $d_b$ corresponds to the base wrapping number and therefore indicates the string charge in the 6d SCFT whereas the other numbers correspond to the flavor fugacity charges. The numbers $n^{(0)}_{d_b,d^{1}_{\hat A_2},d^{2}_{\hat A_2},d^{3}_{\hat A_2}}$ are symmetric in  $d^{1}_{\hat A_2},d^{2}_{\hat A_2},d^{3}_{\hat A_2}$. 
Since the emphasis of this paper is on the strings of the 6d SCFTs, we focus on the BPS invariants corresponding to $ n_b \geq 1 $. For example, the following tables display BPS invariants corresponding to $ n_b = 1,2 $ and small values of $ d^{\,i}_{\hat{A}_2} $.
\begin{table}[H]
\centering
\footnotesize{\begin{tabular} {|c|c|c|c|c|c|c|} 
\hline 
$d^1_{\hat A_2} \backslash d^2_{\hat A_2}$
   &  0 & 1 & 2  & 3  & 4 &5 \\  \hline  
0   &  1& 3& 5& 7& 9& 11\\  \hline
1  & 3& 4& 8& 12& 16& 20 \\  \hline
2  & 5& 8& 9& 15& 21& 27    \\  \hline
3  & 7& 12& 15& 16& 24& 32    \\  \hline 
4  & 9& 16& 21& 24& 25& 35 \\  \hline 
5  & 11& 20& 27& 32& 35& 36\\  \hline 
\end{tabular}} 
\vskip 3pt   $n^{(0)}_{d_b=1,d^1_{\hat A_2},d^2_{\hat A_2},0}$ 
\end{table}

\begin{table}[H]
\centering
\footnotesize{\begin{tabular} {|c|c|c|c|c|c|c|} 
\hline 
$d^1_{\hat A_2} \backslash d^2_{\hat A_2}$
&   1 & 2  & 3  & 4 &5  \\  \hline  
1&   16& 36& 60& 84& 108 \\  \hline  
2  &  36& 56&  96&  144& 192 \\  \hline
3  &  60& 96& 120& 180& 252  \\  \hline
4  &  84& 144& 180& 208& 288  \\  \hline 
5  &  108& 192& 252& 288& 320\\ \hline 
\end{tabular}} 
\vskip 3pt   $n^{(0)}_{d_b=1,d^1_{\hat A_2},d^2_{\hat A_2},1}$ 
\end{table}

\begin{table}[H]
\centering
\footnotesize{\begin{tabular} {|c|c|c|c|c|c|c|} 
\hline 
$d^1_{\hat A_2} \backslash d^2_{\hat A_2}$
&   2  & 3  & 4   &5            \\  \hline    
2  &  149& 288& 465& 651 \\  \hline
3  & 288& 456& 735& 1080  \\  \hline
4  & 465&  735& 954& 1371 \\ \hline
4  & 651&  1080& 1371&  -  \\ \hline
\end{tabular}}
\vskip 3pt   $n^{(0)}_{d_b=1,d^1_{\hat A_2},d^2_{\hat A_2},2}$ \end{table}

\begin{table}[H]
\centering
\footnotesize{\begin{tabular} {|c|c|c|} 
\hline 
$d^1_{\hat A_2} \backslash d^2_{\hat A_2}$ & 3  & 4             \\  \hline    
3  &  1012& 1788 \\  \hline
4  & 1788& -    \\  \hline
\end{tabular}}
\vskip 3pt   $n^{(0)}_{d_b=1,d^1_{\hat A_2},d^2_{\hat A_2},3}$
\end{table}

\begin{table}[H]
\centering
\footnotesize{\begin{tabular} {|c|c|c|c|c|c|c|} 
\hline 
$d^1_{\hat A_2} \backslash d^2_{\hat A_2}$
   &  0 & 1 & 2  & 3  & 4 &5 \\  \hline  
0   & 0& 0& -6& -32& -110& -288\\  \hline
1  &  0& 0& -10& -70& -270& -770\\  \hline
2  & -6& -10& -32& -126& -456& -1330    \\  \hline
3  &  -32& -70& -126& -300& -784& -2052   \\  \hline 
4  & -110& -270& -456& -784& -1584& -3360\\  \hline 
5  & -288& -770& -1330& -2052& -3360& -6076\\  \hline 
\end{tabular}} 
\vskip 3pt   $n^{(0)}_{d_b=2,d^1_{\hat A_2},d^2_{\hat A_2},0}$ 
\end{table}

\begin{table}[H]
\centering
\footnotesize{\begin{tabular} {|c|c|c|c|c|c|c|} 
\hline 
$d^1_{\hat A_2} \backslash d^2_{\hat A_2}$
   & 1 & 2  & 3  & 4 &5 \\  \hline  
1  & -8& -60& -360& -1432& -4280 \\  \hline
2  & -60& -216& -850& -3164& -9720  \\  \hline
3  &  -360& -850& -2176& -6084& -16960 \\  \hline 
4  & -1432& -3164& -6084& -13000& -29526\\  \hline 
5  & -4280& -9720& -16960& -29526& -\\  \hline 
\end{tabular}} 
\vskip 3pt   $n^{(0)}_{d_b=2,d^1_{\hat A_2},d^2_{\hat A_2},1}$ 
\end{table}

\begin{table}[H]
\centering
\footnotesize{\begin{tabular} {|c|c|c|c|c|c|c|} 
\hline 
$d^1_{\hat A_2} \backslash d^2_{\hat A_2}$
   &  0 & 1 & 2  & 3  & 4 &5 \\  \hline  
0   & 0& 0& 0& 27& 286& 1651\\  \hline
1  &  0& 0& 0& 64& 800& 5184\\  \hline
2  &  0& 0& 25& 266& 1998& 11473   \\  \hline
3  & 27& 64& 266& 1332& 6260& 26880   \\  \hline 
4  & 286& 800& 1998&   6260& 21070& 70362\\  \hline 
5  & 1651& 5184& 11473& 26880& 70362& 191424\\  \hline 
\end{tabular}} 
\vskip 3pt   $n^{(0)}_{d_b=3,d^1_{\hat A_2},d^2_{\hat A_2},0}$ 
\end{table}

\subsubsection{${\cal O}(-4)\rightarrow \mathbb{P}^1$ geometry with $\hat D_4$ resolution}
\label{sec:O(-4)}

In this section we describe the elliptic Calabi-Yau which has base $B = \mathbb{F}_4$ and corresponds to the two-dimensional quiver studied in Section \ref{sec:D4Quiver}. Taking the local limit by sending the size of the $\mathbb{P}^1$ fiber of $\mathbb{F}^4$ to infinity one arrives at a local Calabi-Yau which has an affine $\hat D_4$ Kodaira singularity over the $(-4)$ curve. The singularity in the elliptic fiber is resolved by sphere configurations with the affine $D_4$ intersection numbers and multiplicities as depicted below:


\begin{center}
\resizebox{3cm}{!}{\begin{tikzpicture}{xscale=1,yscale=1} 
\draw [fill,thick] (0,0) circle [radius=.1]; 
\node at  (.6,0)   {$c$} ;
\node at  (.3,0)   {$\bf 2$} ;
\draw[thick] (xyz polar cs: angle=45,radius=.1) --(xyz polar cs: angle=45,radius=1.9); 
\node at  (xyz polar cs: angle=39,radius=2.6)   {$1$} ;
\node at  (xyz polar cs: angle=45,radius=2.3)   {$\bf 1$} ;
\draw [fill,thick]  (xyz polar cs: angle=45,radius=2)  circle [radius=.1]; 
\draw[thick] (xyz polar cs: angle=135,radius=.1) --(xyz polar cs: angle=135,radius=1.9); 
\node at  (xyz polar cs: angle=141,radius=2.6)   {$2$} ;
\node at  (xyz polar cs: angle=135,radius=2.3)   {$\bf 1$} ;
\draw [fill,thick]  (xyz polar cs: angle=135,radius=2)   circle [radius=.1]; 
\draw[thick] (xyz polar cs: angle=225,radius=.1) --(xyz polar cs: angle=225,radius=1.9); 
\node at  (xyz polar cs: angle=219,radius=2.6)   {$3$};
\node at  (xyz polar cs: angle=225,radius=2.3)   {$\bf 1$} ;
\draw [fill,thick]  (xyz polar cs: angle=225,radius=2)   circle [radius=.1]; 
\draw[thick] (xyz polar cs: angle=315,radius=.1) --(xyz polar cs: angle=315,radius=1.9); 
\node at  (xyz polar cs: angle=321,radius=2.6)   {$4$} ;
\node at  (xyz polar cs: angle=315,radius=2.3)   {$\bf 1$} ;
\draw [fill,thick]  (xyz polar cs: angle=315,radius=2)   circle [radius=.1]; 
\end{tikzpicture}} 
\end{center}
The toric data are given by a reflexive polyhedron $\Delta^*$ , whose points $\nu$ are the first entries in the table below.

\begin{equation} 
 \label{dataf4} 
 \begin{array}{crrrr|rrrrrrr|rrrrrrrr|} 
    D &\multicolumn{4}{c}{\nu_i^*}     &l^{(1)}& l^{(2)}& l^{(3)} & l^{(4)} & l^{(5)} & l^{(6)} & l^{(7)} &
l_{T^2}& l_{\hat D_4}^{(1)}&  l_{\hat D_4}^{(2)}&l_{\hat D_4}^{(3)}&l_{\hat D_4}^{(4)}& l_{\hat D_4}^{(c)} & l_{de} \\   
    D_0    &          0&   0&   0&   0&         -1&   0&  0&  0&     0&    0&  0&  -6&   0&-2&-2&0& -1 & 0 &     \\ 
    D_1    &        -1&   0&   0&   0&          0&  -2&  0&  0&     0&    0&  1&    2&   0 & 2& 0&0& 0 &  0&\\ 
    D_2    &         0&  -1&   0&   0&          0&   0& -1&  0&     0&    1&  0&    3&  1  &1& 1&0& 0 & 0&\\ 
    D_3    &         1&   1&   0&  -1&          1&   0&  2&  0&     0&   -2&  0&     0&  -2&0&0 &0& 1  &0 &\\ 
    D_4    &         0&   1&   0&  -1&          1&   3&  0&  0&     0&    0& -2&      0& 0&-2&2&0& 0& 0&\\
    D_5    &        1&   2&   0&  -2&         -1&   0&  0&  0&    0&    0&  1&       0&  0&0& -2&0& 1 &0& \\ 
    S'    &        2&   3&   0&  -1&          0&   1& -2&  0&    1&    0&  0&        0& 0&0& 0&-2& 1 & 0&\\ 
    K    &         2&   3&   0&    0&         0&   0&  1&  0&    -2&    0&  0&    1   &0&0&0&   1  &0& -\frac{3}{2}&\\ 
    S''    &         2&   3&   0&   -2&         0&  -2&  0& -2&    0&    1&  0&     0&  1&1& 1&1& -2 & \frac{1}{2}& \\
    F    &         2&   3&  -1&   -4&         0&   0&  0&  1&     0&    0&  0&     0&  0&0&0&0& 0& 0&\\ 
    S &         2&   3&   0&    1&         0&   0&  0&  0&    1&    0&  0&     0&  0&0&0&0& 0 & 1 &\\
    F &         2&   3&   1&    0&         0&   0&  0&  1&     0&    0&  0&    0& 0 &0&0 &0& 0 & 0 &\\
   \end{array} \ . 
\end{equation} 

The Calabi-Yau hypersurface has $\chi(M_3)=-528$ and $h_{11}=7$. The polyhedron $\Delta^*$ 
has $30$ star triangulations and  the $l^{(1)},\ldots, l^{(7)}$ are generators of a simple  
geometrical Mori cone, which are needed to solve the topological string on the global model. 
Note that the evaluation of the K\"ahler classes against the second Chern class are:   
$$\{\int c_2 J_i\}=\{620, 204,140,24, 72, 452,616\}. $$ 
It turns out that $J_4$ appears only linearly in the intersection ring and $\int c_2 J_4=24$. Oguiso's criterion implies that  $M_3$ is a  $K3$ fibration whose base $ \mathbb{P}^1 $ is represented by $l^{(4)}$. This $ \mathbb{P}^1 $ is also the base of the local surface $B = \mathcal{O}(-4)\rightarrow \mathbb{P}^1$ and we will henceforth denote it by $l_b$. Since $J_5$ appears only quadratically in the intersection ring , $l^{(5)}$ represents the base of the elliptic fibration. Also, $l_{T^2}$ represents  the elliptic fiber class and the $l^{(i)}_{D_4}$ correspond to the simple roots of the affine  $\hat D_4$. 
Finally, $l_{de}$ is the class that one can take large to zoom in on the local surface geometry. The relation to the nef 
classes in the global model are 
\be
\begin{array}{rl} 
l_{T^2}&=6 l^{(1)}+2 l^{(2)}+l^{(3)} +4 l^{(6)}+ 6 l^{(7)}, 
l^{(1)}_{\hat D_4}=l^{(6)},\ \  
l^{(2)}_{\hat D_4}=2l^{(1)}+l^{(6)}+2 l^{(7)}, \\   
l^{(3)}_{\hat D_4}&=2l^{(1)}+l^{(6)}, \  \ 
l^{(4)}_{\hat D_4}=l^{(3)}+l^{(6)}, \  \
l^{(c)}_{\hat D_4}=l^{(1)}+l^{(2)}+2 l^{(7)} \ .
\end{array}
\ee
A first check on these identifications is that the $\mathbb{P}^1$ curve classes called $l^{(i)}_{\hat D_4}$ add up to the class of the elliptic fiber with Coxeter labels $a^i$ indicated in the affine Dynkin diagram, i.e.     
\begin{equation} \label{eq:ellD4}
l_{T^2}= {\bf 1}  l^{(1)}_{\hat D_4}+ {\bf 1} l^{(2)}_{\hat D_4}+ {\bf 1} l^{(3)}_{\hat D_4}+ {\bf 1} l^{(4)}_{\hat D_4}+{\bf 2}  l^{(c)}_{\hat D_4} \ ,
\end{equation}
which is obviously the case. Another check is that after transforming to this basis the intersection form takes on a very simple appearance  and is symmetric in  $l^{(i)}_{\hat D_4}$, $i=1,\ldots, 4$:
\be{\cal R}= 
9 J_{T^2}^3+\frac{3}{2} J_4 J_{T^2}^2+6 J_{de} J_{T^2}^2+4 J_{de}^2 J_{T^2}+J_4 J_{de}J_{T^2}-\sum_{i=1}^4 (J_{\hat D_4}^{(i)})^3 -\frac{1}{2} J_4 \sum_{i=1}^4 (J_{\hat D_4}^{(i)})^2.
\ee   

The curve whose volume has to be scaled to infinity to decouple the ${\cal O}(-4)\rightarrow \mathbb{P}^1$  geometry 
from the compact manifold is the K\"ahler class dual to the Mori cone vector  $l^{(5)}$. This  
decompactifies the base of the Cababi-Yau threefold by scaling the fiber of the Hirzebruch 
surface to infinity. The class can be further modified to $l_{de}$ above to make the intersections even simpler.               
  
Let us next come to the evaluation of BPS numbers. We denote the charge associated to the class dual to $l^{(4)}$, which is the base, by $n_b$, the  one dual to $l_{de}$ by $n_{de}$, the ones dual to $l_{\hat D_4}^{(i)}$ by $n_1,n_2,n_3,n_4,n_c$   
and the one dual to $l_{T^2}$ by $n_e$. From the viewpoint of the strings of the 6d SCFT, $n_b$ denotes the string charge; $n_i, i=1, \cdots, 4, n_c$ correspond to the flavor fugacity charges and $n_e$ is the exponent of $Q_{\tau}$ in an expansion of the elliptic genus $Z_{n_b}$. We consider by definition of the local limit only $n_{de}=0$. Due to the relation (\ref{eq:ellD4}) the class $l_{T^2}$ is not an independent class and therefore when labelling BPS states we can omit the dependence on $n_e$. The genus zero invariants are then given by $n^{(0)}_{n_b,n_1,n_2,n_3,n_4,n_c}$ and are fully symmetric in $n_1,\ldots,n_4$. Let us first focus on the BPS states associated to a single string (that is, those corresponding to  $ n_b = 1 $). 
For $n_e=0$ and  $n_c=1$ we get:
\begin{table}[H]
\centering
\footnotesize{\begin{tabular} {|c|c|c|c|c|c|c|c|} 
\hline 
$n_1 \backslash n_2$ & 0 & 1 & 2  & 3  & 4 &5 &6   \\  \hline  
0   & -4& -6& -6& -10& -14& -18& -22  \\  \hline  
1  & -6& -8& -6& -10& -14& -18& -22  \\  \hline
2  & -6& -6& 0& 0& 0& 0& 0\\  \hline
3  & -10& -10& 0& 0& 0& 0& 0 \\  \hline
4  &   -14& -14& 0& 0&   0& 0& 0     \\  \hline 
5 & -18& -18& 0& 0& 0& 0& 0 \\ \hline 
6& -22& -22& 0& 0& 0& 0& 0 \\  \hline
\end{tabular}} 
\vskip 3pt   $n^{(0)}_{1,n_1,n_2,0,0,1}$ 
\end{table}

For $n_e=0$ and  $n_c=2$ we get:
\begin{table}[H]
\centering
\footnotesize{\begin{tabular} {|c|c|c|c|c|c|c|c|} 
\hline 
$n_1 \backslash n_2$ & 0 & 1 & 2  & 3  & 4 &5 &6   \\  \hline  
0&  -6& -10& -12& -12& -18& -24& -30  \\  \hline  
1 &-10& -16& -18& -16& -24& -32& -40 \\  \hline
2  &-12& -18& -18& -12& -18& -24& -30\\  \hline
3  &-12& -16& -12& 0& 0& 0&  0\\  \hline
4  &-18& -24& -18& 0& 0& 0& 0 \\  \hline
5   & -24& -32& -24& 0& 0& 0&   0  \\  \hline 
6  &-30& -40& -30& 0& 0& 0& 0\\ \hline  
\end{tabular}} 
\vskip 3pt   $n^{(0)}_{1,n_1,n_2,0,0,2}$ 
\end{table}

For $n_e=1$ one finds:
\begin{table}[H]
\centering
\footnotesize{\begin{tabular} {|c|c|c|c|c|c|} 
\hline 
$n_1 \backslash n_2$ & 0 & 1 & 2  & 3  & 4   \\  \hline  
0   & -80& -78& -96& 144& -192\\  \hline  
1  &  -78& -48& 32&-& - \\  \hline
2  & -96& -32& - & -& - \\  \hline
3  & -144& -& -& -& - \\  \hline
4  &   -192& -& -&-& -   \\  \hline 
\end{tabular}} 
\vskip 3pt   $n^{(0)}_{1,n_1+1,n_2+1,1,1,2}$ 
\end{table}

Let us now consider the two-string sector. For $n_b=2$, $n_e=0$  and  $n_c=1$ we get:
\begin{table}[H]
\centering
\footnotesize{\begin{tabular} {|c|c|c|c|c|c|c|c|} 
\hline 
$n_1 \backslash n_2$ & 0 & 1 & 2  & 3  & 4 &5 &6   \\  \hline  
0&  -6& -10& -12& -30& -98& -306& -814 \\  \hline  
1 & -10& -16& -18& -40& -112& -324& -836  \\  \hline
2  &-12& -18& -18& -30& -42& -54& -66 \\  \hline
3  &-30& -40& -30& -50& -70& -90& -110 \\  \hline
4  &  -98& -112& -42& -70& -98& -126& -154\\  \hline
5   &  -306&-324& -54& -90& -126& -162& -198 \\  \hline 
6  &-814& -836& -66& -110& -154& -198& -242 \\ \hline  
\end{tabular}} 
\vskip 3pt   $n^{(0)}_{2,n_1,n_2,0,0,1}$ 
\end{table}

For $ n_b = 1,\dots, 4 $ and $n_c = 0$ we also find the following invariants:
\begin{table}[H]
\centering
\footnotesize{\begin{tabular} {|c|c|c|c|c|c|c|c|c|c|c|c|} 
\hline 
$n_b \backslash n_1$ & 0 & 1 & 2  & 3  & 4 &5 &6 & 7 & 8 & 9 & 10 \\ \hline  
1  & -2 & -2 & -4 & -6 & -8 & -10 & -12 & -14 & -16 & -18 & -20 \\ \hline
2  & 0 & 0 & 0 & -6 & -32 & -110 & -288 & -644 & -1280 & -2340 & 4000\\ \hline
3  & 0 & 0 & 0 & 0 & -8 & -110 & -756 & -3556 & -13072 & -40338 & -109120 \\ \hline 
4 & 0 & 0 & 0 & 0 & 0 & -10 & -288 & -3556 & -27264 & -153324 & -690400  \\ \hline 
\end{tabular}} 
\vskip 3pt   $n^{(0)}_{n_b,n_1,0,0,0,0}$ 
\end{table}

Remarkably, and this is the main non-trivial test of the paper, all the invariants listed above can be reproduced from the elliptic genus computation in Section \ref{sec:D4Quiver}!

\subsubsection{${\cal O}(-5)\rightarrow \mathbb{P}^1$ geometry with $\hat F_4$ resolution}

This elliptic singularity corresponds now to a non-simply laced Lie  algebra. Unlike in the simply laced case, the Coxeter labels differ from the dual Coxeter labels. We indicate both Coxeter/dual Coxeter labels in the following diagram:
\begin{center}
\resizebox{4cm}{!}{\begin{tikzpicture}{xscale=1cm,yscale=1cm}
\coordinate[label=below:${\bf 1/1}$,label=above:$0$](A) at (-2,0);
\coordinate[label=below:${\bf 2/2}$,label=above:$1$](B) at (-1,0);
\coordinate[label=below:${\bf 3/3}$,label=above:$2$](C) at (0,0);
\coordinate[label=below:${\bf 4/2}$,label=above:$3$](D) at (1,0);
\coordinate[label=below:${\bf 2/1}$,label=above:$4$](E) at (2,0);
\draw (A)--(B)--(C);
\draw[<-, double] (C) -- (D);
\draw (D)--(E);
\draw (A) circle (.1);
\fill (B) circle (.1);
\fill (C) circle (.1);
\fill (D) circle (.1);
\fill (E) circle (.1);
\end{tikzpicture}} 
\end{center}
The toric polyhedron has 25 star triangulations; we present here the polyhedron together with the simplest choice of 
Mori vectors:
\begin{equation} 
 \label{dataf4} 
 \begin{array}{crrrr|rrrrrrr|rrrrrrrr|} 
    D &\multicolumn{4}{c}{\nu_i^*}     &l^{(1)}& l^{(2)}& l^{(3)} & l^{(4)} & l^{(5)} & l^{(6)} & l^{(7)} 
    &l_{\hat F_4}^{(4)}& l_{\hat F_4}^{(3)}&  l_{\hat F_4}^{(3)}&l_{\hat F_4}^{(1)}&l_{\hat F_4}^{(0)}& \\   
    D_0    &          0&  0&  0&  0&    -2& 0   & 0 &0 &  0     &   0  & 0   &  0 & 0  & 0 & 0& -2& \\ 
    D_1    &        -1&   0&   0&   0&  0& -2   & 0 &0 &  0     &   0  & 1   &  1 & 0  & 0 & 0&  0&  \\ 
    D_2    &         0&  -1&   0&   0&  1&  0   & 0 &0 &  0     &   0  & 0   &  0 & 0  & 0 & 0&  1&  \\ 
    D_3    &         0&   1&   0&  -1&  0& 3   & 0 &0 &   0     &    1 & -2  &  -2& 1  & 0 & 0&  0&  \\ 
    D_4    &         1&   2&   0&  -2&  2& 0   & 0 &0 &   0     &    -2 & 1  &  1 &-2  & 0 & 0&  2&  \\
    S'     &         2&   3&   0&  -1&  0& 1   & -2&0 &   1     &    0 & 0   &  0 & 0  &-2 & 1&  0&  \\ 
    S''    &         2&   3&   0&  -2&  1& -2   &  1&-1&   0    &    0 & 0   &  0 & 0  & 1 &-2&  1&  \\ 
    S'''   &         2&   3&   0&  -3& -2& 0   & 0 &-1&  0      &   1  & 0   &  0 & 1  & 0 & 1& -2&  \\
    K      &         2&   3&   0&   0&  0& 0   & 1 &0 &  -2     &    0 &  0  &  0 & 0  & 1 & 0&  0&  \\ 
    F      &         2&   3&  -1&  -5&  0& 0   & 0 &1 &  0      &   0  & 0   &  0 & 0  & 0 & 0&  0&  \\ 
    S      &         2&   3&   0&   1&  0& 0   & 0 &0 &  1      &    0 & 0   &  0 & 0  & 0 & 0&  0&  \\
    F      &         2&   3&   1&   0&  0& 0   & 0 &1 &  0      &   0  & 0   &  0 & 0  & 0 & 0&  0&  \\
   \end{array} \ . 
\end{equation} 
Evaluation of the second Chern class on the K\"alher forms yields 
$$\{\int c_2 J_i\}=\{336, 240,164,24, 84, 692,708\}\ . $$  
The intersection ring has the property that $J_4$ appears only linearly so that $l^{(4)}$ represents 
the base of a K3 fibration and the base of the local surface $B$, we therefore have $l^{(4)} = l_b$.   $J_5$ appears only quadratically in the intersection ring  
and $l^{(5)}$ represents the base of an elliptic fibration. As before, it is the normal direction to the base of the 
local surface and gets decompactified.   

We find the following basis, which reflect  the curves that represent the Cartan elements of the  affine $\hat F_4$.   
\be
l^{(0)}_{\hat F_4}=l^{(3)},\quad l^{(1)}_{\hat F_4}=2 l^{(7)} + l^{(2)} + l^{(6)},
\quad l^{(2)}_{\hat F_4}=l^{(1)} \quad l^{(3)}_{\hat F_4}=l^{(6)}, \quad l^{(4)}_{\hat F_4}=l^{(7)}\ .
\ee
We see that in this basis
\be 
l_{T^2}=l_{\hat F_4}^{(0)}+2 l_{\hat F_4}^{(1)} + 3 l_{\hat F_4}^{(2)}+ 4 l_{\hat F_4}^{(3)} + 2 l_{\hat F_4}^{(4)},
\ee
as expected (the vector corresponding to a given node is multiplied by the Coxeter label of that node). Note that  the height in the last coordinate of  $\nu_i^*$ is the dual Coxeter 
number of the Dynkin diagram of $\mathbb{F}_4$. This is a consequence of the $F$-theory 
realization of the $G$  bundle moduli space of the heterotic string on an elliptically 
fibered K3 over the same $\mathbb{P}^1$ base as $\mathbb{P}(a_0,\ldots, a_r)$~\cite{Friedman:1997yq}
and will hold for all models below. From these data it is possible to calculate genus zero BPS invariants analogously to the $ n \leq 4 $ cases\footnote{We have calculated thes BPS invariants up to high multi-degree; these numbers are available on request.}.

\subsubsection{${\cal O}(-6)\rightarrow \mathbb{P}^1$ geometry with $\hat E_6$ resolution}
\begin{center}
\resizebox{4cm}{!}{\begin{tikzpicture}{xscale=1cm,yscale=1cm}
\coordinate[label=below:${\bf 1}$,label=above:$6$](A) at (-2,0);
\coordinate[label=below:${\bf 2}$,label=above:$5$](B) at (-1,0);
\coordinate[label=below:${\bf 3}$, label={[label distance=.6mm]45:$4$}](C) at (0,0);
\coordinate[label=below:${\bf 2}$,label=above:$3$](D) at (1,0);
\coordinate[label=below:${\bf 1}$,label=above:$1$](E) at (2,0);
\coordinate[label=left:${\bf 2}$,label=right:$2$](F) at (0,1);
\coordinate[label=left:${\bf 1}$,label=right:$0$](G) at (0,2);
\draw (A)--(B)--(C)--(D)--(E);
\draw (C)--(F);
\draw[thick] (0,1) --(0,1.9);
\fill (A) circle (.1);
\fill (B) circle (.1);
\fill (C) circle (.1);
\fill (D) circle (.1);
\fill (E) circle (.1);
\fill (F) circle (.1);
\draw (G) circle (.1);
\end{tikzpicture}} 
\end{center}

In this hase the hypersurface CY has Euler number  $\chi(M_3)=-624$ and $h_{11}=9$.
The  polyhedron $\Delta^*$ has  199 star triangulations. Again we  choose a simple one  
\begin{equation} 
 \label{dataf6} 
 \begin{array}{crrrr|rrrrrrrrrr|} 
    D &\multicolumn{3}{c}{\nu_i^*}    & &l^{(1)}& l^{(2)}& l^{(de)} & l^{(4)} & l^{(b)} & l^{(6)} & l^{(7)} & l^{(8)} & l^{(9)}& \\   
    D_0    &         0&   0&   0&   0&0  & 0    & 0  &0 & 0   & 0    & 0 & -1 & 0  &    \\ 
    D_1    &        -1&   0&   0&   0&-2 & 0    & 0  &0 & 0   & 0    & 0 &  0 & 1  &   \\ 
    D_2    &         0&  -1&   0&   0&0  &-1    & 0  &0 & 0   & 0    & 0 &  1 & 0  &  \\ 
    D_3    &         0&   0&   0&  -1&0  & 0    & 0  &0 & 0   & 0    & 1 & -1 &-1  &   \\ 
    D_4    &         0&   1&   0&  -1&3  & 0    & 0  &0 & 0   & 0    &-1 &  1 &-1  &   \\
    D_5    &         1&   1&   0&  -2&0  & 2    & 0  &0 & 0   & 1    &-1 &  0 & 1  &   \\
    D_6    &         1&   2&   0&  -2&0  & 0    & 0  &0 & 0   &-2    & 1 &  0 & 0  &  \\ 
    F      &         2&   3&  -1&  -6&0  & 0    & 0  &0 & 1   & 0    & 0 &  0 & 0  &   \\ 
    S'     &         2&   3&   0&  -3&0  & -2   & 0  &0 & -2  & 1    & 0 &  0 & 0  &  \\ 
    S''    &         2&   3&   0&  -2&-2 & 1    & 0  &1 & 0   & 0    & 0 &  0 & 0  &   \\
    S'''   &         2&   3&   0&  -1&1  & 0    & 1  &-2& 0   & 0    & 0 &  0 & 0  &  \\ 
    K      &         2&   3&   0&   0&0  & 0    &-2  &1 & 0   & 0    & 0 &  0 & 0  &   \\ 
    S      &         2&   3&   0&   1&0  & 0    & 1  &0 & 0   & 0    & 0 &  0 & 0  &   \\
    F      &         2&   3&   1&   0&0  & 0    & 0  &0 & 1   & 0    & 0 &  0 & 0\ ,   &   \\
   \end{array} \  
\end{equation} 
which has the property that
$$\{\int c_2 J_i\}=\{276, 360, 96, 188, 24, 764, 1524, 728, 800\}.
$$
By Oguiso's criterion $J_b$ represents the volume of the base of a K3 fibration 
while $J_{de}$ respresents the volume of the base of an elliptic fibration. We have calculated the genus zero BPS invariants up to multi-degree 21.

\subsubsection{The cases with $\hat E_7$ resolution}
\label{E7}

Here we discuss the cases ${\cal O}(-n)\rightarrow \mathbb{P}^1$ with 
$n=7$ and $n=8$. Let us start with $n=8$, that is, the pure $E_7$ gauge 
theory case. The reflexive polyhedron $\Delta^*$ is the convex hull 
of the points $\nu_i^*$ listed below, together with the Mori cone 
that corresponds to a simple of a total of $420$ star triangulations of 
$\Delta^*$.         

\begin{center}
\resizebox{7cm}{!}{\begin{tikzpicture}{xscale=1cm,yscale=1cm}
\coordinate[label=below:${\bf 1}$,label=above:$0$](A) at (-3,0);
\coordinate[label=below:${\bf 2}$,label=above:$1$](B) at (-2,0);
\coordinate[label=below:${\bf 3}$,label=above:$2$](C) at (-1,0);
\coordinate[label=below:${\bf 4}$, label={[label distance=.6mm]45:$3$}](D) at (0,0);
\coordinate[label=below:${\bf 3}$,label=above:$5$](E) at (1,0);
\coordinate[label=below:${\bf 2}$,label=above:$6$](F) at (2,0);
\coordinate[label=below:${\bf 1}$,label=above:$7$](G) at (3,0);
\coordinate[label=left:${\bf 2}$,label=right:$4$](H) at (0,1);
\draw (B)--(C)--(D)--(E)--(F)--(G);
\draw (D)--(H);
\draw[thick] (-2,0) --(-2.9,0);
\draw (A) circle (.1);
\fill (B) circle (.1);
\fill (C) circle (.1);
\fill (D) circle (.1);
\fill (E) circle (.1);
\fill (F) circle (.1);
\fill (G) circle (.1);
\fill (H) circle (.1);
\end{tikzpicture}} 
\end{center}
  
\begin{equation} 
 \label{dataf7} 
 \begin{array}{crrrr|rrrrrrrrrr|} 
    D &\multicolumn{3}{c}{\nu_i^*}   &  &l^{(1)}& l^{(2)}& l^{(3)} & l^{(4)} & l^{(5)} & l^{(6)} & l^{(7)} & l^{(8)} & l^{(9)} & l^{(10)}  \\   
    D_0    &         0&   0&   0&   0&0  & 0    & 0  &0 & 0   & 0    & 0 &  0 & 0  & -1 \\ 
    D_1    &        -1&   0&   0&   0&-1 & 0    & 0  &0 & 0   & 0    & 0 &  0 & 0  & 1 \\ 
    D_2    &         0&  -1&   0&   0&0  &-1    & 0  &0 & 0   & 0    & 0 &  0 & 1  & 0 \\ 
    D_3    &         0&   0&   0&  -1&0  & 0    & 0  &0 & 0   & 0    & 0 &  1 &-2  & 0 \\ 
    D_4    &         0&   1&   0&  -2&0  & 0    & 0  &0 & 0   & 0    & 1 &  1 & 1  & -1 \\
    D_5    &         1&   1&   0&  -2&0  & 2    & 0  &0 & 0   & 0    & 0 &  -1 & 0  & 1 \\
    D_6    &         1&   2&   0&  -3&3  & 0    & 0  &0 & 0   & 0    & -2&  -1 & 0  & 0 \\ 
    S'     &         2&   3&   0&  -4&-3 & 0    & -2 &0 & 0   & 0    &1  &  0 & 0  & 0 \\ 
    S''    &         2&   3&   0&  -3&1  &-2    & 0  &0 & 0   & 1    & 0 &  0 & 0  & 0 \\ 
    S'''   &         2&   3&   0&  -2&0  & 1    & 0  &0 & 1   &-2    & 0 &  0 & 0  & 0 \\
    S''''  &         2&   3&   0&  -1&0  & 0    & 0  &1 &-2   & 1    & 0 &  0 & 0  & 0 \\ 
    F      &         2&   3&  -1&  -8&0  & 0    & 1  &0 & 0   & 0    & 0 &  0 & 0  & 0 \\ 
    K      &         2&   3&   0&   0&0  & 0    & 0  &-2 & 1  & 0    & 0 &  0 & 0  & 0 \\ 
    S      &         2&   3&   0&   1&0  & 0    & 0  &1 & 0   & 0    & 0 &  0 & 0  & 0 \\
    F      &         2&   3&   1&   0&0  & 0    & 1  &0 & 0   & 0    & 0 &  0 & 0  & 0 \\
   \end{array} \  
\end{equation} 
By standard toric methods we calculate $\chi(M_3)=-732$ and $h_{11}=10$, i.e. $h_{21}=376$. For this choice of Mori cone one has:
$$\{\int_{M_3} c_2 J_i\}=\{560, 456, 24, 120, 236, 348, 1724, 1764, 884, 848\}.
$$
In the intersection ring $J_3$ appears only linearly, while $J_4$ appears only 
quadratically indicating a K3- and an elliptic fibration structure 
respectively.    

The case with an additional  $\frac{1}{2} {\bf 56}$ hypermultiplet is obtained by replacing 
the point $\nu^*_F=(2,3,-1,-8)$ with $\nu^*_F=(2, 3,-1,-7)$. The Hodge numbers change to 
$h_{21}= 348$, $h_{11}=10$, and hence $\chi(M_3)=-676$. The only 
change in the Mori generators is the modified element 
$$l^{(3)}=(0,0,0,0,0,0,0,-1,-1,0,0,1,0,0,1),$$
which leads to 
$$\{\int_{M_3} c_2 J_i\}=\{ 524,408,24,108,212,312,1592,1608,806,792\}.
$$
While the intersection ring is modified, the fibration structure with respect to the 
classes $J_3$ and $J_4$ is maintained.  For the $ \hat E_7 $ geometries, we have computed genus zero BPS invariants up to multi-degree 21.

\subsubsection{The cases with $\hat E_8$ resolution }
\label{E8}
\begin{center}
\resizebox{7cm}{!}{\begin{tikzpicture}{xscale=1cm,yscale=1cm}
\coordinate[label=below:${\bf 1}$,label=above:$0$](A) at (-5,0);
\coordinate[label=below:${\bf 2}$,label=above:$8$](B) at (-4,0);
\coordinate[label=below:${\bf 3}$,label=above:$7$](C) at (-3,0);
\coordinate[label=below:${\bf 4}$,label=above:$6$](D) at (-2,0);
\coordinate[label=below:${\bf 5}$,label=above:$5$](E) at (-1,0);
\coordinate[label=below:${\bf 6}$, label={[label distance=.6mm]45:$4$}](F) at (0,0);
\coordinate[label=below:${\bf 4}$,label=above:$3$](G) at (1,0);
\coordinate[label=below:${\bf 2}$,label=above:$1$](H) at (2,0);
\coordinate[label=left:${\bf 3}$,label=right:$2$](I) at (0,1);
\draw (A)--(B)--(C)--(D)--(E)--(F)--(G)--(H);
\draw (F)--(I);
\draw[thick] (-4,0) --(-4.9,0);
\draw (A) circle (.1);
\fill (B) circle (.1);
\fill (C) circle (.1);
\fill (D) circle (.1);
\fill (E) circle (.1);
\fill (F) circle (.1);
\fill (G) circle (.1);
\fill (H) circle (.1);
\fill (I) circle (.1);
\end{tikzpicture}} 
\end{center}

The Calabi-Yau hypersurface has $h_{11}=11$ and $\chi(M_3)=-960$ and is the 
case with maximal absolute value of the Euler number within the class of toric
hypesurfaces. From the 588 star triangulations of $\Delta^*$ we choose one leading to simple Mori cone with 
desired fibration structure:
\begin{equation} 
 \label{dataf12} 
 \begin{array}{crrrr|rrrrrrrrrrrr|}
    D &\multicolumn{3}{c}{\nu_i^*}   &  &l^{(1)}& l^{(2)}& l^{(3)} & l^{(4)} & l^{(5)} & l^{(6)} & l^{(7)} & l^{(8)} & l^{(9)} & l^{(10)}& l^{(11)}  &\\
    D_0    &         0&   0&   0&   0&0  & 0    &0 & 0   & 0    & 0 &  0 & 0&-1& 0& 0& \\ 
    D_1   &         -1&   0&   0&   0&-2 & 0    &0 & 0   & 0    & 0 &  0 & 0& 0& 0& 1& \\ 
    D_2    &         0&  -1&   0&   0&1  & -1    &0 & 0   & 0    & 0 &  0 & 0& 0& 0& 0&  \\ 
    D_3    &         0&   1&   0&  -2&3  & 0    &0 & 0   & 0    & 0 &  0 & 0& 0& 1&-2&   \\
    D_4    &         1&   1&   0&  -3&-2 & 2    &0 & 0   & 0    & 0 &  0 & 0& 1& 0& 0&   \\
    D_5    &         1&   2&   0&  -4&0  & 0    &0 & 0   & 0    & 0 &  0 & 0& 1&-2& 1&   \\
    F      &         2&   3&  -1& -12&0  & 0    &0 & 0   & 0    & 0 &  0 & 1& 0& 0& 0&  \\ 
    S'     &         2&   3&   0&  -6&0  & 0    &0 & 0   & 0    & 0 &  0 &-2&-2& 1& 0&   \\ 
    S''    &         2&   3&   0&  -5&0  & -2    &0 & 0   & 0    & 0 &  1 & 0& 1& 0& 0&   \\ 
    S'''   &         2&   3&   0&  -4&0  & 1    &0 & 0   & 0    & 1 & -2 & 0& 0& 0& 0&   \\
    S''''  &         2&   3&   0&  -3&0  & 0    &0 & 0   & 1    &-2 &  1 & 0& 0& 0& 0&   \\ 
    S''''' &         2&   3&   0&  -2&0  & 0    &0 & 1   &-2    & 1 &  0 & 0& 0& 0& 0&   \\ 
    S''''''&         2&   3&   0&  -1&0  & 0    &1 &-2   & 1    & 0 &  0 & 0& 0& 0& 0&   \\ 
    K      &         2&   3&   0&   0&0  & 0    &-2& 1   & 0    & 0 &  0 & 0& 0& 0& 0&   \\ 
    S      &         2&   3&   0&   1&0  & 0    &1 & 0   & 0    & 0 &  0 & 0& 0& 0& 0&   \\
    F      &         2&   3&   1&   0&0  & 0    &0 & 0   & 0    & 0 &  0 & 1& 0& 0& 0&   \\
   \end{array} \  
\end{equation} 
The evaluations of the K\"ahler forms against the second Chern class 
are: 
$$\{\int_{M_3} c_2 J_i\}=\{ 1496, 948,168, 332, 492, 648, 800, 24, 1092,2228, 3360\}.
$$
We have a K3 fibration over the $\mathbb{P}^1$ represented by $J_8$ and an 
elliptic fibration over the base whose volume is given by $J_3$. As in all 
models with K3 fibrations, the BPS numbers depend only on the square of the curve 
classes in the K3, whose intersection form in the Picard lattice is given by 
coefficients of the ring that is linear in $J_8$, and can be obtained by 
a heterotic one-loop calculation. The genus zero BPS states are available to 
multi-degree 18.

The geometry can be modified by successively adding tensor multiplets 
corresponding to small instantons. This can be done by blow-ups of 
$(-1)$ curves in the base. In the particular case of $E_8$ gauge 
symmetry this corresponds to the blowing up Hirzebruch surfaces 
$\mathbb{F}_n$, $n=11,10,9$. We can construct explicitly models 
in which all divisors have toric representatives, by modifying $\Delta^*$ 
in the following way:
\begin{itemize} 
 \item $n=11$ (one small instanton case): $\nu_F$ gets replaced by 
 two points ($\nu_F\rightarrow \{\{2,3,-1,-11\},\{0,0,0,-1\}\}$). Thus $h_{11} = 12$, and in  
 accord with the 6d anomaly condition we  find $\chi(M_3)=-900$;
 \item $n=10$: one must replace $\nu_F\rightarrow \{\{2,3,-1,-10\},\{0,0,0,-1\},\{2,3,1,-1\} \}$;
 \item $n=9$: here, $\nu_F\rightarrow \{\{2,3,-1,-9\},\{0,0,0,-1\},\{2,3,1,-1\}, \{2,3,1,-2\} \}$.
\end{itemize}

\section*{Acknowledgements}
We are grateful to Clay Cordova, Michele Del Zotto, Abhijit Gadde, Jonathan Heckman, Ben Heidenreich, Min-Xin Huang, Hans Jockers, Seok Kim, Denis Klevers, Maximilian Poretschkin, Martin Ro$\check{\text{c}}$ek, Tom Rudelius and Shu-Heng Shao for discussions. We would like to thank the SCGP for hospitality during the 2014 Simons Summer Workshop. The work of B.H. is supported by the NSF grant DMS-1159412. A.K is supported by the DFG grant KL 2271/1-1 and the NSF grant DMS-1159265. The work of C.V. is supported in part by NSF grant PHY-1067976.

\appendix
\section{Geometry of Landau-Ginzburg models}
\label{LG}
In this part of the Appendix we describe a mirror construction that applies directly to the  $(T^2\times \mathbb{C}^2)/\mathbb{Z}_k$, 
$k=3,4,6$ local orbifold using a Landau-Ginzburg  description of the torus and the LG/CY correspondence. 
We focus on the case $k=3$, which we compare in the next subsection with the constuction of the 
${\cal O}(-3)\rightarrow \mathbb{P}^1$ geometry discussed in section~\ref{f3geom}.

The weighted homogenous Landau-Ginzburg potential $W$ is the relevant term determining the 
CFT that corresponds to the Calabi-Yau sigma model at the infrared fixed point. 
The building blocks of $W$  can be  labeled by a simply laced Lie algebra. The   
$A_{k+1} $ case corresponds to the monomial $ x^{k+2}$; $D_{\frac{k+2}{2}} $ (with $ k $ even) corresponds to $ x_i^{\frac{k}{2}+1}+ x y^2$; finally, one has the following exceptional cases:
 $E_6:x^3+y^4$ (with $k=10$), $E_7:x^3+xy^3$ (with $k=16$), and $E_8:x^3+y^5$ (with $k=28$). 
Here the notation is so that $k+2$ is the Coxeter number of the Lie algebra and the 
contribution of each block to the central charge is
\be 
c=\frac{3 k}{k+2}.
\ee               
Note that quadratic terms do not contribute to the central charge. Each building block is identified 
with a {\sl minimal} $(2,2)$ superconformal theory with this central charge and the label of the Lie algebra 
specifies the way the left and the right sectors of the theory are glued to form a modular invariant  
partition function, a problem that enjoys an $ADE$ classification.  
When one identifies tensor products of LG models or minimal $(2,2)$ SCFT with the CY 
geometry one has to make a consistent projection on integral $U(1)_{L/R} $ 
charges and the correct spin structure. This involves orbifoldizations on 
the LG or minimal model side.      

The $c=3$ cases  correspond to elliptic curves on the Calabi-Yau side. 
There are three cases with only $A$ invariants,  
\bea 
 W_3&=& x_1^3+x_2^3+ x_3^3- 3 \alpha  x_1 x_2 x_3,\\ 
 W_4&=& x_1^4+x_2^4+ x_3^2- 4 \alpha  x_1 x_2 x_3,\\ 
 W_6&=& x_1^6+x_2^3+ x_3^2- 6 \alpha  x_1 x_2 x_3,
\label{ellipticcurves} 
\eea 
which have been identified  with $T^2/\mathbb{Z}_3,T^2/\mathbb{Z}_4$ and  $T^2/\mathbb{Z}_6$ 
orbifolds of the elliptic curve, see for example \cite{Lerche:1989cs}.

If one adds up three copies the above potential  to a $c=9$ LG potential 
and projects to integral charges one gets string vacua that correspond to the  
resolved  $(T^2)^3/(\mathbb{Z}_k\times \mathbb{Z}_k)$  CY manifolds. 
To get the simpler  $(T^2)^3/(\mathbb{Z}_k)$ orbifolds one has to mod out 
on the LG side by a further $\mathbb{Z}_k$. Aspects of the mirror description 
have been described  in~\cite{Candelas:1993nd}.

Let us consider for instance the $k=3$ case. The LG potential is   
\be 
W=\sum_{i=1}^9 x_i^3+ \sum_{i\neq j \neq k}  \alpha_{ijk} x_i x_j x_k ,
\ee       
where we listed the $\left(9\atop 3\right)=84$ independent complex structure 
deformations in the homogeneous degree 3 ring $\mathbb{C}[{\underline x}] /\{\partial_{x_i} W:i=1,\ldots,9\}$. 
The $(T^2)^3/(\mathbb{Z}_k\times \mathbb{Z}_k)$ orbifold has $84$ K\"ahler deformations and the Landau-Ginzburg 
model  can be  viewed as its mirror. Some aspects of the B-model have been analyzed 
in~\cite{Candelas:1993nd}. 
In particular one has a $(5,2)$-form in the sevenfold $W=0$ in $\mathbb{P}^8$ given by 
\be 
\Omega_{5,2}=\frac{1}{2 \pi i} \oint_{S^1} \frac{\mu_8}{W^3}\ ,
\label{g52}
\ee
where
\be 
\mu_{8}= \epsilon_{i_1,\ldots, i_9} x_{i_1} \dd x_{i1}\ldots \dd x_{i_9} \ .
\ee 
is a eight form, well defined under scaling $x_i\rightarrow \lambda x_i$. 
Integrating over the $S^1$ around $W=0$ makes it a $(5,2)$ form.

The orbifold that relates the LG model to the $(T^2)^3/\mathbb{Z}_3$ geometry  acts 
by
\be 
\left( \begin{array}{l}
x_i, i\in I \\
x_j, j\in J \\
x_k, k\in K \end{array}\right) \mapsto
\left( \begin{array}{l}
\alpha x_i, i\in I\\
\alpha^2 x_j, j\in J\\
x_k, k\in K\end{array}\right),
\ee 
where the sets are $I=\{1,2,3\}$,    $K=\{4,5,6\}$ and   $J=\{7,8,9\}$ and $\alpha$ is a 
third root of unity. The invariant monomials are $m_1=x_1 x_2 x_3$, $m_2=x_4 x_5 x_6$ and $m_3=x_7 x_8 x_9$ as well as the 
27 monomials $m_{ijk}=x_ix_jx_k$, where indices $i,j,k$  are in the sets $I,J,K$ respectively. 
Hence, we get the invariant LG potential
\be 
W=\sum_{i=1}^9 x_i^3- 3 \sum_{i=1}^3\alpha_i m_i-3 \sum_{i\in i,j\in J,k\in K}  \alpha_{ijk} x_i x_j x_k .
\ee  

On each $T^2$ of $(T^2)^3/\mathbb{Z}_3$ (which we parametrize by complex coordinates $z_i$, $i=1,2,3$), the $ Z_3 $ action has
three fixed points, corresponding to $z_i=0,\frac{1}{\sqrt{3}} \beta, $ and $ \frac{2}{\sqrt{3}} \beta$, where 
$\beta^{12}=1$, so that the global orbifold has $3^3=27$ 
$\mathbb{Z}_3$ fixed points. Locally it is given by 
$\mathbb{C}^3/\mathbb{Z}_3$, with the action of $ \mathbb{Z}_3 $ as in (\ref{Z3}). 
The resolved geometry near each fixed point looks locally like the total space 
${\cal O}(-3)\rightarrow \mathbb{P}^2$ of the anti-canonical bundle 
over $\mathbb{P}^2$. Here the $\mathbb{P}^2$ is the exceptional divisor $E_i$  
of the blow up. There are nine further K\"ahler classes in the invariant sector: the three  
invariant  $(1,1)$  forms $h_i=d z_i \wedge d\bar z_i$, $i=1,2,3$ and  
the six invariant $(1,1)$ forms $h_{ij}=d z_i  \wedge d\bar z_j$, $i\neq j$, 
$i,j=1,2,3$. If we denote the dual  divisors $H_i$ and $H_{ij}$ respectively 
one has the non-vanishing intersections
\be 
\begin{array}{rl}
H_1 \cdot H_2\cdot H_3 &=\kappa,\quad H_{ij}\cdot H_{ki} \cdot H_{jk}=\kappa,  \\ H_i \cdot H_{jk}\cdot H_{kj}&=-\kappa, \quad {\rm for} \ \  i \neq j, i \neq k \ ,  \quad 
E_i^3=\kappa, \quad i=1,\ldots 27.
\end{array}  
\ee
Here we have taken a normalization $\kappa=9$. Similarly, the intersection numbers of orbifolds 
with fixed tori have been calculated in~\cite{Lust:2006zh}. 

We know from the mirror map of the individual $T^2/\mathbb{Z}_3$'s, given by the cubic 
constraint $W_3=0$ in~(\ref{ellipticcurves}), that $\frac{1}{2 \pi i}\log(\alpha_i)\rightarrow i\infty$ 
corresponds to the large volume limit ${\rm Im} (\tau_i)\rightarrow \infty$ of the $i$'th $T^2$.  Hence the 
limit in which  $(T^2)^3/\mathbb{Z}_3$ becomes the  $T^2\times \mathbb{C}^2$ geometry involves taking, say, $\alpha_2,\alpha_3\rightarrow  \infty $, while keeping $\alpha \equiv \alpha_1$ finite. These 
limits can be taken individually and produce a term $\infty x_4 x_5 x_6$ or $\infty x_7 x_8 x_9$ in $W$, 
which requires taking, say, $x_5\rightarrow 0$ and $x_8\rightarrow 0$. The complex volumes of the $27$ 
$\mathbb{P}^2$'s in the local ${\cal O}(-3)\rightarrow \mathbb{P}^2$ geometies are parametrized in the 
large volume limit of the $\mathbb{P}^2$'s by $t_a\sim \frac{1}{ 2 \pi i} \log(\alpha_{ijk})$, $a=1,\ldots, 27$. 
We want to keep three of these finite, while $24$ of them should be scaled to infinite volume. Again this 
gives infinite terms in $W$, which can be eliminated by setting in addition $x_6=x_8=0$; that is,    
we keep only $x_{1},x_2,x_3,x_4,x_7$ finite. Then, $\frac{1}{2 \pi i} \log(\alpha_{i,4,7})\equiv \frac{1}{2 \pi i}
\log(\beta_i)\sim t_i $, $i=1,2,3$ parametrize the three finite $\mathbb{P}^2$s. Hence, we end up with a potential
\be 
W=x_1^3+x_2^2+x_3^3+x_4^3+x_7^3-3 \alpha x_1 x_2 x_3 - 3 \beta_1  x_1 x_4 x_7-3  \beta_2 x_2 x_4 x_7 -3 \beta_3 x_3 x_4 x_7.   
\label{resctriced} 
\ee 
Let us relabel coordinates and rewrite $W$  as
\be \begin{small}
W=a_0 x_1 x_2 x_3 + \sum_{i=1}^5 a_i x_i^3 + a_6 x_1 x_4 x_5+a_7 x_2 x_4 x_5 + a_8  x_3 x_4 x_5\equiv \sum_{i=0}^8 a_i Y_i. \end{small}
\ee 
We note that there are $k$ relations among the $Y_i$ given by   
\be \prod_i Y_i^{l^{(r)}_i}=1, \qquad r=1,\ldots,k ,
\ee
where $k=4$ and   
\be
\begin{array}{rl} 
l^{(1)}=&(-3;1,1,1,0,0,\pmi 0,\pmi 0,\pmi 0)\\ 
l^{(2)}=&(\pmi 0; 1,0,0,1,1,-3,\pmi 0,\pmi 0)\\
l^{(3)}=&(\pmi 0;0,1,0,1,1,\pmi 0,-3,\pmi 0)\\ 
l^{(4)}=&(\pmi 0;0,0,1,1,1,\pmi 0,\pmi 0,-3)\ . \\ 
\end{array}
\ee
The $Y_i$, $i=0,\ldots,4$ are $\mathbb{C}$ variables while the 
$Y_j$, $j=4,\ldots, 8$ are $\mathbb{C}^*$ variables. All $Y_i$  
are subject to a $\mathbb{C}^*$ action $Y_i\rightarrow \nu Y_i$ 
with  $\nu \in \mathbb{C}^*$. This yields a $(9-4-1-1) = 3$-dimensional 
local Calabi-Yau manifold which is the mirror to the  
$(T^2\times \mathbb{C}^2)/\mathbb{Z}_3$ manifold. 

Its four complex deformations  $z_r=(-1)^{l^{(r)}_0} \prod_{i}  a_i^{l^{(r)}_i}$, $r=1,
\ldots,4$, correspond respectively to the complexified volume of the $T^2$ and three lines 
in different exceptional $\mathbb{P}^2$'s. The local (3,0) form
is given by
\be 
\Omega_{3,0}=\frac{1}{2 \pi i} \oint \frac{\epsilon_{ijk}  x_i d x_j d x_k}{W}\frac{d x_4}{x_4} \frac{d x_5}{x_5}\qquad i,j,k=1,2,3  \ .
\label{l30}
\ee 
From local expression of the (3,0) form in  (\ref{l30}) or a limit of 
(\ref{g52}) one can see that the periods $\tilde\Pi$  in the local 
limit fulfill
\be 
\prod_{l^{(r)}_i<0} \partial^{l^{(r)}_i}_{a_i} \tilde \Pi= \prod_{l^{(r)}_i>0} \partial^{l^{(r)}_i}_{a_i} \tilde \Pi.
\ee 
Also, the periods have the scale invariances acting on the $a_i$ that allow one to eliminate 
the $a_i$ by the invariant combinations $z_r$. The periods $\Pi$ are therefore  
annihilated by the differential operators 
\be
\begin{array}{rl}
{\cal D}_1&=(\theta_1+\theta_2) (\theta_1+\theta_3) (\theta_1+\theta_4)+3 \theta_1 (3 \theta_1-2) (3 \theta_1-1) z_1\\ 
{\cal D}_2&=(\theta_1+\theta_2) (\theta_2+\theta_3+\theta_4)^2+3 (\theta_2-1) (3 \theta_2-1) (3 \theta_2-1) z_2 \\ 
{\cal D}_3&= (\theta_1+\theta_3) (\theta_2+\theta_3+\theta_4)^2+3 (\theta_3-1) (3 \theta_3-2) (3 \theta_3-1) z_3\\ 
{\cal D}_4&= (\theta_1+\theta_4) (\theta_2+\theta_3+\theta_4)^2+3 (\theta_4-1) (3 \theta_4-2) (3 \theta_4-1) z_4,\\ 
\end{array} 
\ee
where $\theta_i=z_i\frac{\dd}{\dd z_i}$. 
Note the following: \begin{itemize}
            \item We normalized the periods to $\Pi=a_0 \tilde \Pi$.
            \item The periods are not completely determined  
by the operators; that is, there are more functions annihilated by the ${\cal D}_i$ than periods.  
           \end{itemize}
Nevertheless, we can identify the relevant solutions by the Frobenius method. In this method, we define a 
$\underline{z}=(z_1,\ldots, k)$ and ${\underline \rho}=(\rho_1,\ldots, \rho_k)$ dependent function 
as 
\be
\omega({\underline  z},{\underline \rho})=
\sum_{\underline{n}} \left(\frac{\Gamma(\sum_{a} l^{(a)}_0(n_a+\rho_a)+1)}{\prod_{i>0} \Gamma(\sum_a l^{(a)}_i (n_a+\rho_a)+1)}\right) z^{\underline n+ \underline \rho}\ .     
\label{omega0} 
\ee
Using the fact that $[\partial_{\rho_a},{\cal D}_k]\sim 0$ and 
$\omega( {\underline  z},{\underline \rho}=0)$ is a solution, we get more 
solutions by taking derivatives with respect to the various $\rho_i$:      
\be 
\begin{array}{rl}
X^0&=\omega( {\underline  z},{\underline \rho})|_{\underline \rho=0}\ ,\\ 
X^r&=\partial_{\rho_r} \omega( {\underline  z},{\underline \rho})|_{\underline \rho=0}\ ,\\ 
\tilde F_r&=\frac{1}{2} c_{rij}  \partial_{\rho_i} \partial_{\rho_j}\omega( {\underline  z},{\underline \rho})|_{\underline \rho=0}\ ,\\ 
\tilde F_0&=\frac{1}{6} c_{ijk}  \partial_{\rho_i} \partial_{\rho_j} \partial_{\rho_k} \omega( {\underline  z},{\underline \rho})|_{\underline \rho=0}\ .\\  
\end{array} 
\label{frobenius} 
\ee
Here $r=1,\ldots, h_{11}(M_3)=h_{21}(W_3)$, and the $c_{ijk}$ are classical intersection numbers. Note that each derivative w.r.t. $\rho_i$ gives a $\log(z_i)$ term 
and that there is only a finite number $b_3(W_3)$ of solutions\footnote{In fact the notation $[\partial_{\rho_a},{\cal D}_k]\sim 0$ means that arbitrary 
derivatives are not all annihilated by the differential operators ${\cal D}_k$; rather, the result will be in general 
proportional to $\log(z_i)$ terms.}, which are given in (\ref{frobenius}).

The single logarithmic solutions define the mirror maps as 
\be 
t^k=\frac{1}{2 \pi i} \frac{X^k}{X^0} \ .
\ee
Mirror symmetry and special geometry implies that an integral sympletic basis of periods is given by
\begin{equation}
\Pi=X^0 \left(\begin{array}{c}
1\\ 
t^a\\ 
\partial_{t^a} F_0\\
2 F_0- t^a \partial_{t^a} F_0
\end{array}\right)\ ,
\label{periodsatinfinty}
\end{equation}
where the prepotential $F_0$ is determined in terms of the C.T.C Wall Data as 
\be 
F_0=X_0^2\left[-\frac{1}{6} c_{ijk} t^i t^j t^k+ \frac{1}{2} A_{ij} t^i t^j + c_{i} t^i - i\chi{\zeta(3)\over 2 (2 \pi)^3}+ \sum_{\underline n} n^{(0)}_{\underline n} { \rm Li}_3(\underline{Q}^{\underline n})\right] \ ,
\ee
with ${\rm Li}_k(x)=\sum_{n=1} \frac{x^n}{n^k}$, $\underline{Q}^{\underline n}=\prod_{i=1}^k \exp( 2 \pi i  t_i n_i)$, $c_i={1\over 24}\int_X \ch_2 J_i$ and 
$\chi$ is the Euler number of $M_3$. In particular, we can read off the genus 0 GV  invariants $n^{(0)}_{\underline n}$ already 
from the double logarithmic solutions
\be 
X^0\partial_{t^r} F_0\ . 
\ee
In the example under consideration, one gets in particular the period $X^0$ and, by taking single derivatives 
with respect to $\rho_i$, one gets $k$ further periods, which are to low orders in $z$ given by   
\be 
\begin{array}{rl}
X^0(z_1)=&1 + 6 z_1 + 90 z_1^2 + 1680 z_1^3 + 34650 z_1^4+{\cal O}(z_1^5)\\ 
X^1(z_1)=& X^0 \log(z_1)+  15 z_1 + (513 z_1^2)/2 + 5018 z_1^3 +  {\cal O}(z_1^4)\\
X^k(z_1,z_k)=& X^0 \log(z_k) -6  (z_1 + z_k) + ( 45 z_k^2-135 z_1^2 - 18 z_1 z_k )\\ & 
+ (90 z_1 z_k^2 -3080 z_1^3 - 180 z_1^2 z_k - 560 z_k^3)+ {\cal O}(z^4), \quad k=2,3,4. \\
\end{array} 
\ee
In the local limit, only the intersections 
$E_i^3\sim \kappa_{lim}$ contribute, with an appropriate 
normalization. We have calculated the genus zero BPS numbers  using  the description above 
and checked that they agree with the ones  calculated in the decompatification limit 
of the  Calabi-Yau threefold  discussed in section~\ref{f3geom} in the basis discussed 
in the next section.

\subsection*{Toric Geometry realization of the ${\cal O}(-3)\rightarrow \mathbb{P}^1$ geometry}

Here we want to obtain the geometry discussed in the last section from the toric polyhedron specified in 
Table \ref{tab:dataf3}. It has $10$ star triangulations. We list only one Mori cone with the intersections   
\be
\begin{array}{rl}
{\cal R}=&1791 J_1 ^3 + 957 J_1 ^2 J_2  + 511 J_1  J_2 ^2 + 272 J_2 ^3 + 180 J_1 ^2 J_3  + \\ &
96 J_1  J_2  J_3  + 51 J_2 ^2 J_3  + 18 J_1  J_3 ^2 + 9 J_2  J_3 ^2 + 360 J_1 ^2 J_4  + \\ &
 192 J_1  J_2  J_4  + 102 J_2 ^2 J_4  + 36 J_1  J_3  J_4  + 19 J_2  J_3  J_4  + 3 J_3 ^2 J_4  + 72 J_1  J_4 ^2 + \\ &
 38 J_2  J_4 ^2 + 7 J_3  J_4 ^2 + 14 J_4 ^3 + 900 J_1 ^2 J_5  + 480 J_1  J_2  J_5  + 256 J_2 ^2 J_5  + 90 J_1  J_3  J_5  + \\ &
 48 J_2  J_3  J_5  + 9 J_3 ^2 J_5  + 180 J_1  J_4  J_5  + 96 J_2  J_4  J_5  + 18 J_3  J_4  J_5  + 36 J_4 ^2 J_5  + 450 J_1  J_5 ^2+\\ &  
 240 J_2  J_5 ^2 + 45 J_3  J_5 ^2 + 90 J_4  J_5 ^2 + 225 J_5 ^3.
\end{array}
\ee
The evaluation of the second Chern class is given by $\int c_2 J_i=\{570,308,60,116,282\}$, and the model has no
particular fibration structure except for the elliptic one.

In relation to the Landau-Ginzburg formulation of the $Z$ orbifold,  the  following classes in the given base,
\begin{equation}
\begin{array}{rl}  
l_{T^2}=&6 l^{(1)}+3 l^{(2)}+l^{(4)} +3 l^{(5)}, \ \  l_{\mathbb{P}^2}^{(1)}=3 l^{(1)}+ l^{(2)}\\
l_{\mathbb{P}^2}^{(2)}=&l^{(2)}, \ \  l_{\mathbb{P}^2}^{(3)}=l^{(2)}+l^{(4)}, \ \ l_{de}=l^{(3)}+\frac{1}{3} (l^{(2)}+l^{(5)}),\\ 
\end{array}
\end{equation}
are of particular interest, because for them the intersection form becomes symmetric in the three classes of 
the $\mathbb{P}^2$'s discussed in Section \ref{LG}:
\be
{\cal R}=\frac{25}{3} J_{T^2}^3-\frac{1}{3}(J_{\mathbb{P}^2_1}^3+J_{\mathbb{P}^2_2}^3+J_{\mathbb{P}^2_3}^3)
+ 5  J_{T^2}^2 J_{de}+ 3 J_{T^2} J^2_{de}\ .
\ee
Here $J_{de}$ is the class that needs to be decompactified to obatain a non-compact geometry.
  
The curve classes, which correspond to the Mori cone  vector  $l_{\mathbb{P}^2}^{(i)}$ generate all the  
BPS states of ${\cal O}(-3)\rightarrow \mathbb{P}^2$. For example, at genus zero  
\begin{equation} 
\{ n^{(0)}_{3k,k,0,0,0}\} = \{ n^{(0)}_{0,k,0,0,0}\}= \{ n^{(0)}_{0,k,0,k,0}\}=\{ 3,-6,27,-192, 1695, -17064,\ldots \}.
\end{equation}
These  curves lie in the divisor classes $D_3,D_4,D_5$, which consist of an exceptional curve 
over the $-3$ curve in the base $S'$ with $(S')^2=-3$, a section of the Hirzebruch surface $\mathbb{F}_3$ fiber. 
The class  $l_{T^2}$ is the class of the elliptic fiber.  We find  that in this direction all 
BPS numbers are $ \{ n^{(0)}_{6k,3k,0,k,3k }\}=\{ 492,492,\ldots \}$.  Note that 492  is the 
Euler number of the Calabi-Yau and this is the first modular direction. This follows from the 
representation of the Eisenstein series $E_4$  as 
\be 
E_4(q)=1+240 \sum_{k=1}^\infty \frac{k^3 q^k}{1-q^k} 
\ee 
and from the multicovering formula for $g=0$, which reads 
\be 
F_0=F_0^{classical} + \sum_{d=1}^\infty n_d^{(0)} {\rm  Li}_3(q^d),
\ee
with ${\rm Li}_k(x)\sum_{n=1}^\infty \frac{x^d}{k^d}$. From this, one gets 
\be 
\partial_t^3 F_0=c +\sum_{d=1}^\infty \frac{n_d d^3 q^d}{1-q^d}=\frac{240 c-\chi}{240} +\frac{\chi}{240} E_4(q)  
\ee 

Let us denote by $d_{T^2}$ the degree of the elliptic curve, and by $d_1,d_2,d_3$ the degrees with respect to 
the  three ${\cal O}(-3)\rightarrow \mathbb{P}^2$ classes $L_i$, $i=1,2,3$, and denote now the BPS invariants by 
$n^{(0)}_{d_{T^2},d_1,d_2,d_3}$.  We have the obvious 
property that  $n^{(0)}_{d_{T^2},d_1,d_2,d_3}$ depends symmetrically on the $d_i$. 
We can study the  mixing of the local $\mathbb{P}^2$ with the fiber. At $d_{T^2}=0$ we have no mixing between the $L_i$ classes:
\begin{table}[H]
\centering
\footnotesize{\begin{tabular} {|c|c|c|c|c|c|} 
\hline 
$d_1 \backslash d_2$
   & 0 & 1 & 2  & 3  & 4   \\  \hline  
0  & $\frac{i 492\zeta(3)}{(2 \pi)^3}$ & 3  & -6  & 27   & -192    \\  \hline
1  & 3 & 0  & 0 & 0 & 0  \\  \hline
2  & -6  & 0  & 0 & 0 & 0    \\  \hline
3  & 27 & 0 & 0 & 0 & 0    \\  \hline 
4  & -192  & 0 & 0 & 0 & 0     \\  \hline 
\end{tabular}} 
\vskip 3pt   $n_{d_{T^2}=0,d_1,d_2,0}$ 
\end{table}

In general $n_{0,d_1,d_2,d_3}=0$ if more then 
two $d_i$ are non zero. This is of obvious 
from the geometry, since the blow up points sit at 
distiguished points of the fiber and are 
uncorrelated as long there is no curve wrapping 
the fiber.     

From $d_{T^2}>1$ the mixing starts. In particular, one finds:

\begin{table}[H]
\centering
\footnotesize{\begin{tabular} {|c|c|c|c|c|c|} 
\hline 
$d_1 \backslash d_2$
   & 0 & 1 & 2  & 3  & 4   \\  \hline  
0  & 492  & 36   & -360   & 4752   & -70560   \\  \hline
1  & 36 & -216   & 2052  &-26082  &  376704 \\  \hline
2  & 360  & 2052 & -17760 & 211140 &  -2912544    \\  \hline
3  &  4752, &-26082& 211140&  -2378484& 31525200      \\  \hline 
4  & -70560&  376704&  -2912544&  31525200& -405029376  \\  \hline 
\end{tabular}} 
\vskip 3pt   $n^{(0)}_{1,0,d_1,d_2}$ 
\end{table}

\begin{table}[H]
\centering
\footnotesize{\begin{tabular} {|c|c|c|c|c|c|} 
\hline 
$d_1 \backslash d_2$ &  1 & 2  & 3  & 4   \\  \hline    
1  &   1458& -12654& 150903& -2087856 \\  \hline
2  &  -12654& 103536& -1177686& 15735024 \\  \hline
3  &  150903& -1177686& 12859560& -1664394480  \\  \hline
4  & -2087856& 15735024& -166439448& 2099613312     \\  \hline 
\end{tabular}} 
\vskip 3pt   $n^{(0)}_{1,1,d_1,d_2}$ 
\end{table}

\begin{table}[H]
\centering
\footnotesize{\begin{tabular} {|c|c|c|c|c|c|} 
\hline 
$d_1 \backslash d_2$ & 2  & 3  & 4   \\  \hline  
2  & -812808& 8923104& -115996032\\  \hline
3  &  8923104& -94862502 & 1201724208 \\  \hline
4  & -115996032& 1201724208&   -14901588864    \\  \hline 
\end{tabular}} 
\vskip 3pt   $n^{(0)}_{1,2,d_1,d_2}$ 
\end{table} 

\begin{table}[H]
\centering
\footnotesize{\begin{tabular} {|c|c|c|c|c|c|} 
\hline 
$d_1 \backslash d_2$ & 0 & 1 & 2  & 3     \\  \hline  
0   & 492& 288& -10656 & 346356 \\  \hline  
1  &  288& -8604 & 225234& -5852520  \\  \hline
2  &  -10656& 225234& -4648248& 102706623\\  \hline
3  &   346356&  -5852520& 102706623& -2009199816\\  \hline 
\end{tabular}} 
\vskip 3pt   $n^{(0)}_{2,0,d_1,d_2}$ 
\end{table}

\begin{table}[H]
\centering
\footnotesize{\begin{tabular} {|c|c|c|c|c|c|} 
\hline 
$d_1 \backslash d_2$ & 0 & 1 & 2      \\  \hline  
0   & 492& 1788& -197568 \\  \hline  
1  &  1788& -76724 & -  \\  \hline
2  &  -197568& -& -\\  \hline
\end{tabular}} 
\vskip 3pt   $n^{(0)}_{3,0,d_1,d_2}$ 
\end{table}


\begin{thebibliography}{10}
\newcommand{\wwwspires}{http://www.slac.stanford.edu/spires/find/hep/www}

\bibitem{Heckman:2013pva} 
  J.~J.~Heckman, D.~R.~Morrison and C.~Vafa,
  ``On the Classification of 6D SCFTs and Generalized ADE Orbifolds,''
  JHEP {\bf 1405}, 028 (2014)
  [arXiv:1312.5746 [hep-th]].
  
\bibitem{Gaiotto:2014lca} 
  D.~Gaiotto and A.~Tomasiello,
  ``Holography for (1,0) theories in six dimensions,''
  arXiv:1404.0711 [hep-th].
  
\bibitem{DelZotto:2014hpa} 
  M.~Del Zotto, J.~J.~Heckman, A.~Tomasiello and C.~Vafa,
  ``6d Conformal Matter,''
  arXiv:1407.6359 [hep-th].

\bibitem{Haghighat:2013gba}
  B.~Haghighat, A.~Iqbal, C.~Kozcaz, G.~Lockhart and C.~Vafa,
  ``M-Strings,''
  arXiv:1305.6322 [hep-th].
  
\bibitem{Haghighat:2013tka} 
  B.~Haghighat, C.~Kozcaz, G.~Lockhart and C.~Vafa,
  ``On orbifolds of M-Strings,''
  Physical Review D 89.4 (2014): 046003
  [arXiv:1310.1185 [hep-th]].
  
\bibitem{Kim:2014dza} 
  J.~Kim, S.~Kim, K.~Lee, J.~Park and C.~Vafa,
  ``Elliptic Genus of E-strings,''
  arXiv:1411.2324 [hep-th].
  
\bibitem{Morrison:2012np} 
  D.~R.~Morrison and W.~Taylor,
  ``Classifying bases for 6D F-theory models,''
  Central Eur.\ J.\ Phys.\  {\bf 10}, 1072 (2012)
  [arXiv:1201.1943 [hep-th]].
  
\bibitem{Witten:1996qb} 
  E.~Witten,
  ``Phase transitions in M theory and F theory,''
  Nucl.\ Phys.\ B {\bf 471}, 195 (1996)
  [hep-th/9603150].

\bibitem{Witten:1995gx} 
  E.~Witten,
  ``Small instantons in string theory,''
  Nucl.\ Phys.\ B {\bf 460}, 541 (1996)
  [hep-th/9511030].
  
\bibitem{Ganor:1996mu} 
  O.~J.~Ganor and A.~Hanany,
  ``Small E(8) instantons and tensionless noncritical strings,''
  Nucl.\ Phys.\ B {\bf 474}, 122 (1996)
  [hep-th/9602120].
  
\bibitem{Seiberg:1996vs} 
  N.~Seiberg and E.~Witten,
  ``Comments on string dynamics in six-dimensions,''
  Nucl.\ Phys.\ B {\bf 471}, 121 (1996)
  [hep-th/9603003].
  
\bibitem{Klemm:1996hh} 
  A.~Klemm, P.~Mayr and C.~Vafa,
  ``BPS states of exceptional noncritical strings,''
  Nucl.\ Phys.\ Proc.\ Suppl.\  {\bf 58}, 177 (1997)
  [hep-th/9607139].

\bibitem{Sen:1997gv} 
  A.~Sen,
  ``Orientifold limit of F theory vacua,''
  Phys.\ Rev.\ D {\bf 55}, 7345 (1997)
  [hep-th/9702165].
  
\bibitem{Huang:2013yta} 
  M.~X.~Huang, A.~Klemm and M.~Poretschkin,
  ``Refined stable pair invariants for E-, M- and $[p, q]$-strings,''
  JHEP {\bf 1311}, 112 (2013)
  [arXiv:1308.0619 [hep-th]].  
  
\bibitem{Vafa:1996xn} 
  C.~Vafa,
  ``Evidence for F theory,''
  Nucl.\ Phys.\ B {\bf 469}, 403 (1996)
  [hep-th/9602022].
  
\bibitem{Morrison:1996na} 
  D.~R.~Morrison and C.~Vafa,
  ``Compactifications of F theory on Calabi-Yau threefolds. 1,''
  Nucl.\ Phys.\ B {\bf 473}, 74 (1996)
  [hep-th/9602114].
  
  \bibitem{Morrison:1996pp} 
  D.~R.~Morrison and C.~Vafa,
  ``Compactifications of F theory on Calabi-Yau threefolds. 2.,''
  Nucl.\ Phys.\ B {\bf 476}, 437 (1996)
  [hep-th/9603161].

  \bibitem{Gadde:2013dda} 
  A.~Gadde and S.~Gukov,
  ``2d Index and Surface operators,''
  JHEP {\bf 1403}, 080 (2014)
  
\bibitem{Benini:2013xpa} 
  F.~Benini, R.~Eager, K.~Hori and Y.~Tachikawa,
  ``Elliptic genera of 2d N=2 gauge theories,''
  arXiv:1308.4896 [hep-th].
  
\bibitem{Benini:2013nda} 
  F.~Benini, R.~Eager, K.~Hori and Y.~Tachikawa,
  ``Elliptic genera of two-dimensional N=2 gauge theories with rank-one gauge groups,''
  Lett.\ Math.\ Phys.\  {\bf 104}, 465 (2014)
  [arXiv:1305.0533 [hep-th]].
  
\bibitem{Candelas:1994hw} 
  P.~Candelas, A.~Font, S.~H.~Katz and D.~R.~Morrison,
  ``Mirror symmetry for two parameter models. 2.,''
  Nucl.\ Phys.\ B {\bf 429}, 626 (1994)
  [hep-th/9403187].

\bibitem{Uranga:1999mb} 
  A.~M.~Uranga,
  ``A New orientifold of C**2 / Z(N) and six-dimensional RG fixed points,''
  Nucl.\ Phys.\ B {\bf 577}, 73 (2000)
  [hep-th/9910155].
  
  \bibitem{Douglas:1996sw} 
  M.~R.~Douglas and G.~W.~Moore,
  ``D-branes, quivers, and ALE instantons,''
  hep-th/9603167.
  
\bibitem{Gimon:1996rq} 
  E.~G.~Gimon and J.~Polchinski,
  ``Consistency conditions for orientifolds and d manifolds,''
  Phys.\ Rev.\ D {\bf 54}, 1667 (1996)
  [hep-th/9601038].
  
  \bibitem{JeffreyKirwan}
 	L.~C.~Jeffrey and F.~C.~Kirwan, 
 	``Localization for nonabelian group actions,”
	Topology 34 no. 2, (1995) 291–327, 
	arXiv:alg-geom/9307001
   	
\bibitem{Gadde:2014ppa} 
  A.~Gadde, S.~Gukov and P.~Putrov,
  ``Exact Solutions of 2d Supersymmetric Gauge Theories,''
  arXiv:1404.5314 [hep-th]].
  
\bibitem{Hosono:1999qc} 
  S.~Hosono, M.~H.~Saito and A.~Takahashi,
  ``Holomorphic anomaly equation and BPS state counting of rational elliptic surface,''
  Adv.\ Theor.\ Math.\ Phys.\  {\bf 3}, 177 (1999)
  [hep-th/9901151].
  
\bibitem{Gopakumar:1998jq}   
  R.~Gopakumar and C.~Vafa,
  ``M theory and topological strings. 2.,''
  hep-th/9812127.
  
\bibitem{Hollowood:2003cv} 
  T.~J.~Hollowood, A.~Iqbal and C.~Vafa,
  ``Matrix models, geometric engineering and elliptic genera,''
  JHEP {\bf 0803}, 069 (2008)
  [hep-th/0310272].
  
\bibitem{Huang:2010kf} 
  M.~x.~Huang and A.~Klemm,
  ``Direct integration for general $\Omega$ backgrounds,''
  Adv.\ Theor.\ Math.\ Phys.\  {\bf 16}, no. 3, 805 (2012)
  [arXiv:1009.1126 [hep-th]].
  
  \bibitem{Klemm:1996ts} 
  A.~Klemm, B.~Lian, S.~S.~Roan and S.~T.~Yau,
  ``Calabi-Yau fourfolds for M theory and F theory compactifications,''
  Nucl.\ Phys.\ B {\bf 518}, 515 (1998)
  [hep-th/9701023].
 
 \bibitem{Green:1984bx} 
  M.~B.~Green, J.~H.~Schwarz and P.~C.~West,
  ``Anomaly Free Chiral Theories in Six-Dimensions,''
  Nucl.\ Phys.\ B {\bf 254}, 327 (1985).

\bibitem{Erler:1993zy} 
  J.~Erler,
  ``Anomaly cancellation in six-dimensions,''
  J.\ Math.\ Phys.\  {\bf 35}, 1819 (1994)
  [hep-th/9304104].
 
   \bibitem{Kachru:1997bz} 
  S.~Kachru, A.~Klemm and Y.~Oz,
  ``Calabi-Yau duals for CHL strings,''
  Nucl.\ Phys.\ B {\bf 521}, 58 (1998)
  [hep-th/9712035].
  
\bibitem{Bizet:2014uua} 
  N.~C.~Bizet, A.~Klemm and D.~V.~Lopes,
  ``Landscaping with fluxes and the E8 Yukawa Point in F-theory,''
  arXiv:1404.7645 [hep-th].
  
  \bibitem{Tate} 
    J.~Tate, 
    "{Algorithm for Determining the Type of a Singular Fibre in an Elliptic Pencil}",
    in {\sl Modular Functions of one variable IV}, Lecture Notes in mathematics {\bf 476},
    Springer-Verlag, Berlin (1975).

\bibitem{CTCWall} C.~T.~C.~Wall, ``Classification problems in differential topology. {V}. {O}n
              certain {$6$}-manifolds'', Invent. Math. 1 (1966), 355-374; corrigendum, ibid.
 
\bibitem{Batyrev:1994hm} 
  V.~V.~Batyrev,
  ``Dual polyhedra and mirror symmetry for Calabi-Yau hypersurfaces in toric varieties,''
  J.\ Alg.\ Geom.\  {\bf 3}, 493 (1994)
  [alg-geom/9310003].

\bibitem{Hosono:1993qy} 
  S.~Hosono, A.~Klemm, S.~Theisen and S.~T.~Yau,
  ``Mirror symmetry, mirror map and applications to Calabi-Yau hypersurfaces,''
  Commun.\ Math.\ Phys.\  {\bf 167}, 301 (1995)
  [hep-th/9308122].

\bibitem{Witten:1993yc} 
  E.~Witten,
  ``Phases of N=2 theories in two-dimensions,''
  Nucl.\ Phys.\ B {\bf 403}, 159 (1993)
  [hep-th/9301042].
  
  \bibitem{Oguiso} K. Oguiso, ``On algebraic fiber space structures on a {C}alabi-{Y}au {$3$}-fold,''
Internat. J. Math. 3 (1993) 439--465.
 
 \bibitem{Friedman:1997yq} 
  R.~Friedman, J.~Morgan and E.~Witten,
  ``Vector bundles and F theory,''
  Commun.\ Math.\ Phys.\  {\bf 187}, 679 (1997)
  [hep-th/9701162].
  
\bibitem{Lerche:1989cs}
  W.~Lerche, D.~Lust and N.~P.~Warner,
  ``Duality Symmetries in $N=2$ Landau-Ginzburg Models,''
  Phys.\ Lett.\ B {\bf 231} (1989) 417.
 
\bibitem{Candelas:1993nd} 
  P.~Candelas, E.~Derrick and L.~Parkes,
  ``Generalized Calabi-Yau manifolds and the mirror of a rigid manifold,''
  Nucl.\ Phys.\ B {\bf 407}, 115 (1993)
  [hep-th/9304045].
  
\bibitem{Lust:2006zh} 
  D.~Lust, S.~Reffert, E.~Scheidegger and S.~Stieberger,
  ``Resolved Toroidal Orbifolds and their Orientifolds,''
  Adv.\ Theor.\ Math.\ Phys.\  {\bf 12}, 67 (2008)
  [hep-th/0609014].
                           
\end{thebibliography}
\end{document}

8